\documentclass[12pt,a4paper]{article}
\usepackage[latin1]{inputenc}
\usepackage{amsmath}
\usepackage{amsfonts}
\usepackage{amssymb}
\usepackage{graphicx,a4wide}
\usepackage{verbatim,cite}
\usepackage{tikz}
\usepackage{float}
\usepackage{enumerate}

\usepackage{color}
\usepackage{mathrsfs}

\usepackage{amsmath}
\usepackage{setspace}

\usepackage[hidelinks]{hyperref}

\def\cO{{\mathcal O}}
\def\cN{{\mathcal N}}
\def\cP{{\mathcal P}}
\def\ula{{\underline{\smash \lambda}}}

\def\hpiccolo{ {\it h}}
\def\Wloop{ {\it W}}
\def\Rem{ {\it R}}

\def\ssigma{ \hat{\sigma} }
\def\ttau{ \hat{\tau}}

\def\cK		{\mathcal{K}}

\newcommand{\beq}{\begin{equation}}
\newcommand{\eeq}{\end{equation}}
\newcommand{\bea}{\begin{eqnarray}}
\newcommand{\eea}{\end{eqnarray}}

\def \be  {\begin{equation}}
\def \ee  {\end{equation}}
\def \ba  {\begin{eqnarray}}
\def \ea  {\end{eqnarray}}

\def\x{x}
\def\xb{\bar{x}}
\def\dbar#1{\bar{D}_{#1}}

\begin{document}

\thispagestyle{empty}

\null\vskip-43pt \hfill
\begin{minipage}[t]{30mm}
\end{minipage}

\null\vskip-12pt \hfill  \\
\null\vskip-12pt \hfill   \\

\vskip2.2truecm
\begin{center}
	\vskip 0.2truecm {\Large\bf
		{\Large One-loop amplitudes in AdS$_5\times$S$^5$ supergravity\\[.2cm] from $\mathcal{N}=4$ SYM at strong coupling}
	}\\
	\vskip 1truecm
	{\bf F.~Aprile${}^{1}$, J.~M. Drummond${}^{2}$, P.~Heslop${}^{3}$, H.~Paul${}^{2}$ \\
	}
	
	\vskip 0.4truecm
	
	{\it
		${}^{1}$ Dipartimento di Fisica, Universit\`a di Milano-Bicocca \& INFN, 
		Sezione di Milano-Bicocca, I-20126 Milano,\\
		\vskip .2truecm }
	\vskip .2truecm
	{\it
		${}^{2}$ School of Physics and Astronomy and STAG Research Centre, \\
		University of Southampton,
		Highfield,  SO17 1BJ,\\
		\vskip .2truecm                        }
	\vskip .2truecm
	{\it
		${}^{3}$ Mathematics Department, Durham University, \\
		Science Laboratories, South Rd, Durham DH1 3LE \vskip .2truecm                        }
\end{center}

\vskip 1truecm 
\centerline{\bf Abstract} 

We explore the structure of maximally supersymmetric Yang-Mills correlators in the supergravity regime. 
We develop an algorithm to construct one-loop supergravity amplitudes of four arbitrary Kaluza-Klein supergravity states, 
properly dualised into single-particle operators.
We illustrate this algorithm by constructing new explicit results for multi-channel correlation functions,  
and we show that correlators which are degenerate at tree level become distinguishable at one-loop. 
The algorithm contains  a number of subtle features which have not appeared until now. 
In particular, we address the presence of non-trivial low twist protected operators 
in the OPE that are crucial for obtaining the correct one-loop results. 
Finally, we outline how the differential operators
$\widehat{\mathcal{D}}_{pqrs}$ and $\Delta^{(8)}$, 
which play a role in the context of the hidden 10d conformal symmetry at tree level, 
can be used to reorganise our one-loop correlators.

\medskip

\noindent

\newpage
\setcounter{page}{1}\setcounter{footnote}{0}
\tableofcontents


\section{Introduction and Summary} \label{Intro}


Recently there has been significant progress in probing the structure of quantum gravity in the context of the AdS/CFT correspondence. 
This has been achieved by combining the effectiveness of the large $N$ expansion and the power of CFT techniques. 
In particular the large $N$ expansion in $\mathcal{N}=4$ super Yang-Mills theory at large 't Hooft coupling 
has been investigated in \cite{Rastelli:2016nze,Alday:2017xua,Aprile:2017bgs,unmixing,Rastelli:2017udc,Alday:2017vkk,
Aprile:2017qoy,Aprile:2018efk,Caron-Huot:2018kta,Alday:2018pdi,Alday:2018kkw,Goncalves:2019znr}. Natural objects of study in this 
context are the four-point functions $\langle \mathcal{O}_{p_1}\mathcal{O}_{p_2}\mathcal{O}_{p_3}\mathcal{O}_{p_4}\rangle$ 
of half-BPS operators $\mathcal{O}_{p}$ which are dual to the scattering processes of four supergravity states of type IIB supergravity on the AdS${}_5\times$S${}^5$ background.

In \cite{Aprile:2017bgs} we were able to obtain the full $1/N^4$ contribution to the correlator 
$\langle \mathcal{O}_2 \mathcal{O}_2 \mathcal{O}_2 \mathcal{O}_2 \rangle$, 
i.e. the one-loop contribution to the four-point amplitude of AdS graviton supermultiplets.\footnote{up to a single ambiguity which was fixed recently in \cite{Chester:2019pvm}. }
This was achieved by promoting the leading logarithmic discontinuity to a crossing-invariant function.
The leading logarithmic singularity itself was deduced in \cite{Alday:2017xua,Aprile:2017bgs} by
the consistency of the operator product expansion (OPE), after
resolving the tree-level mixing of long double-trace operators in the singlet $su(4)$ representation \cite{unmixing}. 
In fact the leading $1/N^2$ corrections to the spectrum of double-trace operators can be completely solved with surprisingly 
simple rational functions of the quantum numbers \cite{Aprile:2018efk}. As observed in \cite{Aprile:2018efk}, the spectrum 
exhibits a partial degeneracy which motivated the discovery of a surprising ten-dimensional conformal symmetry governing tree-level AdS${}_5\times$S${}^5$ supergravity \cite{Caron-Huot:2018kta}.
 In \cite{Aprile:2017qoy} we were able to perform a similar analysis for the amplitude of two graviton 
 supermultiplets and two Kaluza-Klein states, $\langle \mathcal{O}_2 \mathcal{O}_2 \mathcal{O}_3 \mathcal{O}_3 \rangle$. 
Both cases involved surprisingly simple analytic functions based essentially on the two-loop four-dimensional ladder integral.

The approach outlined above does not make any reference to actual one-loop diagrams of IIB supergravity on AdS${}_5\times$S${}^5$, and in fact
this computation in the bulk remains very challenging. 
Instead, scalar theories on AdS at one-loop have been discussed in many references, 
for example, see \cite{Aharony:2016dwx,Yuan:2018qva,Bertan:2018afl,Ghosh:2018bgd,Ponomarev:2019ofr,Carmi:2019ocp}. 
Our approach here uses CFT techniques to extract data in the dual theory, $\mathcal{N}=4$ SYM, and it is complemented with an 
understanding of the possible analytic structure of the one-loop correlators, as functions in position space. 
Similar approaches to half-BPS correlators have been applied also in perturbation theory, 
both from the point of view of particular diagrams (or integrands) e.g. \cite{Bourjaily:2011hi,Eden:2012tu,Chicherin:2015bza,Chicherin:2015edu,Bourjaily:2016evz,Chicherin:2018avq} 
and using the analytic structure of explicitly evaluated loop integrals \cite{Drummond:2013nda,Chicherin:2015edu}. 
It is natural to ask therefore if the large $N$ bootstrap can be applied to arbitrary charge half-BPS operators.


In this paper we solve algorithmically the analytic bootstrap program for the four-point one-loop amplitudes 
of generic single-particle Kaluza-Klein states. This computation presents itself as a significant challenge 
compared to our previous constructions in \cite{Aprile:2017bgs} and \cite{Aprile:2017qoy}. Indeed, 
the one-loop correlators constructed so far had at least two AdS graviton multiplet insertions, and therefore 
had some built-in physical simplicity, stemming from the fact that the OPE of two graviton multiplets
runs over a special set of both protected and long operators. 
In general, this simplicity is absent and we have to face a network of complications, which we will solve in this paper.  

First we recall that the $1/N$ expansion naturally stratifies the four-point amplitude in powers of $\log$s 
of the cross-ratio $u$, and it leads to an expansion of the following form, 
\be
\label{largeNexp}
\langle \mathcal{O}_{p_1}\mathcal{O}_{p_2}\mathcal{O}_{p_3}\mathcal{O}_{p_4}\rangle 
			= \mathcal{G}_{\vec{p};0,0} +  \frac{1}{N^{2}} \sum_{n=0}^1 (\log u)^n\, \mathcal{G}_{\vec{p};1,n} +  \frac{1}{N^{4}} \sum_{n=0}^2 (\log u)^n\, \mathcal{G}_{\vec{p};2,n}+\ldots
\ee
where
\begin{align}
	\vec{p}={(p_1,p_2,p_3,p_4)}
\end{align}
comprises the external charges. 
The expansion in (\ref{largeNexp}) goes together with an expansion 
in the large 't Hooft coupling  $\lambda = g^2 N$. The string corrections to the above expansion have been 
addressed recently in a number of papers \cite{Goncalves:2014ffa,Alday:2018pdi,Binder:2019jwn,Drummond:2019odu} 
but here we will restrict ourselves to the terms of order $\lambda^0$ corresponding to supergravity contributions. 
In very general terms, the consistency of the OPE places strong constraints on the various different functions $\mathcal{G}_{n,m}$. 
We shall now explain how this abstract information, embedded in the $1/N$ expansion, can be used in practice to organise our bootstrap program.

Consider the OPE of single-particle operators $(\mathcal{O}_{p_i} \times \mathcal{O}_{p_j})$, it
contains superconformal primary operators $\mathcal{O}$ of twist $\tau$, spin $l$ and $su(4)$ representation $[a,b,a]$,
\be
\mathcal{O}_{p_i} \times \mathcal{O}_{p_j} = \sum_{\mathcal{O}} C_{p_i p_j}(\mathcal{O}_{\vec{\tau}}) \, \mathcal{O}_{\vec{\tau}} \,.
\ee
where  $\vec{\tau} \equiv  (\tau,l,[aba])$ is a compact notation for the representation labels. 
A key point is that a four-point function, $\langle \mathcal{O}_{p_1}\mathcal{O}_{p_2}\mathcal{O}_{p_3}\mathcal{O}_{p_4}\rangle$ is determined non-perturbatively by summing over 
the OPE coefficients $C_{p_1p_2}(\mathcal{O}_{\vec{\tau}}) C_{p_3p_4}(\mathcal{O}_{\vec{\tau}})$ of common exchanged operators $\mathcal{O}_{\vec{\tau}}$.

Of particular importance for us will be the exchanged two-particle (or double-trace) operators, which have the schematic form, 
\be
\label{doubletrace}
\mathcal{O}_{pq;\vec{\tau}} = \mathcal{O}_p \Box^{\frac{1}{2}(\tau-p-q)} \partial^l \mathcal{O}_q \Big|_{[aba]}\,.
\ee
Such operators fall into different series according to their quantum numbers. 
Half-BPS operators have $l=a=0$ and $\tau=b=p+q$. Semishort operators have $\tau = 2a+b+2=p+q$ and spin $l\ge 0$. 
 In both these cases the $\Box$ is necessarily absent. 
Long operators will generically obey the unitarity bound $\tau \geq 2a+b+2$, but 
long operators of the form (\ref{doubletrace}) actually obey $\tau \geq 2a+b+4$. Notice that $\tau$ might be greater than $p+q$ in this case.

In a given $su(4)$ representation $[aba]$, we can organise semishort and long operators $\mathcal{O}_{pq;\vec{\tau}}$ into a tower, whose levels are labelled by the twist. 
The bottom of the tower corresponds to the unitarity bound. 
For each operator $\mathcal{O}_{pq;\vec{\tau}}$ in this tower we now determine the $N$ counting of the three-point couplings $C_{p_ip_j}(\mathcal{O}_{pq;\vec{\tau}})$.

Only the three-point couplings of the form $C_{p_ip_j}(\mathcal{O}_{p_ip_j;\vec{\tau}})$ will have a leading order 
contribution in the large $N$ expansion (from Wick contractions in supergravity). The true two-particle scaling eigenstates 
with leading order quantum numbers $\vec{\tau}$ will be mixtures containing some contribution from every operator 
$\mathcal{O}_{p_ip_j;\vec{\tau}}$ and hence will have leading order three-point couplings. We conclude that exchanged 
two-particle operators with twist $\tau\ge p_i+p_j$ have leading order three-point couplings $C_{p_ip_j}(\mathcal{O}_{\vec{\tau}})$. 
On the other hand, exchanged two-particle operators with twist in the range $2+2a+b\le \tau<p_i+p_j$ do 
not receive any contribution of the form $\mathcal{O}_{p_ip_j;\vec{\tau}}$ and thus have $1/N^2$ suppressed three-point couplings. 
We conclude that a three-point coupling $C_{p_1p_2}(\mathcal{O}_{\vec{\tau}} )$ has the perturbative expansion : 
\be
C_{p_1p_2}(\mathcal{O}_{\vec{\tau}} )= C^{(0)}_{p_1p_2}(\mathcal{O}_{\vec{\tau}} ) + \frac{1}{N^2} C^{(1)}_{p_1p_2}( \mathcal{O}_{\vec{\tau}} ) +\ldots
\ee
where $C^{(0)}_{p_1p_2}(\mathcal{O}_{\vec{\tau}} )\neq 0$ only for $\tau\ge p_i+p_j$

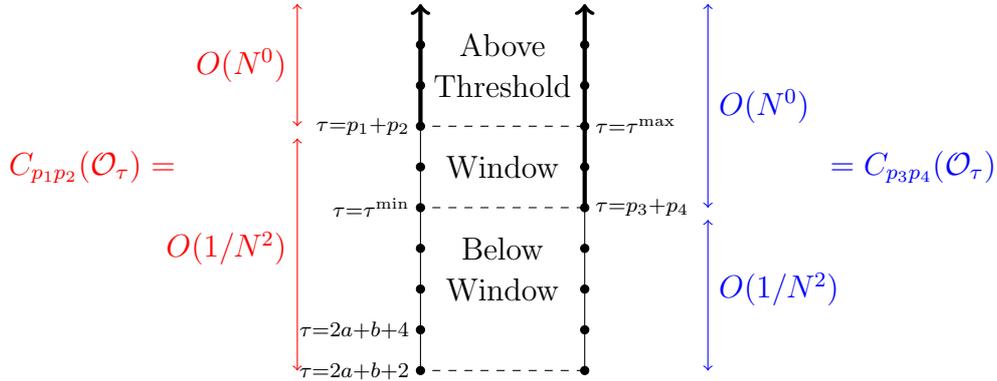
\begin{figure}\label{fig1}
	\begin{center}
		\begin{tikzpicture}[scale=.54]
		\draw[thin] (0,0) -- (0,6);
		\draw[ultra thick,->] (0,6) -- (0,9);
		\draw[thin] (4,0) -- (4,4);
		\draw[ultra thick,->] (4,4) -- (4,9);
		\draw[thin,dashed] (-.1,4) -- (4.1,4);
		\draw[thin,dashed] (-.1,6) -- (4.1,6);
		\draw[thin,dashed] (-.1,0) -- (4.1,0);
		\foreach \x in {0,1,2,3,4,5,6,7,8}
		\draw [fill] (0,\x) circle [radius=0.1]; 
		\foreach \x in {0,1,2,3,4,5,6,7,8}
		\draw [fill] (4,\x) circle [radius=0.1]; 
		\node at (2,8) {Above};
		\node at (2,7) {Threshold};
		\node at (2,5) {Window};
		\node at (2,3) {Below};
		\node at (2,2) {Window};
		\node[left] at (0,0) {$\scriptstyle \tau=2a+b+2$};
		\node[left] at (0,1) {$\scriptstyle \tau=2a+b+4$};
		\node[right] at (4,4) {$\scriptstyle \tau=p_3+p_4$};
		\node[left] at (0,6) {$\scriptstyle \tau=p_1+p_2$};
		\node[left] at (0,4) {$\scriptstyle \tau = \tau^\text{min}$};
		\node[right] at (4,6) {$\scriptstyle \tau = \tau^\text{max}$};
		\draw[red,<->] (-3,6) -- (-3,9); 
		\node[red,left] at (-3,7.5) {$O(N^0)$};
		\draw[red,<->] (-3,0) -- (-3,5.7); 
		\node[red,left] at (-3,3) {$O(1/N^2)$};
		\node[red] at (-8,5) {$C_{p_1 p_2}(\mathcal{O}_\tau)=$};
		\draw[blue,<->] (7,4) -- (7,9); 
		\node[blue,right] at (7,6.5) {$O(N^0)$};
		\draw[blue,<->] (7,0) -- (7,3.7); 
		\node[blue,right] at (7,2) {$O(1/N^2)$};
		\node[blue] at (12,5) {$=C_{p_3 p_4}(\mathcal{O}_\tau)$};
		\end{tikzpicture}
			\end{center}
		\caption{The large $N$ structure of $C_{p_1p_2,\vec{\tau}}\,C_{p_3p_4,\vec{\tau}}$ for two particle operators $\mathcal{O}_{\tau}$ in an $su(4)$ representation $[aba]$, and varying twist.}
	\end{figure}

The exchange of two particle operators in the common OPE of a four point correlator, 
gives a contribution of the form $C_{p_1p_2,\vec{\tau}}\,C_{p_3p_4,\vec{\tau}}$ for different values of twists. 
As before, we associate to each three-point coupling, $C_{p_1p_2,\vec{\tau}}$ and $C_{p_3p_4,\vec{\tau}}$, an infinite tower
representing the semishort and long operators $\mathcal{O}_{pq;\vec{\tau}}$ in the $su(4)$ representation $[aba]$.  
Putting together two of these, we obtain a representation of the common OPE coefficient as in Figure \ref{fig1}.
Referring the $N$ counting of  $C_{p_1p_2,\vec{\tau}}\,C_{p_3p_4,\vec{\tau}}$ to Figure \ref{fig1},
we read off the following pattern.

For $\tau \geq \tau^{\rm max} _{\vec{p}}\equiv {\rm max}(p_1+p_2,p_3+p_4)$,
we find exchanged operators for which both three-point couplings are leading order, i.e. $C_{p_1p_2}^{(0)}$ and $C_{p_3p_4}^{(0)}$ are both non-zero. 
In particular, $\tau^{\rm max}$ is the threshold twist for exchange of two-particle operators in disconnected free theory $\mathcal{G}_{\vec{p};0,0}$. 
In the window $\tau^{\rm max} > \tau\geq \tau^{\rm min}_{\vec{p}}\equiv{\rm min}(p_1+p_2,p_3+p_4)$, 
we find exchanged operators which have leading order three-point couplings with one pair 
of external operators, but $1/N^2$ suppressed three-point couplings with the other pair of external operators, 
e.g. we have $C_{p_1p_2}^{(0)}=0$ but $C_{p_3p_4}^{(0)}$ non-zero.
Finally, below the window $\tau < \tau^{\rm min}$ we have $C_{p_1p_2}^{(0)}=C_{p_3p_4}^{(0)}=0$ and
the OPE contains contributions which only involve products of $1/N^2$ suppressed three-point couplings.
These contributions give rise to a genuine $1/N^4$ effect which enters $\mathcal{G}_{2,0}$. 

For any arrangement of external charges there is always a threshold twist such that a tower of long operators is exchanged. 
The window itself might be empty if $\tau^{\rm min}=\tau^{\rm max}$. 

The location of the unitarity bound in Figure \ref{fig1} depends on the external charges. Generically, the unitarity bound $\tau=2a+b+2$ is below window, 
but there are two other situations which do occur. The unitarity bound can coincide with $\tau^{\rm min}$, i.e  $\tau^{\min}=2a+b+2$, 
in which case there is no below window region. The unitarity bound can coincide with $\tau^{\rm max}$, in which case there is an empty window
and $\tau^{\rm max}=\tau^{\rm min}=2a+b+2$.

The strategy followed in \cite{Aprile:2017bgs,Aprile:2017qoy} to bootstrap the order $1/N^4$ one-loop amplitude 
was to resolve the mixing problem in the long sector from the knowledge of $\mathcal{G}_{0,0}$ and 
$\mathcal{G}_{1,1}$ (focusing on the $su(4)$ representations $[000]$ and $[010]$)
and thereby obtain explicitly the CFT data needed to bootstrap $\mathcal{G}_{2,2}$. 
The double logarithmic discontinuity can also be obtained elegantly by using the hidden 
ten-dimensional conformal symmetry of \cite{Caron-Huot:2018kta}. 

To complete the double logarithmic discontinuities into full amplitudes requires 
additional knowledge about $\mathcal{G}_{2,1}$ and $\mathcal{G}_{2,0}$. 
The CFT data entering $\mathcal{G}_{2,1}$ is obtained only within the long sector. The CFT data entering $\mathcal{G}_{2,0}$ is instead obtained 
from the study of both protected semishort and long operators. In both cases, 
the operators we will consider are two-particle operators.\footnote{Higher multi-trace operators will 
also contribute but only at higher orders in the $1/N$ expansion within the ranges of twists we focus on here.}
Extracting this information in complete generality is a central new result of this paper. In particular, 
the study of the protected semishort sector at order $1/N^4$ has never been addressed before, 
except for the case of $\langle \mathcal{O}_3 \mathcal{O}_3 \mathcal{O}_3 \mathcal{O}_3 \rangle$ in \cite{Doobary:2015gia}.

Let us now project the correlator $\langle \mathcal{O}_{p_1}\mathcal{O}_{p_2}\mathcal{O}_{p_3}\mathcal{O}_{p_4}\rangle$ 
into an $su(4)$ representation $[a,b,a]$, and distinguish between long and protected sector.
Following the logic of Figure \ref{fig1}, we now highlight the main inputs of our bootstrap program. 
These are extensively discussed in Section \ref{free} and \ref{sec:OPEbeyond}.

The leading logarithmic discontinuity $\mathcal{G}_{2,2}$ (or more generally $\mathcal{G}_{n,n}$ for $n\ge 1$) is {\rm only} 
induced by exchanged long two-particle operators with $\tau\ge {\rm max}(p_1+p_2,p_3+p_4)$. 
The CFT data entering $\mathcal{G}_{n,n}$ for $n\ge 1$ comprises the $O(1)$ three-point couplings 
of these long two-particle operators with the external operators, 
and their $O(1/N^2)$ anomalous dimensions to the power $n$. 

Determining $\mathcal{G}_{2,1}$ by definition only involves data from the long sector. 
In particular, the new piece of information is obtained from operators exchanged in the window.  
For this range of twists $\mathcal{G}_{2,1}$ is essentially given by the product of one power of 
the anomalous dimension (of the exchanged operators) with the three-point couplings, 
let's say conventionally, $C^{(1)}_{p_1p_2,\vec{\tau}}$ and $C^{(0)}_{p_3p_4,\vec{\tau}}$. 
The combination $C_{p_1p_2,\vec{\tau}}C_{p_3p_4,\vec{\tau}}$ in the window is 
$O(1/N^2)$ as indicated by the \hyperlink{fig:test}{figure}. 
The physical data in the window determines also $\mathcal{G}_{n,n-1}$, with $n=2$ just the first non trivial case. 
For generic $n$, we simply increase the power of the anomalous dimensions to $n-1$. 

The partial degeneracy of the $1/N^2$ anomalous dimensions found in \cite{Aprile:2018efk} obstructs the explicit determination of 
$C^{(0)}_{pq,\vec{\tau}}$ in general, and consequently of $C^{(1)}_{pq,\vec{\tau}}$.
 Nevertheless, we will show in Section \ref{UnmixingII} that
we can obtain explicit expressions for the SCPW expansion of $\mathcal{G}_{2,2}$ and $\mathcal{G}_{2,1}$, respectively above threshold and in the window, 
from the analysis of $\mathcal{G}_{0,0}$, $\mathcal{G}_{1,1}$ and $\mathcal{G}_{1,0}$ of many different correlators.
This approach is based on the fact that for a given twist and $su(4)$ representation 
we know how many two-particle operators there are  \cite{Aprile:2018efk}. 

Determining $\mathcal{G}_{2,0}$ below the window is more complicated. There are both protected and long contributions, 
and they are all of the form $C^{(1)}_{p_1p_2,\vec{\tau}}C^{(1)}_{p_3p_4,\vec{\tau}}$ for given $\vec{\tau}$ below window. 
We will show that in the long sector, i.e $\tau\ge 4+2a+b$, the SCPW of $\mathcal{G}_{2,0}$ is obtained by rearranging slightly the method used for $\mathcal{G}_{2,1}$. 
At the unitarity bound, $\tau=2+2a+b$, we will have to use a different approach, which we explain in Section \ref{sec:recomb}. 
We will see that the $1/N^4$ semishort contributions to the protected sector 
can also be determined by using the knowledge of the two-particle operators and various different correlators. 
In particular, for a given twist $2+2a+b$, we will use input from $O(1)$ SCPW coefficients for correlators with $\tau^{\rm min}=\tau^{\rm max}=2+2a+b$, 
as well as input from $O(1/N^2)$ SCPW coefficients for correlators with $\tau^{\rm max}>\tau^{\rm min}=2+2a+b$. 
Finally, we emphasize that multiplet recombination at $O(1/N^4)$ will be very much different from multiplet recombination at $O(1/N^2)$.

The functions $\mathcal{G}_{2,1}$ and $\mathcal{G}_{2,0}$, which we bootstrap starting from the 
leading logarithmic discontinuity $\mathcal{G}_{2,2}$, should therefore be such that 
the first can accommodate OPE predictions in the window, and the second can accommodate OPE 
predictions below window i.e. for $2+2a+b\leq\tau < \tau^{\rm min}$.  We recall that the structure of the correlators is 
constrained by the partial non-renormalisation theorem \cite{Eden:2000bk},
\be
\langle \mathcal{O}_{p_1}\mathcal{O}_{p_2} \mathcal{O}_{p_3} \mathcal{O}_{p_4} \rangle = 
\langle \mathcal{O}_{p_1}\mathcal{O}_{p_2} \mathcal{O}_{p_3} \mathcal{O}_{p_4} \rangle_{\rm free} + \mathcal{P}\,\mathcal{I}(x,\bar{x};y,\bar{y}) \mathcal{D}_{\vec{p}}(x,\bar{x};y,\bar{y})\,,
\ee
where $\mathcal{I}(x,\bar{x};y,\bar{y})$ and $\mathcal{P}$ are kinematical factors defined later in \eqref{Pfactor} and (\ref{Iintriligator}).
We find that the large $N$ expansion of the correlator yields a natural structure for the dynamical function $\mathcal{D}$,
\be
\mathcal{D}(x,\bar{x};y,\bar{y}) = \mathcal{T}_{\vec{p}}+ \frac{1}{N^4} \mathcal{H}^{(2)}_{\vec{p}} +  \ldots\,,
\ee
where $\mathcal{T}$ itself admits a large $N$ expansion 
\be
\mathcal{T} = \frac{1}{N^2} \mathcal{T}^{(1)} + \frac{1}{N^4} \mathcal{T}^{(2)} + \ldots\,.
\ee
It follows that the functions $\mathcal{G}_{2,i}$ are given by
\bea
\mathcal{G}_{\vec{p};2,2}&=&\mathcal{P}\, \mathcal{I}(x,\bar{x};y,\bar{y}) \, \mathcal{H}^{(2)}_{\vec{p}}\Big|_{\log^2 u}\\[.2cm]
\mathcal{G}_{\vec{p};2,1}&=&\mathcal{P}\, \mathcal{I}(x,\bar{x};y,\bar{y})\left[\ \mathcal{T}_{\vec{p}}^{(2)} + \mathcal{H}^{(2)}_{\vec{p}}\, \right]_{\log^1 u}
\\[.2cm]
\mathcal{G}_{\vec{p};2,0}&=& 
					\langle \mathcal{O}_{p_1}\mathcal{O}_{p_2} \mathcal{O}_{p_3} \mathcal{O}_{p_4} \rangle_{\rm free}\Big|_{\tfrac{1}{N^4}} +
					\mathcal{P}\, \mathcal{I}(x,\bar{x};y,\bar{y}) \left[\ \mathcal{T}_{\vec{p}}^{(2)} + \mathcal{H}^{(2)}_{\vec{p}}\, \right]_{\log^0 u}
\eea

The function $\mathcal{H}^{(2)}_{\vec{p}}$, which we will refer to as the `minimal' one-loop function, 
meets all the constraints from the $1/N^4$ OPE predictions, both in the long sector, and at the unitarity bound. 
We define $\mathcal{H}^{(2)}_{\vec{p}}$ as the unique solution, up to finite spin ambiguities, 
of our bootstrap algorithm described in Section \ref{sec:one_loop_corr}, where we discuss a number of non-trivial examples.

The function $\mathcal{T}_{\vec{p}}$, studied in more detail in Section \ref{NewRastelli},  is a generalisation of the tree-level function 
of Rastelli and Zhou \cite{Rastelli:2016nze} for all $N$, and it is defined by the property that, 
together with connected free theory, it gives empty contributions to  any exchanged long operators 
with twist $2+2a+b\leq\tau<\tau^{\rm min}$. In this sense, the function $\mathcal{T}_{\vec{p}}$ 
generalises the construction of Dolan, Nirschl and Osborn in \cite{Dolan:2006ec} who 
obtained tree-level results by precisely demanding such a cancellation of low twist 
operators against recombined free theory.  
Because of this property, the minimal loop function  $\mathcal{H}^{(2)}_{\vec{p}}$  
contains all the dynamical information at $O(1/N^4)$. 

Let us point out a finer subtlety about $\mathcal{H}^{(2)}_{\vec{p}}$: The three-point couplings $C_{p_1p_2,\vec{\tau}}$ of exchanged semishort 
operators, 
 which determine a piece of  $\mathcal{G}_{2,0}$,  are obtained only within free theory, since these are not renormalized. 
At the same time, $\mathcal{G}_{2,2}$ and $\mathcal{G}_{2,1}$ are determined only within the long sector. In this sense, 
some inputs in $\mathcal{G}_{2,0}$ are obtained in a completely independent way. 
Nevertheless, $\mathcal{H}^{(2)}_{\vec{p}}$ 
has to be consistent with $\mathcal{G}_{2,i=0,1,2}$,  
and 
the coherence of the whole minimal one-loop function across the various OPE predictions is 
a non-trivial confirmation of the AdS/CFT correspondence within the $\mathcal{N}=4$ bootstrap program.


\section{Free theory of single-particle operators} \label{free}


We are interested in correlation functions of protected half-BPS operators 
which describe scattering of single-particle states in AdS${}_5\times$S${}^5$. 
The first task is thus to determine the operators dual to single-particle states: 
these are not simply single-trace operators but can have multi-trace corrections which we must take into account. 
In~\cite{Aprile:2018efk} we identified the operators dual to single-particle states as those  
{\em half-BPS operators which are orthogonal to all multi-trace operators}. 
In the strict large $N$ limit, our definition reduces to the familiar statement that single-particle states 
correspond to operators in multiplets whose superconformal primaries are given by single-trace 
operators in the $[0,p,0]$ representation of $su(4)$. For finite $N$ instead, our definition automatically 
picks the correct multi-trace admixtures which is needed to uplift half-BPS single-trace operators 
to single-particle operators.\footnote{See also previous discussions in \cite{Arutyunov:1999en,
DHoker:1999jke,Rastelli:2017udc} and more recently \cite{Arutyunov:2018tvn}.}

Single-trace operators in the $[0,p,0]$ rep can be given as
\be
 {\rm tr}( \phi^p)(x,y) = y^{R_1} \ldots y^{R_p} {\rm tr}(\phi_{R_1} \ldots \phi_{R_p})(x) 
 \ee
where the fields $\phi_R$ are the elementary scalars of the $\mathcal{N}=4$ multiplet, 
and the $SO(6)$ null vector $y^R$ is used to project onto the symmetric traceless representation, $\phi(x,y)=y^R\phi_R(x)$. 
The $p=2$ case corresponds to the superconformal primary for the energy-momentum multiplet 
which is dual to the graviton multiplet in AdS${}_5$. The $p=3$ case is the first Kaluza-Klein mode 
arising from reduction of the IIB graviton supermultiplet on $S^5$.  In these two cases, 
the single-particle operator equals the single-trace operator, even at finite $N$, 
since there are no multi-trace operators of charges $p<4$ to mix with.

The  single-particle operators we consider explicitly in this paper 
are:
\bea
\mathcal{O}_2&=& {\rm tr}( \phi^2)\notag\\[.2cm]
\mathcal{O}_3&=& {\rm tr}( \phi^3)\notag\\[.2cm]
\mathcal{O}_4 &=& {\rm tr}( \phi^4) - \frac{2N^2-3}{N(N^2+1)} {\rm tr}( \phi^2)^2
\,.
\eea
The coefficients of the higher multi-trace contributions are determined by the orthogonality conditions, 
according to our definition. For example $\mathcal{O}_4$ is defined by the requirement 
that it is orthogonal to the double-trace operator ${\rm tr}( \phi^2)^2$:
\be
\langle \mathcal{O}_4(x_1,y_1) {\rm tr}( \phi^2)^2(x_2,y_2) \rangle = 0\,.
\ee
Notice that since all operators involved are half BPS, the two-point functions entering the orthogonality conditions
can be computed in free field theory in terms of the elementary propagators
\be
\langle (\phi)_{r}{}^{\bar{r}}(x_1,y_1) (\phi)_{s}{}^{\bar{s}}(x_2,y_2)\rangle = \Bigl(\delta_r^{\bar s} \delta_s^{\bar r} - \frac{1}{N} \delta_r^{\bar r} \delta_s^{\bar s} \Bigr) g_{12} \,.
\ee
where\footnote{
This is just the superpropagator in analytic superspace~\cite{Galperin:1984av,Howe:1995md,Hartwell:1994rp} 
around which much of the following formalism is implicitly based.}
\begin{align} g_{12} = \frac{y_i\cdot y_j}{x_{12}^2}.\label{superprop}
	\end{align}

We will now consider four-point correlators of the single-particle half-BPS operators,
first in free theory, and then in the interacting regime described by supergravity.

\subsection{Free theory four-point functions}

Free field four-point functions of single-particle half-BPS operators can be computed 
simply by performing Wick contractions between the elementary fields. 
The result is a sum over the different allowed superpropagator structures $g_{ij}$ accompanied by their colour factors. 
Graphically, the four external operators $\mathcal{O}_{p_i}$ are represented as 
vertices each with $p_i$ legs, and the propagator $g_{ij}$ is represented as  
a line between point $i$ and point $j$. We arrange the four operators at the corners of a square, 
labelled clockwise from the bottom left.

So for example, for the $\langle \mathcal{O}_3  \mathcal{O}_3  \mathcal{O}_3 \mathcal{O}_3 \rangle_{\rm free}$ correlator
\begin{align}
\label{3333diagrams}
\langle \mathcal{O}_3  \mathcal{O}_3  \mathcal{O}_3 \mathcal{O}_3 \rangle_{\rm free} = & \,\, \phantom{  {+} } 
A_0^0 \,
\begin{minipage}{0.85cm}
\begin{tikzpicture}[scale=0.7]
\draw (0,0) -- (0,1) ;
\draw (1,0) -- (1,1) ;
\draw [bend right=30] (0,0) to (0,1);
\draw [bend left=30] (0,0) to (0,1);
\draw [bend right=30] (1,0) to (1,1);
\draw [bend left=30] (1,0) to (1,1);
\end{tikzpicture}
\end{minipage}\notag\\
&+
A_2^0\,
\begin{minipage}{0.85cm}
\begin{tikzpicture}[scale=0.7]
\draw (0,0) -- (1,1) ;
\draw (1,0) -- (0,1) ;
\draw [bend right=20] (0,0) to (0,1);
\draw [bend left=20] (0,0) to (0,1);
\draw [bend right=20] (1,0) to (1,1);
\draw [bend left=20] (1,0) to (1,1);
\end{tikzpicture}
\end{minipage}
+
A_2^1\,
\begin{minipage}{0.85cm}
\begin{tikzpicture}[scale=0.7]
\draw (0,0) -- (1,0) ;
\draw (0,1) -- (1,1) ;
\draw [bend right=20] (0,0) to (0,1);
\draw [bend left=20] (0,0) to (0,1);
\draw [bend right=20] (1,0) to (1,1);
\draw [bend left=20] (1,0) to (1,1);
\end{tikzpicture}
\end{minipage}
\notag\\
&+
A_4^0\,
\begin{minipage}{0.85cm}
\begin{tikzpicture}[scale=0.7]
\draw (0,0) -- (0,1) ;
\draw (1,0) -- (1,1) ;
\draw [bend right=20] (0,0) to (1,1);
\draw [bend left=20] (0,0) to (1,1);
\draw [bend right=20] (0,1) to (1,0);
\draw [bend left=20] (0,1) to (1,0);
\end{tikzpicture}
\end{minipage}
+
A_4^1 \,
\begin{minipage}{0.85cm}
\begin{tikzpicture}[scale=0.7]
\draw (0,0) -- (1,0) -- (1,1) -- (0,1) -- cycle ;
\draw (0,1) -- (1,0) ;
\draw (0,0) -- (1,1) ;
\end{tikzpicture}
\end{minipage}
+
A_4^2\,
\begin{minipage}{0.85cm}
\begin{tikzpicture}[scale=0.7]
\draw (0,0) -- (0,1) ;
\draw (1,1) -- (1,0) ;
\draw [bend right=20] (0,0) to (1,0);
\draw [bend left=20] (0,0) to (1,0);
\draw [bend right=20] (0,1) to (1,1);
\draw [bend left=20] (0,1) to (1,1);
\end{tikzpicture}
\end{minipage}
\notag \\
&+A_6^0\,
\begin{minipage}{0.85cm}
\begin{tikzpicture}[scale=0.7]
\draw (0,0) -- (1,1) ;
\draw (0,1) -- (1,0) ;
\draw [bend right=30] (0,0) to (1,1);
\draw [bend left=30] (0,0) to (1,1);
\draw [bend right=30] (0,1) to (1,0);
\draw [bend left=30] (0,1) to (1,0);
\end{tikzpicture}
\end{minipage}
+
A_6^1\,
\begin{minipage}{0.85cm}
\begin{tikzpicture}[scale=0.7]
\draw (0,0) -- (1,0) ;
\draw (0,1) -- (1,1) ;
\draw [bend right=20] (0,0) to (1,1);
\draw [bend left=20] (0,0) to (1,1);
\draw [bend right=20] (0,1) to (1,0);
\draw [bend left=20] (0,1) to (1,0);
\end{tikzpicture}
\end{minipage}
+A_6^2\,
\begin{minipage}{0.85cm}
\begin{tikzpicture}[scale=0.7]
\draw (0,0) -- (1,1) ;
\draw (0,1) -- (1,0) ;
\draw [bend right=20] (0,0) to (1,0);
\draw [bend left=20] (0,0) to (1,0);
\draw [bend right=20] (0,1) to (1,1);
\draw [bend left=20] (0,1) to (1,1);
\end{tikzpicture}
\end{minipage}
+
A_6^3\,
\begin{minipage}{0.85cm}
\begin{tikzpicture}[scale=0.7]
\draw (0,0) -- (1,0) ;
\draw (0,1) -- (1,1) ;
\draw [bend right=30] (0,0) to (1,0);
\draw [bend left=30] (0,0) to (1,0);
\draw [bend right=30] (0,1) to (1,1);
\draw [bend left=30] (0,1) to (1,1);
\end{tikzpicture}
\end{minipage}
\notag\\
\end{align}
where $A_\gamma^k$ are the associated colour factors. 
The subscript $\gamma$ is the total number of propagators connecting 
the left half  of the graph to the right half,  whereas $k$ is the number 
of propagators along the top edge of the square.
Of course many colour factors are equal to each other, 
where the corresponding graphs are isomorphic. 
Indeed there are only three independent colour factors in this example and 
explicit computations of the Wick contractions yields the all orders in $N$ factors
\begin{align}\label{3333coeffs}
	A_0^0=A_6^0=A_6^3&=\frac{9(N^2-4)^2(N^2-1)^2}{N^2}\notag \\ 
	A_2^0 = A_2^1=A_4^0=A_4^2=A_6^1=A_6^2&= \frac9{N^2-1} A_0^0\notag \\
	A_4^1 &=   \frac{18(N^2-12)}{(N^2-1)(N^2-4)} A_0^0\ .
\end{align}

For a  general free theory correlator, without loss of generality 
we can arrange the external charges as $p_{43}\ge p_{21}\ge 0$. 
The general free theory  result is then
\be\label{gen_free_theory0}
{\langle \mathcal{O}_{p_1} \mathcal{O}_{p_2} \mathcal{O}_{p_3} \mathcal{O}_{p_4} \rangle_{\rm free}} = {\cP^{}} \times  
\sum_{\substack{ \gamma=p_{43}\\ \gamma-p_{43}\,\in\, 2\mathbb{Z} } }^{ {\rm min}(p_1+p_2,p_3+p_4)}
 \left[ \left(\frac{ g_{13} g_{24} }{ g_{12} g_{34} }\right)^{\frac{\gamma-p_{43}}{2}}  \sum_{k=0}^{  \frac{\gamma-p_{43}}{2} } {A}_{\gamma }^k \left( \frac{ g_{14} g_{23} }{ g_{13} g_{24} } \right)^k \right]\,
\ee
where ${A}_{\gamma }^k$ are color factors, and we defined the prefactor
\bea
\label{Pfactor}
\mathcal{P} = g_{12}^{\frac{p_1+p_2-p_{43}}{2}} g_{14}^{\frac{p_{43}-p_{21}}{2}} g_{24}^{\frac{p_{43}+p_{21}}{2}} g_{34}^{\phantom{\frac{}{}}p_3}.
\eea

Note that the RHS of~\eqref{gen_free_theory0} is $\mathcal P$ times a function of super cross-ratios. 
We define space-time cross ratios $u,v$ (equivalently $x,\bar x$) 
and internal cross-ratios $\ssigma,\ttau$ (equivalently $y,\bar y$) as follows
\begin{align}\label{eq:def_cross_ratios}
	u&= x \bar x =\frac{x_{12}^2x_{34}^2}{x_{13}^2 x_{24}^2}\,,  
	&v&=(1-x)(1-\bar x) =  \frac{x_{14}^2x_{23}^2}{x_{13}^2 x_{24}^2}\,, \notag \\ 
	\frac1\ssigma&= y \bar y =\frac{y_{1}.y_2y_{3}.y_4}{y_{1}.y_3 y_{2}.y_4}\,,  
	&\frac\ttau \ssigma&=(1-y)(1-\bar y) =  \frac{y_{1}.y_4y_{2}.y_3}{y_{1}.y_3 y_{2}.y_4}\,.
\end{align}
Inputting the definition of the superpropagator~\eqref{superprop} 
we find the super cross-ratios 
\bea
\label{supercross}
\frac{ g_{13} g_{24} }{ g_{12} g_{34} }= \frac{x \bar x}{y \bar y} = u\ssigma,\qquad \frac{ g_{14} g_{23} }{ g_{13} g_{24} }=\frac{(1-x) (1-\bar x)}{(1-y) (1-\bar y)}=\frac{\ttau}{v\ssigma} \,.
\eea
which we can substitute directly in \eqref{gen_free_theory0}

For single-particle external operators
the colour factors of extremal and next-to-extremal correlators vanish identically. 
These are correlators whose charges satisfy (with our choice of $p_{43}\ge p_{21}\ge 0$)
\begin{align}
p_4 &= p_1+p_2+p_3 \qquad& &\text{(extremal)}\,,\notag\\
p_4 &= p_1 + p_2 + p_3 - 2 \qquad& &\text{(next-to-extremal)}\,. \label{extremal}
\end{align}
Notice that extremal and next-to-extremal correlators of half-BPS operator 
do not vanish for single trace operators but they do for single-particle operators.

The first single-particle correlators that are non-vanishing are next-to-next-to-extremal, with charges obeying
\be
p_4 = p_1 + p_2 + p_3 - 4\,.
\ee

More generally, we define 
\begin{align}\label{deg_extremality}
	\kappa_{\vec{p}} = \min\left(\tfrac12(p_1+p_2+p_3-p_4),p_3\right) = \text{``degree of extremality"} \ .
\end{align}
and we say that a correlator is a N${}^\kappa$E, according to its degree to extremality. 
Next-to-next-to-extremal correlators have degree of extremality $\kappa_{\vec{p}}=2$.

\begin{figure}
\begin{center}
	\begin{tikzpicture}[scale=.54]
	\draw[thin] (0,0) -- (0,6);
	\draw[thin] (4,1) -- (4,4);
	\draw[thin,dashed] (-.1,4) -- (4.1,4);
	\draw[thin,dashed] (-.1,6) -- (4.1,6);
	\draw[thin,dashed] (-.1,1) -- (4.1,1);
	\foreach \x in {0,1,2,3,4,5,6}
	\draw [fill] (0,\x) circle [radius=0.1]; 
	\foreach \x in {1,2,3,4}
	\draw [fill] (4,\x) circle [radius=0.1]; 
	\node at (2,5) {Window};
	\node at (2,3) {Below};
	\node at (2,2) {Window};
	\node[left] at (0,6) {$\scriptstyle \tau^\text{max}=p_1+p_2$};
	\node[right] at (4,4) {$\scriptstyle  \tau^\text{min}=p_3+p_4$};
	\node[right] at (4,1) {$\scriptstyle  p_4-p_3$};
	\node[left] at (0,0) {$\scriptstyle  p_2-p_1$};
	\draw[thick,red,<->] (-1.5,1) -- (-1.5,4);
	\node[left] at (-1.5,2.5) {$\scriptstyle  2\kappa=2p_3$};
	\end{tikzpicture}
	\qquad
	\begin{tikzpicture}[scale=.54]
	\draw[thin] (0,0) -- (0,4);
	\draw[thin] (4,1) -- (4,6);
	\draw[thin,dashed] (-.1,4) -- (4.1,4);
	\draw[thin,dashed] (-.1,6) -- (4.1,6);
	\draw[thin,dashed] (-.1,1) -- (4.1,1);
	\foreach \x in {0,1,2,3,4}
	\draw [fill] (0,\x) circle [radius=0.1]; 
	\foreach \x in {1,2,3,4,5,6}
	\draw [fill] (4,\x) circle [radius=0.1]; 
	\node at (2,5) {Window};
	\node at (2,3) {Below};
	\node at (2,2) {Window};
	\node[right] at (4,6) {$\scriptstyle \tau^\text{max}=p_3+p_4$};
	\node[left] at (0,4) {$\scriptstyle  \tau^\text{min}=p_1+p_2$};
	\node[right] at (4,1) {$\scriptstyle  p_4-p_3$};
	\node[left] at (0,0) {$\scriptstyle  p_2-p_1$};
	\draw[thick,red,<->] (6,1) -- (6,4);
	\node[right] at (6,2.5) {$\scriptstyle 2\kappa=p_1+p_2+p_3-p_4$};
	\end{tikzpicture}
\caption{Illustration of the possible $su(4)$ representations exchanged in the overlap of the
 $(\mathcal{O}_{p_1} \times \mathcal{O}_{p_2})$ and $(\mathcal{O}_{p_3} \times \mathcal{O}_{p_4})$ OPEs. 
 The vertical axis represents the possible values of $b+2a$.}
\label{fig2}
\end{center}
\end{figure}
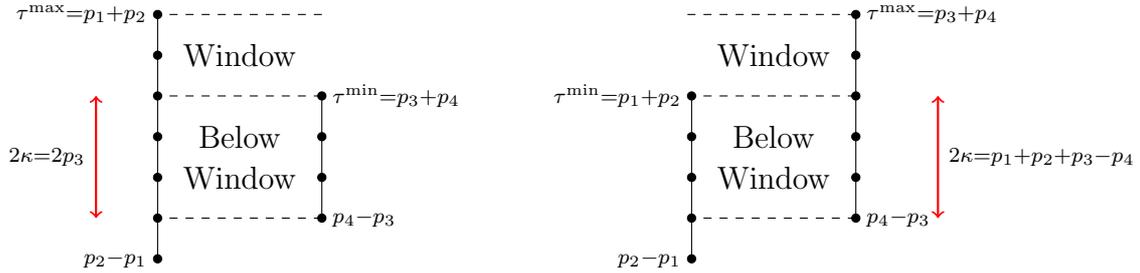

The degree of extremality determines the number of available $su(4)$ representations $[aba]$
in the  overlap of the two OPEs $(\cO_{p_1} \times \cO_{p_2})$ and $(\cO_{p_3} \times \cO_{p_4})$.%
\footnote{
Notice that an $su(4)$ rep $[aba]$ appearing in both OPEs 
will have $b+2a$ lying in the same range of values as $\gamma$ in~\eqref{gen_free_theory0}. 
We can then see that degree of extremality $\kappa_{\vec{p}}$ is equal to the number of values of 
$\gamma$ in~\eqref{gen_free_theory0}, minus 1.} 
For example, N${}^2$E correlators have the feature that the superconformal primaries 
in the long sector have a single possible $su(4)$ representation.
One can visualise the degree of extremality as shown in Fig. \ref{fig2}.
In this Figure, the vertical axis represents the possible values of $b+2a$ in the two OPEs. 
The degree of extremality $\kappa_{\vec{p}}$ then denotes the size of the  overlap in either of the 
two cases $p_1+p_2>p_3+p_4$ or $p_1+p_2<p_3+p_4$.  

Note that, as will be detailed in the next section, the interacting part of the correlator 
has a universal structure which reduces the range of $su(4)$ structures by 2.

We now review the technology that allows us to perform the 
superconformal partial wave expansion (SCPW) of a generic 
$\langle\mathcal{O}_{p_1}\mathcal{O}_{p_2}\mathcal{O}_{p_3}\mathcal{O}_{p_4}\rangle$ correlator.  
We follow the formalism of \cite{Doobary:2015gia}, which is group theoretic, 
manifestly unitary, and has the great advantage of 
dealing with all representations in a uniform way.

\subsection{Review of the SCPW expansion}

To address the SCPW expansion of 
$\langle\mathcal{O}_{p_1}\mathcal{O}_{p_2}\mathcal{O}_{p_3}\mathcal{O}_{p_4}\rangle$
we must first describe conformal blocks for all supermultiplets that might be exchanged in the OPE of half-BPS operators.

Following~\cite{Doobary:2015gia} we label the superconformal primaries 
$\mathcal O_{\gamma,\underline{\lambda}}$ by a number $\gamma$ 
and a finite dimensional representation of $SL(2|2)$ which we specify via a 
Young diagram $\underline{\lambda} \equiv [\lambda_1,\ldots,\lambda_n]$ 
where $\lambda_i$ is the length of the $i$th row.%
\footnote{
The formalism arises from analytic superspace~\cite{Galperin:1984av,Howe:1995md,Hartwell:1994rp} 
which has $SL(2|2) \times SL(2|2) \times \mathbb{C}$ isotropy group. 
A general unitary representation of the $\mathcal N$=4 superconformal group 
is thus specified via two $SL(2|2)$ representations and a weight $\gamma$. 
For four-point functions of half BPS operators, both $SL(2|2)$ representations coincide. 
Remarkably the $SL(2|2)$ representations are always finite dimensional 
and the resulting analytic field is unconstrained~\cite{Heslop:2001zm,Heslop:2002hp}.}
	
The Young diagrams do not have an arbitrary shape but have to fit into a `fat hook' shape, 
which amounts to the additional constraint that the third row (and hence any subsequent rows) 
cannot be longer than length two, i.e. $\lambda_3 \leq 2$. The number of rows also satisfies
$n\leq (\gamma-p_{43})/2$. 
For example a generic such diagram has the form
\begin{align}
	\begin{tikzpicture}[scale=.3,baseline={([yshift=-1ex]current bounding box.center)}]
	\draw (0,8) -- (20,8) -- (20,7) -- (0,7) -- (0,8);
	\node at (10,7.5) {\tiny $\leftarrow\lambda_1\rightarrow$};
	\draw (0,7) -- (16,7) -- (16,6) -- (0,6) -- (0,7);
	\node at (8,6.5) {\tiny $\leftarrow\lambda_2\rightarrow$};
	\draw (0,6) -- (2,6) -- (2,3) -- (0,3) -- (0,6);
	\node at (1,4.5) {\tiny $\begin{array}{c}\uparrow\\ \mu_2\\ \downarrow \end{array}$};
	\draw (0,3) -- (1,3) -- (1,0) -- (0,0) -- (0,3);
	\node at (.55,1.5) {\tiny $\begin{array}{c}\uparrow\\ \mu_1\\ \downarrow \end{array}$};
	\end{tikzpicture} = [\lambda_1,\lambda_2,2^{\mu_2},1^{\mu_1}]
\end{align}
with first row of length $\lambda_1$, second row of length $\lambda_2$ and then $\mu_2$ 
rows of length 2 (denoted $2^{\mu_2}$) and $\mu_1$ rows of length 1 (denoted $1^{\mu_1}$). 
Such a generic Young tableau corresponds to a long multiplet.

 Short multiplets instead have row 2 of length 
$1$ or $0$ and so have the shape of a `thin hook'. 
The parameters $\gamma$ and $\underline{\lambda}$ determine the usual quantum 
numbers of spin $l$, dimension $\Delta$ (or twist $\tau \equiv \Delta - l  $) and $su(4)$ representation, 
which here always takes the form $[aba]$. The dictionary is 
summarized by the following table
\vspace{-0.1cm}
	\begin{align}
	\begin{array}{|c||c|c|c|c|}
	\multicolumn{5}{c}{ 
		\begin{array}{l}
		\rule{0pt}{0cm}
		\end{array}
	}\\
	\hline
	GL(2|2) \text{ rep }\ula     					&    \tau = \Delta{-}l			& l     			&  su(4) \text{ labels } 			  	& \text{multiplet type} \\\hline
	[\emptyset]                                     					&    \gamma          			&0    				& [0,\gamma,0]                            	& \text{half BPS}       \\\hline
	\left[1^\mu\right]                                 		 	& \gamma             			&0				&   [\mu,\gamma{-}2\mu,\mu]		&\text{quarter BPS} \\ 
	\left[\lambda, 1^\mu\right]\ (\lambda\geq 2)   	& \gamma             			&\lambda{-}2      	&  [\mu,\gamma{-}2\mu{-}2,\mu] 	&  \text{semi-short}  \\\hline
	{ [\lambda_1,\lambda_2,2^{\mu_2},1^{\mu_1}]\  (\lambda_2\geq 2)} 
	& \gamma{+}2\lambda_2{-}4	&\lambda_1{-}\lambda_2&[\mu_1,\gamma{-}2\mu_1{-}2\mu_2{-}4,\mu_1]&\text{long} \\ \hline
	\end{array}
		\label{table}
	\end{align}
Note that the YT representation of a long multiplet is invariant up to the 
shift-symmetry,
\be\label{shift_symm_long}
\lambda_1 \rightarrow \lambda_1 +1, \ 	\lambda_2 \rightarrow \lambda_2 +1, \ \mu_2\rightarrow \mu_2-1,\  \gamma\rightarrow \gamma-2,
\ee
under which twist $\tau$, spin $l$, and $su(4)$ rep $[a,b,a]$ remain fixed. 
On the contrary, protected operators require both $\gamma$ and the Young tableau to be fully specified.

We denote the superconformal block corresponding to the contribution 
of an operator $\cO_{\gamma,\underline{\lambda}}$ to the four-point correlator 
$\langle \mathcal{O}_{p_1}\mathcal{O}_{p_2}\mathcal{O}_{p_3}\mathcal{O}_{p_4}\rangle$ 
as
\begin{align}
\text{superblock}: \qquad	\mathbb{S}_{\vec{p};\gamma,\underline{\lambda}}\ .
\end{align}

Long superblocks (those with $\lambda_2=2,3,...$) will occur often 
and we will also denote them by $\mathbb{L}^{}_{\vec{p};\vec{\tau}}$.  
They have the following factorised structure, 
\be\label{LONGSCPW}
\mathbb{L}^{}_{\vec{p};\vec{\tau}} \equiv 
\mathbb{S}^{}_{\vec{p};\gamma,\underline{\lambda}} = 
\mathcal{P} \times \mathcal{I}\times {\widetilde{\mathbb{L}} }_{\vec{p};\vec{\tau}}  \, 
\qquad 
{\widetilde {\mathbb{L}}}^{}_{\vec{p};\vec{\tau}} =
 \frac{ \mathcal{B}^{(2+\frac{\tau}{2},l)} }{ u^{2+ \frac{ p_{43}}{2} } } \ \Upsilon_{[aba]} \,,
\ee
where $\mathcal{P}$ is given in (\ref{Pfactor}), and $\mathcal{I}$ by
\be \label{Iintriligator}
\mathcal{I}(x,{\bar x},y,{\bar y})= \frac{(x-y)(x-{\bar y})({\bar x}-y)({\bar x}-{\bar y})}{(y {\bar y})^2} \,.
\ee 
Here $\mathcal{B}^{\,t,l}$ and $\Upsilon_{[aba]}$ are ordinary bosonic blocks for conformal and internal symmetries. 
Explicitly,
\beq
\mathcal{B}^{(t,l)}(x,{\bar x}) = (-1)^l\, u^t \left[ \frac{ x^{l+1} F_{t+l}({\bar x}) F_{t-1}({\bar x}) - (x \leftrightarrow {\bar x})}{x-{\bar x}}\right]\,,
\eeq
and
\bea
\Upsilon_{[aba]}(y,{\bar y}) &=& - \frac{P_{n+1}(y)P_m({\bar y}) - (y \leftrightarrow {\bar y})}{y-{\bar y}}\,,\qquad
\left\{\begin{array}{l}
n=m+a,\\[.2cm] \displaystyle m=\frac{b-p_{43}}{2}\,,\end{array}\right.
\eea
where
\be
F_t(x) = {}_2F_1\bigl(t-\tfrac{p_{12}}{2},t+\tfrac{p_{34}}{2},2t;x\bigr)\,,\quad
P_n(y) = \frac{n! \, y}{(n+1+p_{43})_n} {\rm JP}^{(p_{43}-p_{21}|p_{43}+p_{21})}\bigl(\tfrac{2}{y}-1\bigr)
\ee
The notation JP stands for Jacobi polynomial. 

Explicit formulae for semishort, $\tfrac{1}{4}$-BPS and $\tfrac{1}{2}$-BPS superblocks 
were obtained in~\cite{Doobary:2015gia} and can be found in  appendix~\ref{superblocks}.
Especially in these cases, the superblock formalism naturally provides manifestly unitary representations.

Since the parameters $\lambda_i$ are defined by a Young diagram, 
they are a priori integer valued.  
For long superblocks however in the interacting theory, 
the scaling dimension $\Delta$ (or equivalently the twist $\tau$) of an operator 
becomes anomalous and hence non-integer. We can thus allow an analytic 
continuation of $\lambda_1$ and $\lambda_2$ such that the spin 
$\lambda_1-\lambda_2=l$ remains integer.  In such cases we even allow 
for continuations such that $\lambda_2<2$. This means that the labels of 
such continued long superblocks can coincide with those of short superblocks 
when $\lambda_2 \rightarrow 1,\mu_2=0$. 
To avoid this potential confusion therefore we simply use the notation for long superblocks,  
$\mathbb{L}_{\vec{p};\vec{\tau}}$, on the LHS of~\eqref{LONGSCPW} 
and allow $\tau\geq2a+b+2$ to be non-integer valued.

When long supermultiplets sit exactly on the unitarity bound, $\tau = 2 + 2a + b$, 
they become reducible and can be expressed as a sum of short multiplets
\be
\label{Lonbound}
\mathbb{L}_{\vec{p};\vec{\tau}} = \mathbb{S}_{\vec{p};\tau,[l+2,1^{a}]} + \mathbb{S}_{\vec{p};\tau+2,[l+1,1^{a+1}]}\,\qquad \tau = 2 + 2a + b\,.
\ee
The first term on the RHS of (\ref{Lonbound}) is a semi-short superblock of spin $l$ 
while the second is a semi-short superblock of spin $l-1$  or a quarter-BPS superblock (if $l=0$). We will make use of this reducibility in Section \ref{sec:recomb}.

\subsection{The SCPW expansion of the free theory}

The SCPW of free theory correlators naturally stratifies by the label 
$\gamma=p_{43},p_{43}+2,\ldots, \tau^{\rm min} = {\rm min}(p_1+p_2,p_3+p_4)$ 
introduced in~\eqref{gen_free_theory0}. As mentioned in that context, $\gamma$ counts the number of
propagators connecting operators inserted at points 1 and 2 to 
operators inserted at points 3 and 4. 
In the SCPW expansion, 
$\gamma$ simply corresponds to the number of fundamental 
fields appearing in  the operator, $\mathcal O_{\gamma,\underline{\lambda}}$ 
being exchanged in the OPE. Note that this is a good quantum number only for 
free theory, and simply reflects the number of Wick contractions which have occurred in the OPE: 
\begin{align}
	\gamma  &= \# \text{ fundamental fields defining } \mathcal O_{\gamma,\underline{\lambda}} \notag\\
		       &= p_1+p_2 - (\# \text{ Wick contractions in } \cO_{p_1} \cO_{p_2}\sim \mathcal O_{\gamma,\underline{\lambda}} \text{ OPE } ) \notag\\
		        &= p_3+p_4 -( \# \text{ Wick contractions in } \cO_{p_3} \cO_{p_4}\sim \mathcal O_{\gamma,\underline{\lambda}} \text{ OPE })
\end{align}
 The general free theory correlator~\eqref{gen_free_theory0} then decomposes as 
\be\label{SCPW_expa}
  \langle \mathcal{O}_{p_1} \mathcal{O}_{p_2} \mathcal{O}_{p_3} \mathcal{O}_{p_4} \rangle_{\rm free} 
  = \sum_{\substack{ \gamma=p_{43}\\ \gamma-p_{43}\,\in\, 2\mathbb{Z} } }^{ {\rm min}(p_1+p_2,p_3+p_4)} 
  \sum_{\underline{\lambda}} {A}_{\vec{p};\gamma, \underline{\lambda} }\ \mathbb{S}_{\vec{p};\gamma, \underline{\lambda} }
\ee
where each term in the sum over $\gamma$ represents the expansion in SCPW
of the analogous terms in~\eqref{gen_free_theory0}.
Furthermore the Young tableau $\underline{\lambda}$ have at most $(\gamma-p_{43})/2$ rows. 
Note also that in free theory all Young Tableau are proper, 
having both integer rows and correct shape. 
Thus the decomposition \eqref{SCPW_expa} is unambiguous. 

But we do not consider the free theory in isolation, 
rather we will consider it as the limit of the interacting theory as the coupling vanishes.
In the interacting theory, the OPE of two half-BPS operators contains both operators 
in short supermultiplets, whose dimensions are protected, 
and long operators which have anomalous dimensions. 
Therefore we will split the SCPW expansion~\eqref{SCPW_expa} accordingly, 
and we will distinguish between the short sector {\it which by definition remains short in the interacting theory}, 
and a {\it free long} sector which will then acquire an anomalous dimension in the interacting theory. 
For the short sector we sum over superblocks with the specific form $\mathbb{S}_{\gamma,[\lambda,1^\mu]}$,
and for the long sector we sum over superblocks $\mathbb{L}_{\vec{\tau}}$, 
\be\label{equa_shortlong}
\langle \mathcal{O}_{p_1}  \mathcal{O}_{p_2}  \mathcal{O}_{p_3} \mathcal{O}_{p_4} \rangle_{\rm free}= 
\langle \mathcal{O}_{p_1}  \mathcal{O}_{p_2}  \mathcal{O}_{p_3} \mathcal{O}_{p_4} \rangle_{\rm short} 
+ \langle \mathcal{O}_{p_1}  \mathcal{O}_{p_2}  \mathcal{O}_{p_3} \mathcal{O}_{p_4} \rangle_{\rm free\, long}\,.
\ee
More explicitly, we introduce the SCPW coefficients $S_{\gamma,[\lambda,1^\mu]}$ and $L^f_{\vec{\tau}}$ as follows 
\begin{align}\label{free1234comb}
\langle \mathcal{O}_{p_1}  \mathcal{O}_{p_2}  \mathcal{O}_{p_3} \mathcal{O}_{p_4} \rangle_{\rm short} &=
\sum_{\substack{{\gamma=p_{43}}\\\gamma-p_{43}\in2\mathbb{Z}}}^{\tau^{\text{min}}} \biggl[
S_{\gamma,\emptyset} \mathbb{S}_{\gamma,\emptyset} + \sum_{\lambda=1}^\infty\sum_{\mu=0}^{\frac12(\gamma-p_{43})-1} S_{\gamma,[\lambda,1^\mu]}\ \mathbb{S}_{\gamma,[\lambda,1^\mu]} \biggr]\notag \\
\langle \mathcal{O}_{p_1}  \mathcal{O}_{p_2}  \mathcal{O}_{p_3} \mathcal{O}_{p_4} \rangle_{\rm free\, long} &= 
\sum_{\substack{a,b\in2\mathbb{Z}\\  4+b+2a \leq \tau^\text{min}}} \sum_{l=0}^\infty \sum_{\substack{\tau > 2a+b\\\tau-b \in 2\mathbb{Z}}}   {L}^{f}_{\vec{\tau}}\ \mathbb{L}_{\vec{\tau}} \,,
\end{align}
This split is non-trivial due to multiplet recombination; in the free limit a long multiplet whose twist lies on the unitary bound is indistinguishable from the direct sum  of certain short multiplets. 
A consequence of this is the identity of superblocks \eqref{Lonbound}.
The challenge then is to relate the SCPW coefficients 
$S_{\gamma,[\lambda,1^\mu]}$ and ${L}^{f}_{\vec{\tau}}$ 
to the original ones ${A}_{\gamma, \underline{\lambda} }$ in~\eqref{SCPW_expa}.

The simplest SCPW coefficients to identify are the coefficients of half BPS ops 
($\underline\lambda =\emptyset$) which are unchanged. Thus
\begin{align}\label{hbps}
	S_{\gamma,\emptyset}=A_{\gamma,\emptyset}\ .
\end{align}

The next simplest to deal with are the long representations above the unitary bound.
Here we take into account the fact that  $\gamma$ ceases to be a good 
quantum number for long operators. This is because long operators 
with different numbers of fundamental fields mix. For example 
$\cO_3\cO_3$ ($\gamma= 6$) mixes with $\cO_2 \Box \cO_2$  ($\gamma=4$) 
which both  have twist 6. This is the origin of the ambiguity 
in the description of long operators~\eqref{shift_symm_long}. Thus 
we need to collect together all SCPW coefficients with the same quantum numbers 
$\vec{\tau}$ (but different values of $\gamma$)
using the shift symmetry \eqref{shift_symm_long}. Thus
\be\label{Lcoefs}
L^{f}_{\vec{\tau}}= \sum_{\gamma=b+2a+4 }^{ {\rm min}(p_1+p_2,p_3+p_4)} 
{A}_{\gamma,[2+\frac{\tau-\gamma}{2}+l,2+\frac{\tau-\gamma}{2},2^{ \frac{\gamma-b}{2}-a-2  } ,1^a]},\qquad \tau \geq 4+2a+b\ .
\ee

The most difficult SCPW coefficients to identify in~\eqref{free1234comb} are the 
(non half-BPS) short coefficients $S_{[\lambda,1^\mu]}$ 
with non-zero $\lambda$ or $\mu$ and the related long coefficients 
at the unitary bound~${L}_{\vec{\tau}}$ with $\tau=2a+b+2$.    
This is because as we deform away from the free theory, 
some semi-short blocks combine to become long (as in~\eqref{Lonbound}), whereas others remain semi-short. 
Thus, a single SCPW coefficient  ${A}$ for a semi-short block at the unitarity bound, 
can actually contain the contribution of both short and long multiplets of the interacting theory. 

Our next task will be to explain how to properly disentangle physical
semishort contributions from the SCPW coefficients of free theory, and find $S_{[\lambda,1^\mu]}$ .
Let us motivate this problem further by mentioning that separating the coefficients $S$ from $L$ 
at the unitary bound is actually straightforward at $O(1/N^2)$.  
In particular we will show that apart from the case $S_{\gamma,[\lambda,1^\mu]}$ with $\gamma = \min({p_1+p_2,p_3+p_4})$, i.e when $\tau = \tau^{\rm min}$, all other 
the coefficients $S_{\gamma,[\lambda,1^\mu]}$ vanish.  Thus the values of $L$ will be trivially fixed by multiplet recombination.
This feature at $O(1/N^2)$ has lead various people to the assumption that  the same would be true for all $N$ 
(see~\cite{Bianchi:2006ti} for a discussion of this point). However, beyond $O(1/N^2)$ the separation of coefficients $S$ from $L$ is a  
non-trivial problem. We will solve this problem to $O(1/N^4)$ using knowledge about the form of the semi-short operators.

\subsection{Multiplet Recombination}
\label{sec:recomb}

We now show how to determine, up to order $1/N^4$, the genuine semishort sector of the single particle correlators 
$\langle \mathcal{O}_{p_1}  \mathcal{O}_{p_2} \mathcal{O}_{p_3} \mathcal{O}_{p_4} \rangle$ 
in the full interacting theory, purely using free theory correlators. 
In particular we provide formulae for all SCPW coefficients -- split according 
to operators which remain short in the interacting theory and those which 
are long~\eqref{free1234comb} -- in terms of the coefficients  $A_{\vec{p},\gamma,\underline{\lambda}}$ in~\eqref{SCPW_expa}. 

Recall that for long blocks at the unitary bound $\tau =2a+b+2$ we need to 
resolve the ambiguity which follows from the reducibility condition~\eqref{Lonbound}, 
i.e.  that a long SCPW is a sum of two semishort SCPWs 
\be
\label{Lonbound2}
\mathbb{L}_{\vec{\tau}} = \mathbb{S}_{\tau,[l+2,1^{a}]} + \mathbb{S}_{\tau+2,[l+1,1^{a+1}]} \qquad \tau=2a+b+2\,.
\ee
Comparing the two pieces of the SCPW expansion~\eqref{free1234comb}, 
and equating the coefficient of $\mathbb{S}^{}_{\tau,[l+2,1^{a}]}$, using \eqref{Lonbound2}, yields
\begin{align}\label{asll}
{A}_{\tau,[l+2,1^{a}]} = S_{\tau,[l+2,1^{a}]}+ {L}^f_{\vec{\tau}}+ {L}^f_{\tau-2,l+1,[a-1,b,a-1]} \qquad \tau=2a+b+2\ .
\end{align}

One of the key points allowing us to resolve the ambiguity at the unitarity bound, and correctly distinguish 
CPW coefficients of long and semi-short operators, is the following (already tacitly assumed in (\ref{free1234comb})):
 {\emph{a long operator at the unitarity bound necessarily has twist less than 
 $\tau_{\text{min}}$={\rm min}$(p_1+p_2,p_3+p_4)$, i.e. $L^f_{\vec{\tau}} = 0$ if $\tau = 2a+b+2 \geq \tau^{\rm min}$.}}
This is a non-perturbative statement, a non-trivial consequence of superconformal 
symmetry for the corresponding three-point functions~\cite{Arutyunov:2001qw,Heslop:2001dr}.

This fact allows us to use equation (\ref{asll}) to  determine the CPW coefficients 
of semi-short operators of twist $\tau^\text{min}=\min(p_1+p_2,p_3+p_4)$ in terms of lower twist coefficients
\begin{align}\label{asll2}
S_{\vec{p};\tau,[l+2,1^{a}]} &= A_{\vec{p};\tau,[l+2,1^{a}]}- {L}^f_{\vec{p};\tau-2,l+1,[a-1,b,a-1]} \quad &\text{for} \quad  \tau&=2a{+}b{+}2=\tau^\text{min}, \quad a\geq1\notag\\
S_{\vec{p};\tau,[l+2]} &= A_{\vec{p};\tau,[l+2]} \quad &\text{for} \quad  \tau&=b{+}2=\tau^\text{min} \ .
\end{align}

It is useful to understand the $1/N$ expansion\footnote{
 Note that here and below, `order $1/N^{k}$', really means 
 $N^{\frac12(p_1+p_2+p_3+p_4)}O(1/N^k)$ because we have not normalised our external operators.} 
of $S_{\vec{p};\tau^{\rm min},[l+2,1^{a}]}$ first, since it will play a role in our later formulas. 
Referring to figure~\ref{fig1}, when $\tau^\text{min}=2+2a+b$  two lines coincide, 
i.e. the lower dashed line sits on top of the middle dashed line,
thus we find that $S_{\vec{p};\tau^{\rm min},[l+2,1^{a}]}$ in \eqref{asll2} is non trivial at $O(1/N^2)$. 
In particular it gets a contribution from leading order connected propagator structures.  
In the special case of correlators $\langle \mathcal{O}_{p}  \mathcal{O}_{q} \mathcal{O}_{p} \mathcal{O}_{q} \rangle$, 
$\tau^\text{min}=\tau^\text{max}$ and free theory starts with an $O(1)$ contribution from disconnected diagrams. 
For all representations $[a,b,a]$ such that $\tau^\text{min}=2+2a+b$ we find then that all three 
dashed lines of the Figure \ref{fig1} 
coincide and $S_{\vec{p};\tau^\text{max},[l+2,1^{a}]}$ indeed has an $O(1)$ 
contribution from disconnected free theory diagrams.

What about CPW coefficients of semi-short operators of twist less than $\tau^\text{min}$? 
Semi-short operators generically will sit in the range of twists $\tau\leq{\rm min}(p_1+p_2,p_3+p_4)$,
therefore at the bottom dashed line in Figure \ref{fig1} below the window. 
It follows that the corresponding SCPW coefficient is $O(1/N^4)$,
\begin{align}\label{assum}
 S_{\vec{p};\tau,[l+2,1^a]}=O(1/N^4)\qquad  \tau=2a+b+2<\tau^\text{min}\ .
\end{align}

This is the well known statement that at $O(1/N^2)$ there are no semishort 
operators in the  spectrum below the window, which implies a cancellation between free theory and the interacting part.
Using this information we can solve $S_{\vec{p};\tau^{\rm min},[l+2,1^{a}]}$ in \eqref{asll2} and ${L}^f_{\vec{\tau}}$ in \eqref{asll} explicitly up to order $1/N^2$. 
First we solve~\eqref{asll} recursively, thus obtaining the long SCPW coefficients 
\begin{align}\label{long1on4}
	L^f_{\vec{\tau}}= \sum_{k=0}^a (-1)^k {A}_{\tau-2k,[l+2+k,1^{a-k}]} +  O(1/N^4)  \qquad  \tau=2a+b+2<\tau^\text{min}\ ,
\end{align} 
Then, we plug this result into~\eqref{asll2} to give the genuine semi-short coefficients at threshold
\begin{align}\label{Spi}
	S_{\vec{p};\tau,[l+2,1^{a}]} =\sum_{k=0}^a (-1)^k {A}_{\tau-2k,[l+2+k,1^{a-k}]} +  O(1/N^4) \qquad  \tau=2a+b+2=\tau^\text{min}\ .
\end{align} 
When $a=0$, we obtain correctly $S_{\vec{p};\tau,[l+2]}$ given above. 

Now, can we determine the $1/N^4$ CPW coefficients of semi-short operators of twist less than $\tau^\text{min}$? 
The answer is affirmative. We first need to use some non-trivial information about the spectrum of semi-short operators, and then 
we can determine these CPW coefficients unambiguously using data from many different correlators! 

The key point here is that we know the explicit form of the double trace semi-short operators - 
or more importantly the number of them. They are twist $\tau$, spin $l$ operators in the $[aba]$ $su(4)$ rep of the form 
\begin{equation}
\cO_{q\tilde{q}}=\cO_q\partial^l\cO_{\tilde{q}} \label{opq}
\end{equation}
as in eq. (\ref{doubletrace}) with
$\tau=q+\tilde{q}=2a+b+2$. For fixed twist and $su(4)$ structure we can enumerate the independent operators as 
\begin{align}
\label{pofr}
q_r&= a+1+r,&\qquad \tilde{q}_r &=a+1+b-r \qquad r=\delta_{a,0},\dots,\mu{-}1
\end{align}
where
\begin{align}
\mu \equiv   \left\{
\begin{array}{ll}
\bigl\lfloor{\frac{b+2}2}\bigr\rfloor \quad &a+l \text{ even,}\\[.2cm]
\bigl\lfloor{\frac{b+1}2}\bigr\rfloor  \quad &a+l \text{ odd.}
\end{array}\right.
\end{align}

Unlike the case of long operators, semishort operators receive no anomalous dimension. The operators enumerated in (\ref{pofr}) 
are therefore degenerate and we may freely take the $\mathcal{O}_{q\tilde{q}}$ themselves as our basis. 
The SCPW coefficients of such operators are then expressed in terms of the products of three-point couplings as follows,
\be
\label{SisCproducts}
S_{\vec{p};\tau,[l+2,1^a]} = \sum_{r,s} C_{p_1 p_2}(\mathcal{O}_{q_r \tilde{q}_r}) (M^{-1})_{rs} C_{p_3p_4}(\mathcal{O}_{q_s \tilde{q}_s})\,,
\ee
where $M$ is the matrix of two-point functions (which is diagonal at at leading order in large $N$),
\be
\label{twopointfns}
M_{rs} = \langle \mathcal{O}_{q_r\tilde{q}_r} \mathcal{O}_{q_s \tilde{q}_s}\rangle = Y_r \delta_{rs} + O(1/N^2)\,.
\ee

We also recall the fact, discussed in Section \ref{Intro}, that the only couplings with a leading order contribution in the large 
$N$ expansion are the ones of the form $C_{pq}(\mathcal{O}_{pq})$. From this it follows that at leading order in large $N$ 
we have a diagonal structure for the following three-point couplings,
\be
\label{diagform}
C_{q_r \tilde{q}_r}(\mathcal{O}_{q_s \tilde{q}_s}) = \delta_{rs} X_r + O(1/N^2)\,.
\ee

Armed with this information we can now predict the CPW coefficients of semishort operators, 
$S_{\vec{p};\tau,[l+2,1^a]}$,  of twist $\tau < \tau^\text{min}$ in terms of SCPW coefficients 
of correlators with either $\tau=\tau^\text{min}$.
These SCPW are known through \eqref{Spi}.
The formula for $S_{\vec{p};\tau, [l+2,1^a]}$, correct up to and including order $1/N^4$, is given by 
\begin{align}\label{form}
	S_{\vec{p};\tau, [l+2,1^a]} = \sum_{r=0}^{\mu-1} \frac{S_{p_1p_2q_r \tilde{q}_r}S_{ q_r\tilde{q}_rp_3p_4 }}{S_{ q_r \tilde{q}_r q_r \tilde{q}_r}} + O(1/N^6)\qquad  \tau=2a+b+2<\tau^\text{min}\ .
\end{align}
For simplicity, we have suppressed labels $\tau$ and $[l+2,1^a]$ in the SCPW coefficients on the RHS above.

The two factors giving the numerator of \eqref{form} in the RHS are both $O(1/N^2)$ whereas the factor in  
the denominator is leading in large $N$, thus the RHS is $O(1/N^4)$ as we stated already in~\eqref{assum}.
The formula (\ref{form}) may be proven by simply using (\ref{SisCproducts}) on both sides and 
then using (\ref{diagform}) and (\ref{twopointfns}) on the RHS to cancel the denominator.

Finally, with the knowledge of \eqref{form} to hand, we can improve 
$L^f_{\vec{\tau}}$ in~\eqref{long1on4} and $S_{\tau^{\rm min},[l+2,1^a]}$ in~\eqref{Spi} up to order $1/N^4$. The results are 
\begin{align}\label{longfree}
L^f_{\vec{\tau}}= \sum_{k=0}^a (-1)^k {A}_{\tau-2k,[l+2+k,1^{a-k}]} - \sum_{k=0}^a (-1)^k {S}_{\tau-2k,[l+2+k,1^{a-k}]}+  O(1/N^6)  \notag \\ \qquad \qquad \qquad  \tau=2a+b+2<\tau^\text{min}\ ,
\end{align} 
\begin{align}\label{threshold}
S_{\tau,[l+2,1^a]}= \sum_{k=0}^a (-1)^k {A}_{\tau-2k,[l+2+k,1^{a-k}]} - \sum_{k=1}^a (-1)^k {S}_{\tau-2k,[l+2+k,1^{a-k}]}+  O(1/N^6)  \notag \\ \qquad \qquad \qquad  \tau=2a+b+2=\tau^\text{min}\ .
\end{align}

Concluding, all SCPW coefficients of 
$\langle \mathcal{O}_{p_1}  \mathcal{O}_{p_2}  \mathcal{O}_{p_3} \mathcal{O}_{p_4} \rangle_{\rm short}$ 
and $\langle \mathcal{O}_{p_1}  \mathcal{O}_{p_2}  \mathcal{O}_{p_3} \mathcal{O}_{p_4} \rangle_{\rm long}$ 
in \eqref{free1234comb} have been obtained to $O(1/N^4)$ and therefore we have successfully 
split the free theory correlators into a protected contribution and an unprotected one. In general 
we can not go further in $1/N$ since to do so would require input from triple-trace (and higher multi-trace) operators.

We conclude this section by illustrating our formulas \eqref{form} and \eqref{threshold} for the semishort sectors of
$\langle \mathcal{O}_3 \mathcal{O}_3 \mathcal{O}_3 \mathcal{O}_3 \rangle$, which has been already examined in detail in \cite{Doobary:2015gia}, 
and $\langle \mathcal{O}_4 \mathcal{O}_4 \mathcal{O}_4 \mathcal{O}_4 \rangle$, which is new.

In the case of $\langle \mathcal{O}_3 \mathcal{O}_3 \mathcal{O}_3 \mathcal{O}_3 \rangle$ we have below threshold twist $2$ and $4$ semishort predictions. 
This semishort sector is special because no multi-trace mixing occurs in the large $N$ expansion. Therefore we can give formulas exact in $N$.
Very explicitly we find that, 
\begin{align}\label{s4l}
	S^{\langle 3333 \rangle}_{2,[\lambda]}&=0\notag \\
	S^{\langle 3333 \rangle}_{4,[l+2]} &= \frac{\bigl(S_{4,[\lambda]}^{\langle 2233\rangle }\bigr)^2}{S_{4,[\lambda]}^{\langle 2222 \rangle}} 
	= \frac{ 288   ((l+3)!)^2 }{(2l+6)!((l+3)(l+4)+\frac{4}{(N^2-1)})} \frac{A_0^0}{ (N^2-1)} \notag \\ 
	S^{\langle 3333 \rangle}_{4,[l+2,1]} &= \frac{\bigl(S_{4,[\lambda,1]}^{\langle 2233\rangle}\bigr)^2}{S_{4,[\lambda,1]}^{\langle 2222 \rangle}} 
	= \frac{576  ((l+3)!)^2  }{(2l+6)!((l+2)(l+5)-\tfrac{12}{(N^2-1)} )} \frac{A_0^0}{ (N^2-1)}
\end{align}
where $A_0^0=(3(N^2{-}1)(N^2{-}4)/N)^2$. 
The structure of the CPW coefficients of operators at threshold, i.e. twist $6$, follow straightforwardly by applying~\eqref{threshold}, 
\begin{align}
	S_{6,[\lambda]}&=A_{6,[\lambda]}\notag \\
	S_{6,[\lambda,1]} &= A_{6,[\lambda,1]}  - A_{4,[\lambda+1]}+S_{4,[\lambda+1]}\,, \notag\\
	S_{6,[\lambda,1,1]} &= A_{6,[\lambda,1,1]} -A_{4,[\lambda+1,1]}+ A_{2,[\lambda+2]} +S_{4,[\lambda+1,1]}\ .
\end{align}

In the case of $\langle \mathcal{O}_4 \mathcal{O}_4 \mathcal{O}_4 \mathcal{O}_4 \rangle$ we have twist $2$, $4$ and $6$ predictions. 
The computations at twist $2$ and twist $4$ are analogous to the case of  $\langle \mathcal{O}_3 \mathcal{O}_3 \mathcal{O}_3 \mathcal{O}_3 \rangle$. We find, 
\begin{align}\label{s4_4444}
	S^{\langle 4444 \rangle}_{2,[\lambda]}&=0\notag \\
	S^{\langle 4444 \rangle}_{4,[l+2]} &
	= \frac{ 16\times 1152 }{(l+3)(l+4)}\frac{((l+3)!)^2}{  (2l+6)! } \frac{1+(-)^l}{2}  \frac{1}{N^4}   , \notag \\
	S^{\langle 4444 \rangle}_{4,[l+2,1]} &
	= \frac{ 6\times 1600 }{  (l+2)(l+5) } \frac{((l+3)!)^2}{  (2l+6)! }\frac{1-(-)^l}{2}   \frac{1}{N^4} 
\end{align}
The twist $6$ results are new, 
\begin{align}\label{s6_4444}
	S^{\langle 4444 \rangle}_{6,[l+2]}&= \frac{16\times 384 ( {29} + 3(2l+9)^2 ) }{(l+3)(l+6)} \frac{(l+4)!^2}{(2l+8)!}\frac{1+(-)^l}{2}   \frac{1}{N^4}\notag \\
	S^{\langle 4444 \rangle}_{6,[l+2,1]} & = \frac{ 16\times 72(401+174(2l+9)^2 + (2l+9)^4)}{(l+2)(l+4)(l+5)(l+7)} \frac{(l+4)!^2}{(2l+8)!}\frac{1-(-)^l}{2}   \frac{1}{N^4}\notag  \\
	%
	S^{\langle 4444 \rangle}_{6,[l+2,1,1]} & = \frac{16\times 2400 (l+2)(l+7) }{(l+3)(l+6)} \frac{(l+4)!^2}{(2l+8)!}\frac{1+(-)^l}{2}   \frac{1}{N^4}
\end{align}

We insisted on $\langle \mathcal{O}_3 \mathcal{O}_3 \mathcal{O}_3 \mathcal{O}_3 \rangle$ and 
$\langle \mathcal{O}_4 \mathcal{O}_4 \mathcal{O}_4 \mathcal{O}_4 \rangle$  since these two correlators 
capture generic features of our discussion about the semishort sector, 
and furthermore because they will be investigated in Section \ref{sec:one_loop_corr}, 
where we will construct explicitly their one-loop completion. 
We will see then how crucial it is the information from the semishort sector for our bootstrap program.


\section{OPE in AdS$_5\times$S$^5$: Beyond Tree-Level}\label{sec:OPEbeyond}


We now turn to the study of correlation functions of single-particle operators in the interacting theory. 
In particular, we consider $\mathcal{N}=4$ SYM in the regime of large 't Hooft coupling
$\lambda \equiv g_{YM}^2 N$ with $N \gg \lambda$ and $\lambda$ fixed.
In this interacting corner of $\mathcal{N}=4$ SYM, the theory sits at the boundary of a classical AdS${}_5\times$S${}^5$. 
The size of the holographic space-time is controlled by $L^4/\alpha'^{\,2}=4\pi \lambda$, 
and the action of IIB supergravity is weighted by $N^2$. 
Quantum corrections are then organised in a double expansion in $1/N^2$ and $\lambda^{-{1}/{2}}$.

The partial non-renormalization theorem \cite{Eden:2000bk} is a non perturbative 
statement about superconformal symmetry, and restricts the most general form of 
the four-point correlator into the sum of free theory, and a particular form for the dynamical function, 
\bea
\langle \mathcal{O}_{p_1}\mathcal{O}_{p_2} \mathcal{O}_{p_3} \mathcal{O}_{p_4} \rangle& =& 
\langle \mathcal{O}_{p_1} \mathcal{O}_{p_2} \mathcal{O}_{p_3} \mathcal{O}_{p_4} \rangle_{\rm free} 
+ \mathcal{P}\, \mathcal{I}(x,\bar{x};y,\bar{y})\, { \mathcal{D}}_{p_1p_2p_3p_4}(x,\bar{x};y,\bar{y};\lambda)
\eea
where $\mathcal{I}(x,\bar{x};y,\bar{y})$ is the same rational function characterizing 
long superblocks in \eqref{Iintriligator}, and $\mathcal{P}$ is the prefactor \eqref{Pfactor}. 

Contrary to free theory, the dynamical function depends on both $N$ and $\lambda$. 
Here we will be focussing on the order zero terms in the large $\lambda$ expansion, 
and consequently we will drop the $\lambda$ dependence in our discussion. 
Stringy corrections have been considered in \cite{Goncalves:2014ffa,Alday:2018pdi,Alday:2018kkw,Binder:2019jwn,Drummond:2019odu}.

In the previous section we studied the SCPW decomposition of free theory. 
In particular we cleanly split the free theory correlator into the piece with only (semi)-short operators 
in the CPW expansion and a piece with only long operators 
$\langle \mathcal{O}_{p_1}  \mathcal{O}_{p_2}  \mathcal{O}_{p_3} \mathcal{O}_{p_4} \rangle_{\rm free\, long}$ 
in \eqref{equa_shortlong}. We will now incorporate the dynamical function, $\mathcal{D}_{p_1p_2p_3p_4}$ and 
specialize to the long sector.  
It will be convenient to distinguish the two $1/N$ expansions, 
\bea
\langle \mathcal{O}_{p_1} \mathcal{O}_{p_2} \mathcal{O}_{p_3} \mathcal{O}_{p_4} \rangle_{\rm free\, long}&=&
\langle \mathcal{O}_{p_1} \mathcal{O}_{p_2} \mathcal{O}_{p_3} \mathcal{O}_{p_4} \rangle^{0}_{\rm free\, long}+ \frac{1}{N^2}\langle \mathcal{O}_{p_1} \mathcal{O}_{p_2} \mathcal{O}_{p_3} \mathcal{O}_{p_4} \rangle^{(1)}_{\rm free\, long}+\ldots\\
\mathcal{D}_{\vec{p}}&=&\frac{1}{N^2} \mathcal{D}^{(1)}_{\vec{p}}+ \frac{1}{N^4} \mathcal{D}^{(2)}_{\vec{p}}+\ldots
\eea
The notation we will use to refer to the SCPW expansion of the long sector
of $\langle \mathcal{O}_{p_1} \mathcal{O}_{p_2} \mathcal{O}_{p_3} \mathcal{O}_{p_4} \rangle$, (i.e. the long sector of free theory together with the dynamical  part) up to order $1/N^4$, is 
\bea
&&
\langle \mathcal{O}_{p_1} \mathcal{O}_{p_2} \mathcal{O}_{p_3} \mathcal{O}_{p_4} \rangle\Big|_{\rm long}= \notag\\[.2cm]
&&
\label{log0proj}
\quad \log^0 u \sum_{\vec{\tau}} \left( L^{(0)}_{\vec{p};\vec{\tau}}+ \frac{1}{N^2} L^{(1)}_{\vec{p};\vec{\tau}} + \frac{1}{N^4} L^{(2)}_{\vec{p};\vec{\tau}} \right) \mathbb{L}_{\vec{p};\vec{\tau}}  + \ldots \\ 
 &&
 \label{log1proj}
\quad \log^1 u \sum_{\vec{\tau}}\ \left( 
\frac{1}{N^2} M^{(1)}_{\vec{p};\vec{\tau}} + \frac{1}{N^4} M^{(2)}_{\vec{p};\vec{\tau}} \right) \mathbb{L}_{\vec{p};\vec{\tau}} + \ldots \\ 
&&
 \label{log2proj}
\quad \log^2 u \sum_{\vec{\tau}}\ \, \left( 
\frac{1}{N^4} N^{(2)}_{\vec{p};\vec{\tau}} \right) \mathbb{L}_{\vec{p};\vec{\tau}} 
\eea
In the above formulae we are omitting terms which are accompanied by derivatives of the blocks with respect to $\tau$ since these are not important for our purpose here.

The $\log^{m\ge 1}$ terms receive contributions only from the dynamical function, $\mathcal D$. 

The $\log^0$ projection \eqref{log0proj} is subject to non trivial interplay between free theory and the 
dynamical function ${\mathcal{D}}$, since beyond the leading order, both contribute in the $1/N$ expansion, 
\bea
 L^{(0)}_{\vec{p};\vec{\tau}}&=&L^{f(0)}_{\vec{p};\vec{\tau}}\\ 
 \label{Libigger0}
 L^{(i)}_{\vec{p};\vec{\tau}}&=&L^{f(i)}_{\vec{p};\vec{\tau}}+L^{\mathcal{D}(i)}_{\vec{p};\vec{\tau}}\qquad i=1,\ldots
\eea

In  \eqref{log0proj}-\eqref{log2proj}
we clustered together various contributions within each $\log$ strata, and we did not specify the range of summation. 
In fact, understanding the range of summation for different contributions needs extra explanations, which we make precise
in Sections~\ref{Unmix_section_sugra} and~\ref{UnmixingII}.
We will summarize all the relevant results in Section~\ref{back_to_future} . 

Every SCPW coefficient in \eqref{log0proj}-\eqref{log2proj}, is predicted by the OPE, 
however in order to have control on these predictions we should first have control on the spectrum of the theory: 
The spectrum of supergravity consists of single particle
half-BPS operators $\mathcal{O}_p$ and multi-particle operators built out of single particle operators. 
Multi-particles operators can be either protected or long, but regardless of this, 
multi-particle operators labelled by more than two particles do not have a leading order three-point 
function with the (normalized) external operators,  therefore cannot appear in the leading order OPE.
It follows that in the strict large $N$ limit, supergravity describes a free theory of single- and two-particles 
states of integer twist $\tau$, which we can then classify.

Recall then, that a basis of {\it long} two-particle superconformal primary operators of twist $\tau$, 
spin $\ell$ and $su(4)$ channel $[aba]$, has the schematic form \cite{Aprile:2018efk}, 
\be \label{Kpq}
\mathcal{O}_{pq;\vec{\tau}}=\mathcal{O}_{p} \partial^l \Box^{\frac12(\tau-p-q)}  \mathcal{O}_{q} \,, \qquad (p \leq q)\,.
\ee
where the pairs $(p,q)$ are in the set ${R}_{\vec{\tau}}$
\begin{align}\label{ir}
 {R}_{\vec{\tau}}:=\left\{(p,q):\begin{array}{l}
	p=i+a+1+r\\q=i+a+1+b-r\end{array}\text{ for } \begin{array}{l}
	i=1,\ldots,(t-1)\\ r=0,\ldots,(\mu -1)\end{array}
	\right\}
\end{align}
There are in total $d=\mu(t-1)$ allowed values, where 
\beq\label{multiplicity}
t\equiv \frac{(\tau-b)}{2}-a\,,\quad 
\mu \equiv   \left\{\begin{array}{ll}
\bigl\lfloor{\frac{b+2}2}\bigr\rfloor \quad &a+l \text{ even,}\\[.2cm]
\bigl\lfloor{\frac{b+1}2}\bigr\rfloor \quad &a+l \text{ odd.}
\end{array}\right.
\eeq
Plotting the set ${R}_{\vec{\tau}}$ in the $(p,q)$ plane helps visualizing that the pairs fill in a rectangle, where
the $+\pi/4$ direction contains $(t-1)$ pairs and the $-\pi/4$ direction contains $\mu$ pairs.  
For example, we (re)draw here below ${R}_{24,2\mathbb{N},0,6}$, 
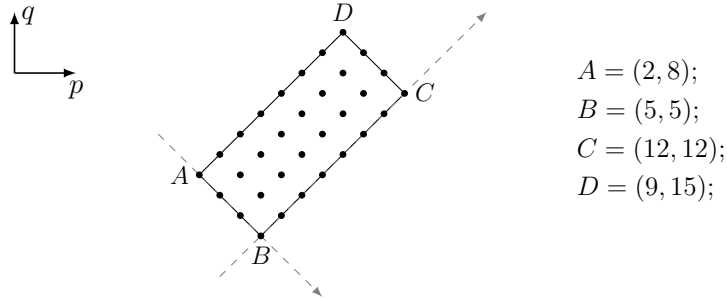
\beq
\begin{tikzpicture}[scale=.54]
%
%
\def\prop{.5}
\def\shifthor{\prop*2}
\def\ptuno{(\prop*2-\shifthor,\prop*8)}
\def\ptdue{(\prop*5-\shifthor,\prop*5)}
\def\pttree{(\prop*9-\shifthor,\prop*15)}
\def\ptquattro{(\prop*12-\shifthor,\prop*12)}
%
\draw[-latex, line width=.6pt]		(\prop*1   -\shifthor-4,         \prop*14          -0.5*\shifthor)    --  (\prop*1  -\shifthor-2.5  ,   \prop*14-      0.5*\shifthor) ;
\node[scale=.8] (oxxy) at 			(\prop*1   -\shifthor-2.5,  \prop*16.5     -0.5*\shifthor)  {};
\node[scale=.9] [below of=oxxy] {$p$};
%
\draw[-latex, line width=.6pt] 		(\prop*1   -\shifthor-4,     \prop*14       -0.5*\shifthor)     --  (\prop*1   -\shifthor-4,        \prop*17-      0.5*\shifthor);
\node[scale=.8] (oxyy) at 			(\prop*1   -\shifthor-2,   \prop*16.8   -0.5*\shifthor) {};
\node[scale=.9] [left of= oxyy] {$q$};
%
\draw[] 								\ptuno -- \ptdue;
\draw[black]							\ptuno --\pttree;
\draw[black]							\ptdue --\ptquattro;
\draw[]								\pttree--\ptquattro;
\draw[-latex,gray, dashed]					(\prop*0-\shifthor,\prop*10) --(\prop*8-\shifthor,\prop*2);
\draw[-latex,gray, dashed]					(\prop*3-\shifthor,\prop*3) --(\prop*16-\shifthor,\prop*16);
%
%
\foreach \indeyc in {0,1,2,3}
\foreach \indexc  in {2,...,9}
\filldraw   					 (\prop*\indexc+\prop*\indeyc-\shifthor, \prop*6+\prop*\indexc-\prop*\indeyc)   	circle (.07);
%
%
\node[scale=.8] (puntouno) at (\prop*4-\shifthor,\prop*8) {};
\node[scale=.8]  [left of=puntouno] {$A$};   
\node[scale=.8] (puntodue) at (\prop*5-\shifthor,\prop*6+.5) {};
\node[scale=.8] [below of=puntodue]  {$B$}; 
\node[scale=.8] (puntoquattro) at (\prop*13-\shifthor,\prop*15) {};
\node[scale=.8] [below of=puntoquattro] {$C$};
\node[scale=.8] (puntotre) at (\prop*9-\shifthor,\prop*13) {};
\node[scale=.8] [above of=puntotre] {$D$}; 
%
%
\node[scale=.84] (legend) at (11,5) {$\begin{array}{l}  
													\displaystyle A=(2,8); \\[.1cm]
													\displaystyle B=(5,5); \\[.1cm]
													\displaystyle C=(12,12); \\[.1cm]
													\displaystyle D=(9,15); \\[.1 cm] \end{array}$  };
\end{tikzpicture}
\notag
\eeq
In general the operators $\mathcal{O}_{pq,\vec{\tau}}$ will mix. We will denote the true eigenstates (with well-defined scaling dimensions) by $\mathcal{K}_{pq}$.

Considering now a four point function $\langle \mathcal{O}_{p_1} \mathcal{O}_{p_2} \mathcal{O}_{p_3} \mathcal{O}_{p_4} \rangle$, 
the OPE predicts the following form of the SCPW coefficients for the indicated ranges of the twist of the exchanged operators 
\bea
\label{summary1}
\tau \geq \tau^{\rm max}: \quad L^{(0)}_{\vec{p};\vec{\tau}}&=& \ \ \sum_{(pq)\in R_{\vec{\tau}}} C^{(0)}_{p_1p_2 
\cK_{pq} } C^{(0)}_{p_3p_4 
\cK_{pq}}\\
\label{summary2}
M^{(1)}_{\vec{p};\vec{\tau}}&=& \tfrac{1}{2} \sum_{(pq)\in R_{\vec{\tau}}} C^{(0)}_{p_1p_2 
\cK_{pq} }\, \eta_{
\cK_{pq} }\, C^{(0)}_{p_3p_4
\cK_{pq} }\\
\label{summary3}
N^{(2)}_{\vec{p};\vec{\tau}}&=& \tfrac{1}{8}\sum_{(pq)\in R_{\vec{\tau}}} C^{(0)}_{p_1p_2
\cK_{pq} }\, \eta^2_{
\cK_{pq} }\, C^{(0)}_{p_3p_4 
\cK_{pq} }\\[.3cm]
\label{summary4}
\tau^{\rm max} > \tau \geq \tau^{\rm min}: \quad L^{(1)}_{\vec{p};\vec{\tau}}&=& 
\ \ \sum_{(pq)\in R_{\vec{\tau}}} C^{(1)}_{p_1p_2 
\cK_{pq} } C^{(0)}_{p_3p_4 
\cK_{pq} }+C^{(0)}_{p_1p_2 
\cK_{pq} } C^{(1)}_{p_3p_4
\cK_{pq} }\\
\label{summary5}
M^{(2)}_{\vec{p};\vec{\tau}}&=& \tfrac{1}{2}\sum_{(pq)\in R_{\vec{\tau}}} C^{(1)}_{p_1p_2 
\cK_{pq} } \, \eta_{
\cK_{pq} }\, C^{(0)}_{p_3p_4 
\cK_{pq} }+C^{(0)}_{p_1p_2 
\cK_{pq} } \, \eta_{
\cK_{pq} }\, C^{(1)}_{p_3p_4 
\cK_{pq} }\\[.3cm]
\label{summary6}
\tau < \tau^{\rm min}:\quad L^{(2)}_{\vec{p};\vec{\tau}}&=& \ \ 
\sum_{(pq)\in R_{\vec{\tau}}} C^{(1)}_{p_1p_2 
\cK_{pq} } C^{(1)}_{p_3p_4 
\cK_{pq} }
\eea
where $C_{p_ip_j 
\cK} \equiv C_{p_ip_j}(\mathcal{K})$ denotes the three-point coupling, 
\be
C_{p_ip_j 
\cK}=C^{(0)}_{p_ip_j 
\cK}+\frac{1}{N^2}  C^{(1)}_{p_ip_j 
\cK}+\ldots 
\ee
of an exchanged pure (and unit normalised) operator $
\cK_{pq;\vec{\tau}}$ w.r.t to the external operators $\mathcal{O}_{p_i}\mathcal{O}_{p_j}$, and $\eta_{
\cK}$ is its anomalous dimensions, 
\be
\Delta_{
{\cK} }= \tau+\ell +\frac{1}{N^2}  \eta_{
\cK} + \ldots \ .
\ee 

The pure operators $\cK_{pq;\vec{\tau}}$ (i.e. those with well-defined anomalous dimensions) are simply certain linear combinations  of  the operators $\mathcal{O}_{pq;\vec{\tau}}$ in~\eqref{Kpq}. At leading order the explicit change of basis is  given in terms of the above leading three-point functions as
{ 
	\begin{align}
	\cK_{pq} = \sum_{(p'q')\in R_{\vec{\tau}} }^{} \mathcal{O}_{p'q'} \frac{C^{(0)}_{p'q'\cK_{pq}}}{C^{(0)}_{p'q'\mathcal{O}_{p'q'}}} + O(1/N^2)\ ,
	\end{align}
	where the normalisation $\mathcal{N}_{pq}$ is fixed (up to a sign) by insisting on unit two-point function. As before we note that the leading three-point functions with the naive operators are diagonal, i.e. $C^{(0)}_{pq\mathcal{O}_{p'q'}} = 0$ unless $(pq)=(p'q')$, as can be easily verified via explicit Wick contractions.
}

Our next task is to leverage data mined from tree-level four-point functions, 
specifically the CPW coefficients $L^{(0)}, M^{(1)}$ and $L^{(1)}$, in order to 
obtain information about the one-loop four-point function, in particular the entire 
double log discontinuity, $N^{(2)}$, but also pieces of the single log part $M^{(2)}$ and analytic part  $L^{(2)}$.
 The strategy used in our previous works \cite{Aprile:2017bgs} \cite{unmixing} \cite{Aprile:2017qoy} 
 was to fully solve the mixing problem and obtain complete data for the leading three-point 
 couplings and anomalous dimensions of long operators in $[a,0,a]$ and $[a,1,a]$, 
 by considering the leading ( i.e. disconnected free theory) four-point functions and 
 the (leading discontinuity of the) $1/N^2$ correction of Rastelli and Zhou \cite{Rastelli:2016nze, Rastelli:2017udc}. 
However, for more general correlators, specifically when $b>1$, 
the anomalous dimensions exhibit degeneracy \cite{Aprile:2018efk}. 
This degeneracy means it is not possible to fully repeat this program 
in the same way and determine all leading three-point functions. 
Nevertheless we will now discuss how to overcome this problem 
and bootstrap one-loop data from tree-level correlators.

In order to have better control over the various phenomena taking place in \eqref{summary1}-\eqref{summary6},
we will distinguish between three types of contributions:
\begin{itemize}
\item {\bf Above threshold, $\tau\geq \tau_{\vec{p}}^{\rm max}$}. 

We define the threshold twist $\tau^{\rm max}_{\vec{p}}\equiv {\rm max}(p_1+p_2,p_3+p_4)$. 
The  leading log CPW coefficients, $L^{(0)}$, $M^{(1)}$, and $N^{(2)}$,
have contributions above the threshold only, 
and we use $L^{(0)}$ and $M^{(1)}$ (for different correlators) to bootsrap 
the double discontinuity $N^{(2)}$.  This analysis 
 is similar to our previous works \cite{Aprile:2018efk}. 
For general correlators the leading log discontinuity alone
 is not sufficient to fully fix the one-loop dynamical function consistenly. 
 There are also important pieces of the one-loop functions which will be fully determined by data below threshold.

\item {\bf The window $ \tau^{\rm min} \leq \tau  <\tau^{\rm max}$}. 

We define the ``window'' as the range of twists strictly below $\tau_{\vec{p}}^\text{max}=\max(p_1{+}p_2,p_3{+}p_4)$ 
and above $\tau_{\vec{p}}^\text{min}=\min(p_1{+}p_2,p_3{+}p_4)$. Clearly the window 
is absent when $p_1+p_2 = p_3+p_4$. The significance of this region is that the leading 
three-point function is absent on one side but not on the other (see diagram below).
This allows us to use tree-level SCPW coefficients $L^{(1)}$ to predict  part of the single 
discontinuity of the one-loop correlator,  $M^{(2)}$. 

\item {\bf Below window, $\tau<\tau^{\rm min}$}. 

Finally, for each $su(4)$ channel $[aba]$ we define the region 
``below window", for which the twist is strictly below $\tau_{\vec{p}}^\text{min}=\min(p_1{+}p_2,p_3{+}p_4)$. 
For any $su(4)$ rep we also require $\tau \geq 2+2a+b$, i.e. the unitarity bound relative to the rep $[aba]$. 
In this range of twists one can predict a piece of the analytic (in small $u$) part of the one-loop 
correlator, $L^{(2)}$, using lower order data, specifically results from  $L^{(1)}$ for other correlators. 
\end{itemize}

The three regions described above are shown pictorially in Figure~\ref{fig1}.

The precise details about how to obtain these predictions are described in the next subsections.

\subsection{Predicting the Leading Log}\label{Unmix_section_sugra}

Let us begin from the $\log^2u$ discontinuity $N^{(2)}$. 
We shall use many different correlators and it is thus useful to  
package together the CPW coefficients, and three-point functions, into matrices. Similarly we want to  
rewrite the equations~\eqref{summary1}-\eqref{summary6} in matrix form. 
To this end we view the space of operators with given twist, spin and $su(4)$ 
labels as a $\mu(t-1)$ dimensional vector space denoted with the (multi-) index $(pq) \in R_{\vec{\tau}}$. 

Define the $\mu(t-1)\times \mu(t-1)$ square matrix of leading order three-point functions 
$\mathbf{C}^{(0)}_{\vec{\tau}}$ which has entries
\be\label{C0}
\left(\mathbf{C}^{(0)}_{\vec{\tau}}\right)_{(rs),(pq)}=  
						C^{(0)}_{ rs \mathcal{K}_{pq;\vec{\tau}}} \qquad \text{with} \qquad (pq), (rs)\in R_{\vec{\tau}} \ ,
\ee 
and the matrix of leading CPW coefficients ${\bf L}^{(0)}_{\vec{\tau}}$ which has entries
\be\label{data_disco}
\left({\bf L}^{(0)}_{\vec{\tau}}\right)_{(p_1p_2),(p_3p_4)}=  
						{L}^{(0)}_{\vec{p};\vec{\tau}} \qquad \text{with} \qquad (p_1p_2), (p_3p_4)\in R_{\vec{\tau}}\ .
\ee
As argued in Section \ref{Intro} we only obtain a non-zero result at leading order for twists $\tau \geq \tau^{\rm max}$, i.e. above threshold.

Then, the equation arising from the OPE and the SCPW expansion in \eqref{summary1} 
is written in matrix form simply as:
\be\label{data_disco2}
\mathbf{C}^{(0)}_{\vec{\tau}} \, \mathbf{C}^{(0)T}_{\vec{\tau}}
						={\bf L}^{(0)}_{\vec{\tau}}
\ee
(where for notational convenience we have also dropped the subscripts 
denoting the quantum numbers of interest with $\vec{\tau}$).
By construction this matrix is symmetric. Moreover, it is diagonal, because   
only correlators on the diagonal have disconnected diagrams. Very explicitly  
\cite{Aprile:2018efk}~\footnote{A similar formula was given in \cite{Caron-Huot:2018kta}.}
\bea\label{formula_free_disc_FA}
{\bf L}^{(0)}_{\vec{\tau}}={\rm diag}\left[
\tfrac{ (a+1)(a+2+2r+q-p) (l+1)(l+2+\tau)  (1+r+q-p)_{r+1}(2+a+r+q-p)_{2+a+r}}{ (1- { }\delta_{p,q}/2) (p-1-r)!(p-2-r-a)!(q+r)!(q+r+1+a)! r! (r+1+a)!}\Pi_{\tfrac{\tau}{2}} \Pi_{l+1+\tfrac{\tau}{2} }\right]_{
\substack{ p=2+a+i+r \\ q=p+b-2r\ \ } }
\eea
where $(i,r)$ label the pairs $(pq)\in R_{\tau,l,[aba]}$, as explained \eqref{ir}, and the function $\Pi_s$ can be given in the form, 
\be
\Pi_s=\frac{ \left( s-\tfrac{q-p}{2} \right)!\left(s+\tfrac{q+p}{2}\right)!}{(2s)!} \times 
\frac{ \prod_{k=0}^{b+2a+2+i} \left(s+\tfrac{p+q}{2}-k \right) }{(s+\tfrac{4+2a+b}{2})(s+\tfrac{2+b}{2})(s-\tfrac{b}{2}) } \times  \prod_{m=1}^{i} \left(s-\tfrac{p+q}{2}+m\right).
\ee
The three factors $(s+\tfrac{4+2a+b}{2})(s+\tfrac{2+b}{2})(s-\tfrac{b}{2})$ cancel against the numerator 
for the values of $k=i,i+a+1,i+a+b+2$ respectively. Then, for fixed $\vec{\tau}$, the spin dependent factors of 
the elements of the disconnected free theory matrix ${\bf L}^{(0)}_{\vec{\tau}}$ are the following
\bea
&&
{L}^{(0)}_{pqpq;\vec{\tau}}\ \sim\quad
\Bigl[\tfrac{ \left(\frac{2+2l+\tau+q-p}{2}\right)!\left(\frac{2+2l+\tau-q+p}{2}\right)! }{  (2+2l+\tau)!}\Bigr]\times (l+1)(l+\tau +2) 
\prod_{k=0}^{a-1} \tfrac{2l+2+\tau-b-2a+2k}{2}\times \notag \\
&&
\prod_{k=1}^b \tfrac{2l+2+\tau-b+2k}{2}\ \prod_{k=2}^{a+1} \tfrac{2l+2+\tau+b+2k}{2} 
\prod_{k=1}^{\tfrac{1}{2}(p+q-(b+2a+4))} (\tfrac{ 2l+2+\tau-b-2a-2-2k}{2} ) (\tfrac{2l+2+\tau+b+2a+4+2k}{2} )\qquad
\label{facto_dfree}
\eea
Aside from the factorials in the square brackets, the dependence is a polynomial in spin of degree $(p+q-2)$ 
which is fully factorised into linear factors. It follows that for fixed $\vec{\tau}$, the highest degree polynomial factor 
in ${\bf L}^{(0)}_{\vec{\tau}}$ occurs for correlators with $p+q=\tau$. For fixed $\vec{\tau}$ the polynomial depends 
only on the combination $p+q$. The factorials in ${L}^{(0)}_{pqpq;\vec{\tau}}$ instead depend on both $p$ and $q$.

Define also the matrix ${\bf M}^{(1)}_{\vec{\tau}}$ of leading log tree-level CPW coefficients 
of operators with given quantum numbers $\vec{\tau}$, at order $1/N^2$,  
and the (diagonal) matrix of their anomalous dimensions
$\pmb{\eta}_{\vec{\tau}}$. Then~\eqref{summary2} becomes
\bea\label{data_rastelli}
\tfrac{1}{2} \mathbf{C}^{(0)}_{\vec{\tau}}
							\, \pmb{ \eta}_{\vec{\tau}}
							\, \mathbf{C}^{(0)T}_{\vec{\tau}}
														={\bf M}^{(1)}_{\vec{\tau}}\ .
\eea

In matrix form it is straightforward to see that the CPW coefficients 
contributing to the $\log^2 u$ discontinuity at  one-loop are given by
\begin{align}\label{N2}
	\mathbf{N}^{(2)}_{\vec{\tau}} = \tfrac18 \mathbf{C}^{(0)}_{\vec{\tau}}
	\, \pmb{ \eta}^2_{\vec{\tau}}
	\, \mathbf{C}^{(0)T}_{\vec{\tau}} = \tfrac12{\bf M}^{(1)}_{\vec{\tau}} \left({\bf L}^{(0)}_{\vec{\tau}}\right)^{-1}   {\bf M}^{(1)}_{\vec{\tau}}\ ,
\end{align}
i.e. purely in terms of tree-level CPW coefficients. 
The second equality is obtained straightforwardly from \eqref{data_disco2} and \eqref{data_rastelli}.

In order to have ${\bf M}^{(1)}_{\vec{\tau}}$ explicitly, we first need an  
expression for the dynamical function $\mathcal{D}^{(1)}_{\vec{p}}$ at $O(1/N^2)$. 
We obtain $\mathcal{D}^{(1)}_{\vec{p}}$ by using the Mellin Amplitude of Rastelli and Zhou~\cite{Rastelli:2016nze}, 
together with the normalisation derived in~\cite{Aprile:2018efk}. 
Then, we use this to read off its partial wave 
decomposition and construct ${\bf M}^{(1)}_{\vec{\tau}}$. The details of this procedure are discussed in Appendix \ref{AppTree2}. 

Finally we use~\eqref{N2} to obtain the full double log discontinuity at $O(1/N^4)$,
\bea\label{ddisc_fa}
\mathcal{D}^{(2)}_{\vec{p}}\Big|_{\log^2 u}&=& 
				\sum_{\tau\ge\tau^\text{max},l,a,b } 
				\left({\bf N}_{\vec{\tau}}^{(2)}\right)_{(p_1p_2),(p_3p_4)} \widetilde{\,\mathbb{L}}_{\vec{p};\vec{\tau}}\,,
\eea 
where we recall that the long blocks take the form $\mathbb{L} = \mathcal{P} \times \mathcal{I} \times \widetilde{\mathbb{L}}$ as in (\ref{LONGSCPW}).

Note that the above method does not require us to find the leading 
anomalous dimensions $\pmb \eta$ or 3-point functions $ \bf C^{(0)}$ themselves. 
The anomalous dimensions $\pmb \eta_{\vec{\tau}}$ are just the eigenvalues of
\begin{align}\label{cetac}
 \tfrac{1}{2}
	 \mathbf{C}^{(0)}_{\vec{\tau}}
	\, \pmb{ \eta}_{\vec{\tau}}
	\, (\mathbf{C}^{(0)}_{\vec{\tau}})^{-1} ={\bf M}^{(1)}_{\vec{\tau}} \left({\bf L}^{(0)}_{\vec{\tau}}\right)^{-1} 
\end{align}
and eigenvalues can always be found unambiguously. {The eigenvectors however,
are ambiguous if there are repeated eigenvalues, as turns out to be the case here  \cite{Aprile:2018efk}. 
This is the aforementioned degeneracy of anomalous dimensions, which is 
consequence of a surprising physical statement about tree-level physics: 
There is a hidden ten-dimensional conformal symmetry which prevents the spectrum from being fully unmixed at tree-level \cite{Caron-Huot:2018kta}. 
}

The general solution for the anomalous dimensions is \cite{Aprile:2018efk}
\be\label{eta_anom_dim}
			\eta_{\mathcal{K}_{pq}}=
						- \frac{ 1 }{N^2} \frac{ \delta^{(4)}_{t} \delta^{(4)}_{t+l+1}}{ \left(l+2p-2-a - \frac{1+(-)^{a+l}}{2} \right)_6}
\ee
where $(\ldots)_6$ is the Pochhammer symbol, $t\equiv \frac{(\tau-b)}{2}-a$ 
was defined in \eqref{multiplicity}, and\footnote{Compared to \cite{Aprile:2018efk} we 
added a factor of $2$ in the definition of $\delta^{(4)}_t$. }
\be
			\delta^{(4)}_t\equiv 
							2(t-1) (t+a) (t+a+b+1)(t+2a+b+2) \,.
\ee
Since $\eta_{pq}$ depends only on $p$ and not $q$, the anomalous dimensions are in general partially degenerate. 
States which all lie on the same vertical line in the figure $R_{\vec{\tau}}$ have the same anomalous dimension.

Notice that the r.h.s of \eqref{cetac} is rational in spin because it is obtained from ${\bf M}^{(1)}_{\vec{\tau}}$ and ${\bf L}^{(0)}_{\vec{\tau}}$ which are both rational in spin. 
Indeed, we emphasise that the original eigenvalue problem set up in \cite{unmixing} is more sophisticated than \eqref{cetac}, since it 
was set up to have a direct correspondence between eigenvectors and three-point couplings.

\subsection{Below threshold predictions}\label{UnmixingII}

A feature of four-point correlators of half-BPS single-particle operators with generic charges 
is that one can bootstrap one-loop pieces of the $\log u$ 
and analytic part, below threshold. 
As well as the double log discontinuity, which is entirely above threshold, there is information from within the window and below the window which further constrains the four-point function. Remarkably using all of this available lower order data always fixes the one-loop four-point function up to certain well understood ambiguities which only have finite spin dependence in the SCPW expansion.

To begin with consider long SCPW coefficients of the analytic part of the 
tree-level correlator  $L^{(1)}$ arising from operators in the window region, 
$ \tau^{\rm min} \leq \tau  <\tau^{\rm max}$ (see Fig. \ref{fig1}). 
For simplicity assume $p_1+p_2 \geq p_3+p_4$ (the other case is similar), then~\eqref{summary4}  becomes
\be\label{equa_cprime}
L^{(1)}_{\vec{p};\vec{\tau}}=
						 \sum_{ (pq)\in{R}_{\vec{\tau}}  } 
						C^{(1)}_{p_1p_2 \cK_{pq} }  C^{(0)}_{p_3p_4 \cK_{pq}} \qquad 
						\tau^{\rm min} \leq \tau  <\tau^{\rm max},\quad p_1+p_2 \geq p_3+p_4\ .
\ee
The key point here is that there are new, leading three-point functions 
at $O(1/N^2)$ -- $C^{(1)}_{p_1p_2 \cK_{pq} } $ -- with below threshold twist $\tau < p_1+p_2$.

Fixing $(p_1 p_2)$ and $\vec \tau$, let us consider all values of $(p_3p_4)\in R_{\vec{\tau}}$ and 
rewrite~\eqref{equa_cprime} as a vector equation
\be\label{boot_V}
\mathbf{L}^{(1)}_{(p_1p_2);{\vec{\tau}}}=
\mathbf{C}^{(1)}_{(p_1p_2);{\vec{\tau}}} \mathbf{C}^{(0)T}_{\vec{\tau}}\ .
\ee
Here we have defined the  row
 vector  
$\mathbf{C}^{(1)}_{(p_1p_2);\vec{\tau}}$ with entries
\be\label{unmix_3ptfunc}
\left(\mathbf{C}^{(1)}_{(p_1p_2);\vec{\tau}}\right)_{(pq)} 
				=  {C}^{(1)}_{p_1p_2 \cK_{(pq);\vec{\tau}} }  
				\qquad 
									{\rm \forall }(pq)\in R_{\vec{\tau}} 
									 \ 
\ee
and the row vector of no-log  $O(1/N^2)$ CPW coefficients, ${\bf L}^{(1)}_{(p_1p_2);\vec{\tau}}$, with entries
\be
\left({\bf L}^{(1)}_{(p_1p_2);\vec{\tau}}\right)_{(p_3p_4)} =  
					L^{(1)}_{\vec{p};\vec{\tau}}\qquad {\rm \forall }(p_3p_4)\in R_{\vec{\tau}} \ .
\ee

The other ingredient is the matrix of leading three-point couplings $\mathbf{C}^{(0)}_{\vec{\tau}}$ defined in~\eqref{C0}. 

Consider now the $\log u$ part of the one-loop four-point function 
with SCPW coefficients $M^{(2)}$~\eqref{summary5}. 
In direct analogy to ${\bf L}^{(1)}_{(p_1p_2);\vec{\tau}}$ above,
define the vector ${\bf{M}}^{(2)}_{(p_1p_2);\vec{\tau}}$. 
The OPE in the window~\eqref{summary5} (for varying $(p_3p_4)\in R_{\vec \tau}$), becomes the vector equation
\begin{align}
	{\bf{M}}^{(2)}_{(p_1p_2);{\vec{\tau}}}= \tfrac12
							\mathbf{C}^{(1)}_{(p_1p_2);{\vec{\tau}}} {\pmb \eta_{\vec{\tau}}}
							\mathbf{C}^{(0)T}_{\vec{\tau}}
\ .
\end{align}
If we know $\mathbf{C}^{(0)}_{\vec{\tau}}$ we can explicitly solve for $\mathbf{C}^{(1)}_{(p_1p_2);{\vec{\tau}}}$ 
using~\eqref{boot_V} and plug in here to get the one-loop SCPW coefficients ${\bf{M}}^{(2)}_{(p_1p_2);{\vec{\tau}}}$.
However even if we don't, because of the degeneracy of the anomalous dimensions, 
we see that by using~\eqref{data_disco2},\eqref{data_rastelli} and \eqref{boot_V} we obtain ${\bf{M}}^{(2)}_{(p_1p_2);{\vec{\tau}}}$ purely in terms of tree-level SCPW data as
\begin{align}\label{bM2}
	{\bf{M}}^{(2)}_{(p_1p_2);{\vec{\tau}}} = 
		{\bf{L}}^{(1)}_{(p_1p_2);{\vec{\tau}}}({\bf{L}}^{(0)}_{\vec{\tau}})^{-1} {\bf{M}}^{(1)}_{\vec{\tau}}\ .
\end{align}
We thus bootstrap a piece of the single log coefficient of the one-loop correlator from tree-level data.

In a very similar way, 
pieces of the analytic part of the one-loop four-point function, namely the coefficients $L^{(2)}$ for twists below the window, 
can be determined purely in terms of tree-level SCPW coefficients. 
From~\eqref{summary6} we find
\begin{align}\label{L2}
	L^{(2)}_{\vec{p};{\vec{\tau}}} = 
	\mathbf{C}^{(1)}_{(p_1p_2);{\vec{\tau}}} \mathbf{C}^{(1)T}_{(p_3p_4);{\vec{\tau}}} = 
	\mathbf{L}^{(1)}_{(p_1p_2);{\vec{\tau}}}({\bf{L}}^{(0)}_{\vec{\tau}})^{-1}\mathbf{L}^{(1)T}_{(p_3p_4);{\vec{\tau}}} 
	\qquad 
	4{+}2a{+}b\leq\tau <p_3+p_4 
	\ .
\end{align}
%
%
%
%
Recall that the SCPW coefficients $L^{(1)}$ appearing in \eqref{L2} are determined by summing contributions of
$\mathcal{D}^{(1)}_{\vec{p}}$, and connected free theory, as in \eqref{Libigger0}.
A general formula for connected free theory at order $1/N^2$ 
can be obtained using results in \cite{Aprile:2018efk}, 
and was presented already in \cite{Caron-Huot:2018kta}. 
In our notation, it is recorded in Appendix \ref{AppTree1}. 

In the above discussion we suppressed the dependence on the spin $l$ of the exchanged operator. 
Let us now be more concrete and describe how the quantities ${\bf{M}}^{(2)}_{(p_1p_2);{\vec{\tau}}}$ 
and $L^{(2)}_{\vec{p};{\vec{\tau}}}$, obtained from matrix multiplication, depend on spin.
In fact, such a spin dependence follows from a)
the knowledge of the spin dependence of disconnected free theory, 
b) the spin dependence of tree-level SCPW coefficients  \cite{Aprile:2018efk}, 
given explicitly in  \eqref{fitcalM} Appendix \ref{Apptree3prima},
and c) reciprocity symmetry $l\leftrightarrow -l-\tau-3$. 
Proofs of the following formulas are collected in Appendix \ref{App_structure_spin}.

When $b$ is even we can treat even and odd spins separately. In each of these cases we find,
\bea\label{test_M2}
M^{(2)}_{\vec{p};\tau,l,[aba]}&=& 
\tfrac{ \left(\frac{2+2l+\tau+p_{43}}{2}\right)!\left(\frac{2+2l+\tau-p_{21}}{2}\right)! }{  (2+2l+\tau)!}\times
\frac{ num_{M^{(2)}}(2l+\tau+3) }{ den_{M^{(2)}}(2l+\tau+3) } \\
\label{test_L2}
L^{(2)}_{\vec{p};\tau,l,[aba]}&=& 
\tfrac{ \left(\frac{2+2l+\tau+p_{43}}{2}\right)!\left(\frac{2+2l+\tau-p_{21}}{2}\right)! }{  (2+2l+\tau)!}\times 
\frac{ num_{L^{(2)} }(2l+\tau+3) }{ {den}_{L^{(2)}}(2l+\tau+3) }
\eea
The polynomials $num$ and $den$ are even in the variable $(2l+\tau+3)$, with coefficients which depend on $\tau$.
The denominators are fully predicted by the formula \eqref{facto_dfree} for $p+q=\tau$. In particular they have degree $\tau-2$. 
The numerators have degree,
\bea\label{estimates_M2_spin}
{\rm degree}~num_{M^{(2)}}(l)&=& (\tau-4) + 2(\kappa_{\vec{p}}-2) +p_{21}\\
\label{estimates_L2_spin}
{\rm degree}~num_{L^{(2)}}(l)&=& 2(\tau-4)-(p_{43}-p_{21})
\eea

When $b$ is odd, the symmetry $l\leftrightarrow -l-\tau-3$ exchanges even and odd spin. By picking the sector of even spins as representative, we find
\bea
M^{(2)}_{\vec{p};\tau,even,[aba]}&=&
\tfrac{ \left(\frac{2+2l+\tau+p_{43}}{2}\right)!\left(\frac{2+2l+\tau-p_{21}}{2}\right)! }{  (2+2l+\tau)!}\times
\frac{ (l+\tau+2)^2 num^{even}_{M^{(2)}}(l) }{ den^{even}_{M^{(2)}}(l) } \\
L^{(2)}_{\vec{p};\tau,even,[aba]}&=& 
\tfrac{ \left(\frac{2+2l+\tau+p_{43}}{2}\right)!\left(\frac{2+2l+\tau-p_{21}}{2}\right)! }{  (2+2l+\tau)!}\times 
\frac{ (l+\tau+2)^2 num_{L^{(2)} }(l) }{ {den}_{L^{(2)}}(l) }
\eea 
where the denominators are again predicted by the formula \eqref{facto_dfree} for $p+q=\tau$, and 
\bea
{\rm degree}~num_{M^{(2)}}(l)&=& (\tau-4) + 2(\kappa_{\vec{p}}-2) +p_{21}-2\\
{\rm degree}~num_{L^{(2)}}(l)&=& 2(\tau-4)-(p_{43}-p_{21})-2
\eea

Summarizing, from all the results given above we can determine the following pieces of the $O(1/N^2)$ four-point functions.

$\bullet$ $\log^1u$ stratum obtained from a finite number of twists, $\forall$ spin,\\
\bea\label{logu1stratum}
\mathcal{D}^{(2)}_{\vec{p}}\Big|_{\log^1 u} =  \tfrac{1}{2}\,\log^1 u \sum_{l,a,b}  \sum_{ \tau=\tau^\text{min} }^{\tau^\text{max}-2}  \left({\bf M}_{(p_1p_2);\vec{\tau}}^{(2)}\right)_{(p_3p_4)} \widetilde{\,\mathbb{L}}_{\vec{p};\vec{\tau}} + \ldots 
\eea
where ${\bf M}^{(2)}$ is given in~\eqref{bM2} and we are omitting here terms contributing to twist $\tau \geq \tau^{\rm max}$.

$\bullet$  $\log^0u$ stratum obtained from a finite number of twist, $\forall$ spin,\\
\bea\label{analitic_prediction_uno}
\mathcal{D}^{(2)}_{\vec{p}}\Big|_{\log^0 u}
							=   \mathcal{D}^{(2)}_{\rm bound} +  \sum_{l,a,b } \sum^{ \tau^{\rm min}-2}_{  \tau= 4+2a+b }  L^{(2)}_{\vec{p};\vec{\tau}}\ 
										\widetilde{\mathbb{L}}_{\vec{p};\vec{\tau}} + \ldots 
\eea
with $L^{(2)}$ given in~\eqref{L2} and we are omitting here terms contributing to twist $\tau \geq \tau^{\rm min}$.
There is an extra subtlety which needs to be tackle in order to determine fully the $\log^0 u$ stratum. 
It enters the contribution called $\mathcal{D}^{(2)}_{\rm bound}$, and has to do with multiplet recombination at the unitarity bound, $\tau=2+2a+b$, in each channel. 
We will discuss this in the next section.

Equations~\eqref{bM2} and~\eqref{L2} show how to obtain 
SCPW coefficients of the one-loop four-point functions directly in terms of
SCPW coefficients of tree-level functions. Unmixing of the CFT data is not necessary to achieve this.

There are cases in which we can unmix explicitly the three-point couplings ${\bf C}^{(0)}$. Referring these cases to rectangle ${R}_{\vec{\tau}}$ given in \eqref{ir},
they correspond to operators in $su(4)$ representations  $[n,0,n]$ and $[n,1,n]$, label by $\mu=1$ $\forall t$, and operators
$\cK\in {R}_{4+2a+b,l,[aba]}$, labeled by $t=2$ $\forall \mu$, for 
which the rectangle $R_{\vec{\tau}}$ collapses into a line, and therefore the degeneracy has no space yet to show up. 
Given the knowledge of ${\bf C}^{(0)}$ we can proceed to obtain the subleading three-point couplings.  
A number of examples is given in Appendix~\ref{Apptree3}, where we also comment on the so called derivative relation.  

{Despite the fact that the explicit three-point functions and anomalous dimensions are not needed 
to produce the one-loop OPE predictions described above, we emphasise that they do follow a \emph{significantly} simpler pattern, compared to
the partial wave coefficients they are obtained from. 
The beauty of the pattern is manifest in the structure of the anomalous dimension \eqref{eta_anom_dim}, but also in the three-point couplings unmixing when possible. 
As it was found in \cite{unmixing} and \cite{Aprile:2017qoy}, 
the three-point couplings unmixing always reduces to the problem of finding unitarity matrices with predicted spin dependence.}

\subsection{Semi-short sector predictions}
\label{UnmixingIII}

As anticipated in Section \ref{sec:recomb}, we now come back 
to the delicate point of multiplet recombination at the unitarity bound.   
In  \eqref{analitic_prediction_uno} we gave the one-loop $\log^0u$ predictions 
which originate from twists {\em above} the unitary bound, 
i.e $4+2a+b\leq \tau< \tau^{\rm min}_{\vec{p}}$.
In addition, we claim that the dynamical one-loop function 
must contain a contribution {\em at} the unitarity bound $\tau=2+2a+b$, 
which we also predicted. Namely,
\bea\label{master_recom}
L^{(2)}_{\vec{p};{\vec{\tau}}}&=& 
					L^{(2)f}_{ \vec{p};\vec{\tau} }
				     + L^{(2)\mathcal{D}^{} }_{\vec{p};\vec{\tau}} =0, 
				\qquad \qquad \tau=2+2a+b\\
\label{L2fromPaul}
L^{(2)f}_{ \vec{p};2+2a+b,l,[aba] }&=&
					\sum_{k=0}^a (-1)^k {A}_{[l+2+k,1^{a-k}]} \Big|_{\frac{1}{N^4}}
					- \sum_{k=0}^a (-1)^k {S}_{[l+2+k,1^{a-k}]}
\eea
The coefficient $L^{(2)f}_{\tau=2+2a+b}$ was obtained in \eqref{longfree}. 
Its first term is given by the CPW of connected free theory 
${A}_{2+2a+b-2k,[l+2+k,1^{a-k}]}$ restricted at order $1/N^4$. 
Its second term is given by summing over the new coefficients 
$S_{2+2a+b-2k,[l+2+k,1^{a-k}]}$, and it follows non trivially from the 
analysis of the semishort sector, which by construction is of order $1/N^4$. 
The contributions to the analytic, $\log^0 u$ part, of $\mathcal{D}^{(2)}$, which come
from twists at the unitarity bound, combine to give the function $\mathcal{D}^{(2)}_{\rm bound}$ in (\ref{analitic_prediction_uno}) in the form
\be
\mathcal{D}^{(2)}_{\rm bound} = - \sum_{l,a,b} L^{(2)f}_{\vec{p};\vec{\tau}} \widetilde{\mathbb{L}}_{\vec{p};\vec{\tau}}\,.
\ee

The reason for \eqref{master_recom}  is the following: the OPE of 
$\mathcal{O}_{p_i}\mathcal{O}_{p_j}$ in free theory runs, by definition, 
over all operators of $\mathcal{N}=4$ SYM, but supergravity states correspond 
only to operators built from half-BPS operators, i.e. they are either half-BPS operators 
themselves or multi-particle operators. Other single-trace operators at the unitarity bound, 
which are present in free theory, correspond thus to excited string states, and should 
be absent from the OPE in the supergravity regime.  

Simple examples of operators which correspond to excited string states are the Konishi 
operator ${\rm tr} (\phi^2)$ in the $[000]$ representation, and twist $3$ superconformal 
primaries of the form ${\rm tr}( \phi^3)$ in the $[010]$ representation.
However, these two cases are special because there are no other existing operators  with such quantum numbers. 
In particular, there will be no $S$-type contribution in \eqref{L2fromPaul}. 
Beyond twist $3$, instead, we have to distinguish carefully between multi-trace semishort 
operators, which do remain in the spectrum of supergravity, 
and excited string states, as was done in Section \ref{sec:recomb}.

It is very instructive to compare the new features of $1/N^4$ physics with corresponding tree-level terms. 
Let us begin from the analogue of equation \eqref{master_recom} at tree level. It reduces to 
\be\label{master_recom_tree}
L^{(1)}_{\vec{p};2+2a+b,l,[aba]}= 
		\underbrace{ \sum_{k=0}^a (-1)^k {A}_{[l+2+k,1^{a-k}]} \Big|_{\frac{1}{N^2}} }_{= L^{(1)f}_{\vec{p};2+2a+b,l,[aba]} }
		+ L^{(1)\mathcal{D}}_{\vec{p};2+2a+b,l,[aba]} =0
\ee
The difference compared to \eqref{master_recom}, 
is precisely the difference between performing multiplet recombination with 
CPW coefficients of connected free theory -- assuming all  below threshold 
($\tau=2+2a+b<\tau^\text{min}$) semishort operators are absent -- and 
performing multiplet recombination with remaining below threshold semishort operators. 
This is just because the semishort three-point functions are all of $O(1/N^2)$ 
and so are only visible in the SCPW decomposition at $O(1/N^4)$.

Indeed, the leading three-point function $C^{(0)}_{p_ip_j K_{pq} }=0$ whenever $p_i+p_j>\tau$, thus  
this vanishing condition extends to the non-semishort  ``below window'' sector, $\tau<\tau^\text{min}$ 
at tree level.
Thus 
\be\label{split}
L^{(1)}_{\vec{p};\vec{\tau}}= {L^{(1)f}_{\vec{\tau}}}
+ L^{(1)\mathcal{D}}_{\vec{p};\vec{\tau}} =0
  \qquad \qquad \forall \tau<\tau^\text{min}\ ,
\ee
with the free theory part, $ L^{(1)f}$ given in~\eqref{Lcoefs} when $\tau$ is above the 
unitary bound $\tau \geq 4+2a+b$ and~\eqref{master_recom_tree} when at the bound $\tau=2+2a+b$.

\subsection{Back to the Bootstrap}\label{back_to_future}

In the previous three sections we have explained how to bootstrap, 
from tree-level results, predictions about the dynamical one-loop 
function of $\langle \mathcal{O}_{p_1} \mathcal{O}_{p_2} \mathcal{O}_{p_3} \mathcal{O}_{p_4} \rangle$. Summarizing, 
we have obtained the leading $\log^2 u$ discontinuity (see discussion around \eqref{N2}-\eqref{ddisc_fa}). Then, we have obtained pieces of the single $\log^1 u$ 
from exchanged operators in the window (see discussion around \eqref{bM2} and \eqref{logu1stratum}), and also pieces of the analytic $\log^0 u$ part of the correlator
from below window data (see discussion around \eqref{analitic_prediction_uno} and \eqref{L2}). 
Finally, we understood how to deal with the SCPW coefficient of long operators at the unitarity bound 
in \eqref{L2fromPaul}. We emphasize that even though the
leading log discontinuity can be obtained more elegantly by using the hidden symmetry of \cite{Caron-Huot:2018kta}, 
our approach here 
allows us  to go beyond that, and compute $M^{(2)}$ and $L^{(2)}$, which are very important pieces of our bootstrap program.

The OPE predictions introduced so far were organised according to the $\log u$ 
stratification of the correlators given in \eqref{log0proj}-\eqref{log2proj}.  
We now point out that the structure of the $O(1/N^4)$ dynamical function admits a further refinement. 

Consider first the following observation: Looking at below threshold physics at tree level 
we found that the analytic sector of the dynamical function $\mathcal{D}^{(1)}$ is subject to the constraints~\eqref{split}, i.e
\be\label{requirementD1}
L^{(1)\mathcal{D}}_{\vec{p};\vec{\tau}} =- {L^{(1)f}_{\vec{\tau}}}
 \qquad \qquad \forall 4+2a+b\leq\tau<\tau^\text{min}\ . 
\ee
augmented by a similar constraint at the unitarity bound, given in \eqref{master_recom_tree}. 
We claim (and we will show in section \ref{NewRastelli}) that $\mathcal{D}^{(1)}$ 
is entirely fixed by these constraints, together with the requirement that 
its $\log u$ discontinuity has threshold twist $\tau^{\rm max}$.

Consider now the analytic sector at one-loop, we find instead  
\be
L^{(2)\mathcal{D}}_{\vec{p};\vec{\tau}} =- {L^{(2)f}_{\vec{\tau}}} +L^{(2)}_{\vec{p};\vec{\tau}}
\qquad \qquad  4+2a+b\leq \tau<\tau^\text{min}\ .
\ee
where $L^{(2)}$ is the new $O(1/N^4)$ prediction \eqref{L2} arising from the tree-level data via the OPE. 
It is clear then that the analytic part of $\mathcal{D}^{(2)}$ has two separate contributions, 
one cancelling free theory contribution, i.e $- {L^{(2)f}_{\vec{\tau}}}$, and another one linked 
to predictions from tree-level data  $L^{(2)}_{\vec{p};\vec{\tau}}$.
Furthermore, at the unitarity bound we find a similar split into a piece depending directly on free theory SCPW coefficients
and a non-trivial prediction arising from correlators of different charges~\eqref{L2fromPaul}.
Since the $\log^2$ and $\log^1$ strata of $\mathcal{D}^{(2)}$ are determined uniquely 
by tree-level data via the OPE (see sections~\ref{Unmix_section_sugra} and~\ref{UnmixingII}),
and have no free theory contribution, it is natural to split  the one-loop function into
\be\label{d=t+h}
\mathcal{D}^{(2)}=\mathcal{T}^{(2)}+ \mathcal{H}^{(2)}\,,
\ee 
where $\mathcal{T}^{(2)}$ and $\mathcal{H}^{(2)}$ have a different interplay with connected free theory.

The function $\mathcal{T}^{(2)}$ generalises the tree-level function\, $\mathcal{D}^{(1)}$, and it will be defined 
by the properties that it has $\log^1 u$ discontinuity with threshold twist $\tau^{\rm max}$, no $\log^2 u$ double discontinuity, 
and it fully cancels below window long contributions coming from recombined free theory, 
hence the name of generalised tree-level function.
Indeed,  for any $4+2a+b\leq \tau < \tau^\text{min}$, i.e. strictly above the unitary bound, we expect
\bea
\label{LT} 
	L^{(2)\mathcal{T}}_{\vec{p};\vec{\tau}}=-L^{(2)f}_{\vec{p};\vec{\tau}} 
\eea
with $L^{(2)f}$ the $O(1/N^4)$ part of~\eqref{Lcoefs}, and at the unitarity bound
\be\label{LT2}
	L^{(2)\mathcal{T}}_{\vec{p};2+2a+b,l,[aba]}=-\sum_{k=0}^a (-1)^k {A}_{[l+2+k,1^{a-k}]} \Big|_{\frac{1}{N^4}}
\ee

It follows that the one-loop OPE predictions \eqref{L2}, below the window, 
will be encoded only in the function $\mathcal{H}^{(2)}$, i.e.
\bea \label{newL2due}
L^{(2)\mathcal{H}}_{\vec{p};\vec{\tau}}&=&
					L^{(2)}_{\vec{p};\vec{\tau}} \qquad \qquad 4+2a+b\leq \tau <\tau^\text{min}\ .
\eea
and at the unitarity bound
\bea
\label{newL2uno}
L^{(2)\mathcal{H}^{} }_{\vec{p};2+2a+b,l,[aba]} &=& +\sum_{k=0}^a (-1)^k {S}_{[l+2+k,1^{a-k}]}
\eea

Our task now is to construct the one-loop correlator $\mathcal{D}^{(2)}$ consistently with the OPE predictions. 
We will see that the splitting $\mathcal{D}^{(2)}=\mathcal{H}^{(2)}+\mathcal{T}^{(2)}$ is also 
strongly motivated by features of the $\log^2u$ discontinuity.  In fact, we will discover that 
$\mathcal{H}^{(2)}$ is the minimal one-loop function which consistently emanates top-down from the leading $\log^2u$ discontinuity. 
Furthermore, we will find that $\mathcal{T}$ can be constructed as an exact function of $N$. 
The interplay of $\mathcal{H}^{(2)}$ with the semishort prediction \eqref{newL2uno} is very remarkable. 
When we think of it as descending from the double logarithmic discontinuity, 
it is a tangible triumph of supergravity within our $\mathcal{N}=4$ bootstrap program.


\section{One-loop Correlators}\label{sec:one_loop_corr}


The discussion throughout Section \ref{sec:OPEbeyond}  provided us with predictions for the one-loop function
that we now make operative by introducing an ansatz which will suit them all. 
To understand this ansatz and impose as many constraints as possible, we will first
consider the consequence of crossing symmetry and those of the OPE  on the structure of one-loop correlators, 
and secondly we will obtain a (two-variable) resummation of the 
leading double log discontinuity. 
We then go on to assemble this and the information below threshold to obtain the one-loop correlator.

\subsection{Structure of One-loop Correlators}\label{structure_oneLoop}

From the OPE we expect different parts of the correlator to possess contributions 
from operators of different twists (see the previous section). 
The $\log^2 u$ discontinuity has contributions only from operators above threshold $\tau \geq \tau^\text{max}$. 
The $\log^1 u$ part can have contributions from the window, $ \tau \geq  \tau^\text{min}$. 
Finally, the analytic $\log^0 u$ part, can have  contributions starting from the semishort operators with $\tau \geq p_{43}+2$ (see figure~\ref{fig1}.)

Because a long operator of twist $\tau$ gives a contribution to the correlator 
which goes like  $\mathcal P\times\mathcal{I}\times u^{\frac12(\tau-p_{43})}$ for small $u$, 
the OPE then dictates that 
\begin{align}\label{smallu}
	\mathcal{D}^{(2)}_{\vec{p}}|_{\log^2 u} &= O(u^{\frac12(\tau^\text{max}-p_{43})})\notag \\
	\mathcal{D}^{(2)}_{\vec{p}}|_{\log ^1u} &= O(u^{\frac12(\tau^\text{min}-p_{43})})\notag \\
	\mathcal{D}^{(2)}_{\vec{p}}|_{\log^0 u} &= O(u)\ .
\end{align}
where
\begin{align}
\tfrac{1}{2}(\tau^\text{max}-p_{43})&={\rm max}(p_3,\tfrac{p_1+p_2+p_3-p_4}{2} )\notag\\
\tfrac{1}{2}(\tau^\text{min}-p_{43})&={\rm min}(p_3,\tfrac{p_1+p_2+p_{3}-p_4 }{2}) =\kappa_{\vec{p}}
\end{align}
The latter is precisely the degree of extremality.

Consider now the split  $\mathcal{D}^{(2)}=\mathcal{T}^{(2)}+\mathcal{H}^{(2)}$. 
We claim that only $\mathcal{T}^{(2)}$  has a contribution at $O(u)$ whereas $\mathcal{H}^{(2)} = O(u^2)$. 
The reason for this follows again from the detailed understanding of the semishort sector:
The contributions at $O(u)$ arise from semishort operators with twist $p_{43}+2$ 
in the $[a=0,p_{43},0]$ $su(4)$ representation.  In this case there is a single  $A$-type 
contribution in the sum of~\eqref{longfree}, which is to be dealt with by $\mathcal{T}^{(2)}$, 
and a single  $S$ contribution, to be dealt with by $\mathcal{H}^{(2)}$. Recall that we deal with 
the split $\mathcal{D}^{(2)}=\mathcal{T}^{(2)}+\mathcal{H}^{(2)}$ by using \eqref{LT2} and \eqref{newL2uno}.
Then notice that the $S$ contribution itself is obtained in~\eqref{form} in terms of 
SCPW coefficients $S_{q_r \tilde{q}_r p_3p_4}$ where $q_r+\tilde{q}_r=p_{43}+2$. 
But these correlators are next to extremal, and they completely vanish 
when we use the correct definition of single-particle operators -- as discussed 
below~\eqref{extremal} -- so the $S$ contribution vanishes at this twist.
Thus 
\begin{align}\label{difftd}
	\mathcal{T}^{(2)}|_{\log^0 u} = O(u) \qquad \qquad  \mathcal{H}^{(2)}|_{\log^0 u} = O(u^2)\ .
\end{align}

Under crossing $u\leftrightarrow v$, the analysis of the small $u$ expansion in \eqref{smallu} 
translates into predictions for the small $v$ expansion, which is then useful to understand how 
to constrain the ansatz for the full function.

For the correlator itself crossing symmetry very simply implies that 
\begin{align}
	\langle\mathcal{O}_{p_1}(x_1)\mathcal{O}_{p_2}(x_2)\mathcal{O}_{p_3}(x_3)\mathcal{O}_{p_4}(x_4)\rangle =\langle\mathcal{O}_{p_{\sigma_1}}(x_{\sigma_1})\mathcal{O}_{p_{\sigma_2}}(x_{\sigma_2})\mathcal{O}_{p_{\sigma_3}}(x_{\sigma_3})\mathcal{O}_{p_{\sigma_4}}(x_{\sigma_4})\rangle\ ,
\end{align}
for any permutation $\sigma \in S_4$. 
The implications of this taking into account the prefactor $\mathcal P$ requires a little care.  
When defining the prefactor we always made the choice $0\leq p_{21} \leq p_{43}$ 
which should therefore be maintained under the permutation, whilst sending $u \leftrightarrow v$. 
This requires considering  a number of different cases for the relative values of the charges $p_i$. 
In all cases however there is a unique permutation $\sigma$ satisfying the above requirements and one finds that for this permutation
\begin{align}
	\mathcal D^{(2)}_{p_1p_2p_3p_4}(u,v) = \left(\frac {u\tau}v \right)^{\kappa_{\vec{p}}}\mathcal D^{(2)}_{p_{\sigma_1}p_{\sigma_2}p_{\sigma_3}p_{\sigma_4}}(v,u) \ .
\end{align} 
The small $u$ behaviour of $D^{(2)}_{p_{\sigma_1}p_{\sigma_2}p_{\sigma_3}p_{\sigma_4}}(u,v)$ given in~\eqref{smallu} then yields  the following small $v$ behaviour of $\mathcal{D}^{(2)}_{\vec{p}}(u,v)$
\begin{align}\label{smallv}
\mathcal{D}^{(2)}_{\vec{p}}|_{\log^2 v} &= O(v^{\frac12(p_1+p_4-p_2-p_3)})\notag \\
\mathcal{D}^{(2)}_{\vec{p}}|_{\log^1 v} &= O(v^0)\notag\\
\mathcal{D}^{(2)}_{\vec{p}}|_{\log^0 v} &= O( 1/v^{\kappa_{\vec{p}}\,-1} )\ .
\end{align}
Further, the different small $u$ behaviour of $\mathcal T^{(2)}$ and $\mathcal H^{(2)}$ in~\eqref{difftd} implies a different small $v$ limit, namely
\begin{align}\label{consideration_smallv}
	\mathcal{T}^{(2)}_{\vec{p}}|_{\log^0 v} = O(1/v^{\kappa_{\vec{p}}\,-1}) \qquad \qquad \mathcal{H}^{(2)}_{\vec{p}}|_{\log^0 v} = O(1/v^{\kappa_{\vec{p}}\,-2})\ .
\end{align}

The differences in the $v$ behaviour between $\mathcal{H}$ and $\mathcal{T}$ are crucial in the determination of  our ansatz.

\subsection{Resummation of Leading Logs}\label{resummationLL}

Only $\mathcal{H}^{(2)}$ carries the $\log^2 u$ discontinuity in the splitting $\mathcal{D}^{(2)}=\mathcal{T}^{(2)}+\mathcal{H}^{(2)}$, by definition.  
We can then use the arguments of the previous section to infer that in the small $u$ and $v$ expansion we expect
\be
\label{expectedH2log2}
\mathcal{H}^{(2)}_{\vec{p}} \Big|_{\log^2 u} = u^{ -\tfrac{p_{43}}{2}+\tfrac{ {\rm max}({p_1+p_2},{p_3+p_4} )}{2} }\left[ O(v^{ \frac{p_{43}-p_{21}}{2} }) \log^2 v +O(v^{0}) \log v +O(1/v^{\kappa_{\vec{p}}\,-2}) \right] 
\ee
As explained in section \ref{Unmix_section_sugra},
the leading log discontinuity is defined by the sum
\bea\label{ddisc_fa2}
\mathcal{D}^{(2)}_{\vec{p}}\Big|_{\log^2 u}&=& 
\sum_{\vec{\tau}: \tau\ge\tau^\text{max} } \left({\bf N}_{\vec{\tau}}^{(2)}\right)_{(p_1p_2),(p_3p_4)} \widetilde{\,\mathbb{L}}_{\vec{p};\vec{\tau}}
\eea
where $\mathbf{N}^{(2)}=\tfrac12{\bf M}^{(1)} \left({\bf L}^{(0)}\right)^{-1}   {\bf M}^{(1)}$ is the matrix assembled from tree level data.

By explicit computation of \eqref{ddisc_fa2} to higher order in $\tau$  we obtained the resummation of the leading log discontinuity 
in a number of cases, and deduced that, as function of the external charges, it always has the structure
\bea
\label{anatomy_ddisc}
\begin{array}{ccl}
\mathcal{D}^{(2)}_{\vec{p}}\Big|_{\log^2 u}&=&
			\frac{ {P}_{2,1}(x,{\bar x},\sigma,\tau) }{ \ (x-{\bar x})^{\mathbf{d}_{\vec{p}}+8 } }\ \Big[ {\rm Li}_2(x)-{\rm Li}_2({\bar x}) \Big] 
			+ \frac{ {P}_{1,2}(x,{\bar x},\sigma,\tau) }{ \  (x-{\bar x})^{\mathbf{d}_{\vec{p}}+8} } \Big[ {\rm Li}^2_1(x)-{\rm Li}^2_1({\bar x}) \Big]+ \\[.3cm]
			& &\frac{ {P}_{1,-}(x,{\bar x},\sigma,\tau) }{ \ (x-{\bar x})^{\mathbf{d}_{\vec{p}}+8 } }\,\Big[ {\rm Li}_1(x)-{\rm Li}_1({\bar x}) \Big] 
			+ \frac{ {P}_{1,+}(x,{\bar x},\sigma,\tau) }{ \  (x-{\bar x})^{\mathbf{d}_{\vec{p}}+7} } \log(v) 
			+ \frac{ {P}_{0}(x,{\bar x},\sigma,\tau) }{ \  (x-{\bar x})^{\mathbf{d}_{\vec{p}}+7} }\frac{1}{v^{\kappa_{\vec{p}}-2}} 
\end{array}
\eea
where
\be
\mathbf{d}_{\vec{p}}=p_1+p_2+p_3+p_4-1,
\ee
for certain polynomials ${P}$ depending implicitly on the external charges $\vec{p}$. 
 These polynomials are obtained 
by matching the series expansion in $x$ and $\bar{x}$ of \eqref{anatomy_ddisc}, 
with that on the r.h.s. of \eqref{ddisc_fa2}. The latter is obtained
by considering that each conformal block, of twist $\tau$ and spin $l$, has a series
representation of the form $u^{\tau}(1-v)^l f(x,\bar{x})$ where $u=x\bar{x}$, $v=(1-x)(1-\bar{x})$, 
and $f$ is an analytic symmetric function in $x$ and $\bar{x}$. 
We call an expression of the form (\ref{anatomy_ddisc}) a \emph{two-variable resummation}.

After a case-by-case inspection of \eqref{anatomy_ddisc} we indeed verify its consistency 
with the general structure \eqref{expectedH2log2}.  In particular, we always find an 
overall $u^{(\tau^\text{max}-p_{43})/2}$ factor, which gives the leading term in the small $u$ expansion. 
Then, in the small $v$ expansion, the $\log^2v$ behaviour is dictated by ${P}_{1,2}$, 
and goes like $ v^{ {(p_{43}-p_{21})}/{2} }$, the $\log^1 v$ behaviour is given by the limit 
of $-{P}_{1,-}+{P}_{1,+}$, and goes like $v^{0}$, and finally
only the ${\log^0 v}$ contribution has a singularity of the form $1/v^{\kappa_{\vec{p}}\,-2}$.

In fact a ten-dimensional conformal structure observed  in~\cite{Caron-Huot:2018kta} 
was found to give a direct formula for these leading logs. 
We checked in many cases that our results agree, and we 
postpone to Section \ref{exploration10d} a more detailed description of this 
ten-dimensional structure.

\subsection{Minimal One-Loop functions}\label{minimal_loop_sec}

We now have all the relevant information  to write an ansatz for the minimal one-loop function $\mathcal{H}^{(2)}$, 
which is consistent with crossing symmetry, and matches the two-variable resummation of the leading $\log^2u$ discontinuity.

We consider single-valued transcendental functions up to weight-4 functions antisymmetric in $x\leftrightarrow \bar{x}$. 
The weight counting follows from the resummation \eqref{anatomy_ddisc} in which we find an overall $\log^2 u$ 
paired at most with weight-2 anti-symmetric transcendental functions. Therefore we need a basis for weight-4 
antisymmetric transcendental functions, and their lower weight completion. We can make it very explicit, 
by introducing the series of ladder integrals,
\be
\phi^{(\ell)}=\sum_{r=0}^{\ell} \frac{(-)^r (2\ell-r)!}{l!(l-r)!r!} (\log u)^r ( {\rm Li}_{2\ell-r}(x)-{\rm Li}_{2\ell-r}({\bar x}) )
\ee
Then, the basis has the following form
\bea\label{basis1}
\Wloop_{4-}&=&\hpiccolo_1\, \phi^{(2)}(x'_1,x'_2) + \hpiccolo_2\,  \phi^{(2)}(x,{\bar x}) + \hpiccolo_3\, \phi^{(2)}(1-x,1-{\bar x})
\notag\\[.3cm]
\Wloop_{3-}&=&  \hpiccolo_4\, x \partial_{x} \phi^{(2)}(x,{\bar x}) +\hpiccolo_5\, (x-1)\partial_{x}  \phi^{(2)}(1-x,1-{\bar x})  - (x \leftrightarrow {\bar x} )
\notag\\[.2cm]
\Wloop_{3+}&=& (x-{\bar x})\Big(  \hpiccolo_6\, \partial_v  \phi^{(2)}(x,{\bar x})+ \hpiccolo_7\, \partial_u  \phi^{(2)}(1-x,1-{\bar x})\Big) + \hpiccolo_8\, \zeta_3
\notag\\[.2cm]
\Wloop_{2+}&=&\hpiccolo_9 \log(u)\log(v)  +\hpiccolo_{10} \log^2 v +\hpiccolo_{11}\log^2 u
\eea
and
\beq\label{basis2}
\begin{array}{ll}
\Wloop_{2-}=\hpiccolo_{\Box}\, \phi^{(1)}(x,{\bar x})\qquad& \,\Wloop_{0}\,= \hpiccolo_0\\[.2cm]
\Wloop_{1u}\,=\hpiccolo_{u} \log u \qquad&  \Wloop_{1v}=\hpiccolo_{v} \log v\\
\end{array}
\eeq
The weight -4 and -3 basis have been written in terms of the double box function, which is the $\ell=2$ integral in the ladder series.
The weight-2 anti-symmetric element is instead the $\ell=1$ box function. 
Each coefficient function $\hpiccolo_{i=1,..,11,\Box,u,v,0}$ will be polynomial 
in the variables $x,{\bar x},\ssigma,\ttau$.

From considerations about crossing in \eqref{consideration_smallv}, 
and the structure of the two-variable resummations \eqref{anatomy_ddisc}, 
we conclude that the ansatz for the minimal one-loop function is given by
\bea
\label{G2uvresult}
\mathcal{H}_{\vec{p}}^{(2)} &=&
				\frac{ \Wloop_{4-} +\Wloop_{3-} }{(x-{\bar x})^{\mathbf{d}_{\vec{p}}+8}} 
				+ \frac{1}{(x-{\bar x})^{\mathbf{d}_{\vec{p}}+7}}\left[\Wloop_{3+} 
				+ \frac{ \Wloop_{2+}}{v^{\kappa_{\vec{p}} -2}} \right]
\notag \\[.2cm]
& & 
			+\frac{1}{ v^{\kappa_{\vec{p}} -2} }\left[
						\frac{  \Wloop_{2-} }{ (x-{\bar x})^{\mathbf{d}_{\vec{p}}+8} }
						+ \frac{ \Wloop_{1v} + \Wloop_{1u}  }{(x-{\bar x})^{\mathbf{d}_{\vec{p}}+7} }+
							\frac{ \Wloop_{0}}{ (x-{\bar x})^{ \mathbf{d}_{\vec{p}}+5}} \right]\qquad
\eea
where recall
\be
\qquad \mathbf{d}_{\vec{p}}=p_1+p_2+p_3+p_4-1,\qquad \kappa_{\vec{p}}=\tfrac{ {\rm min}({p_{1}+p_2},{p_3+p_4})}{2} -\tfrac{p_{43}}{2} .
\ee

A smaller basis made of just the box function $\phi^{(1)}$, together with its weight one and weight zero completions will be referred to as {\it tree-like}. 
For example, any $\overline{D}$ function can be decomposed in such a basis.  However, 
consistently with our splitting of the one-loop function as $\mathcal{D}^{(2)}=\mathcal{T}^{(2)}+\mathcal{H}^{(2)}$, 
we will point out in which way the tree-like coefficient functions for $\Wloop_{2-},\Wloop_{1u},\Wloop_{1v},\Wloop_0$ really encode physics beyond tree level.

In the following we describe our bootstrap algorithm, 
going through the sequence of steps that need to be performed in order to obtain $\mathcal{H}^{(2)}_{\vec{p}}$.

\subsubsection*{Crossing Symmetry and Leading Log matching}


For any orientation of the external charges $\vec{p}$, we consider the $\log^2 u$ projection 
of the ansatz and match with the explicit two-variable resummation described in~\ref{resummationLL}.
This fixes combinations of coefficient functions from $W_{4-},W_{3\pm},W_{2+}$. Note that the power $v^{\kappa_{\vec{p}}-2}$ 
in the denominator of $W_{2+}$~\eqref{G2uvresult} is consistent with the weight-0 part of the leading log~\eqref{anatomy_ddisc}. 
Matching all independent leading log discontinuities actually  fixes completely the polynomials $h_{1,2,..,7,9,.. 11}$. 
When $\kappa_{\vec{p}}=2$, the correlators are next-to-next-to extremal, for example $2222$, and $2233$, 
and there is no singular $v$ behavior in the ansatz.  In these cases our ansatz \eqref{G2uvresult} reduces 
to the ansatz considered in our previous works \cite{Aprile:2017bgs,Aprile:2017qoy}.

\subsubsection*{Absence of unphysical poles}


Any leading log discontinuity has itself no poles at $x=\bar{x}$.
However, that only counts the $\log^2 u$ projection of the function $\mathcal{H}^{(2)}$. 
In order for the ansatz to produce a well defined function we have to ensure that globally there are no unphysical poles.  
In this way, lower weight coefficient functions become entangled with those at weights -4, -3 and -2 symmetric. In particular,
both the power of $(x-\bar{x})$ in the denominators, and the coefficient functions 
of $\Wloop_{2-}, \Wloop_{1u},\Wloop_{1v}$ and $\Wloop_0$, have the right structure 
such that all $x=\bar{x}$ poles coming from weight -4, -3 and -2 symmetric coefficient functions can be cancelled. 
For this reason the `tree-like' coefficients functions of $\mathcal H^{(2)}$, $h_\Box,h_u,h_v,h_0$, have quite different features compared to their counterparts at tree level.
In this process we can keep $v^{\kappa_{\vec{p}}-2}$ as the maximum singular power in the denominator.

\subsubsection*{Matching the OPE prediction in and below window}


At this stage of the algorithm we have found a well defined ansatz with the correct $\log^2 u$ discontinuities. 
It differs from $\mathcal{H}^{(2)}_{\vec{p}}$ because we have not yet imposed the remaining predictions 
in and below window, which we have to compute explicitly by using the strategy outlined in section \ref{UnmixingII} and \ref{UnmixingIII}. Such OPE predictions come as SCPW coefficients at fixed twist, and varying spin, i.e. from a sum like 
\be
\label{windowsum}
\sum_{l} c^{\tau_0}_l \mathcal{B}^{\tau_0,l} + \ldots + \sum_{l} c^{\tau_k}_l \mathcal{B}^{\tau_k,l}\,,
\ee 
where $c^{\tau}_l$ stands for $M^{(2)}_{\tau,l}$ or $L^{(2)}_{\tau,l}$, and $k<\tau^{\rm max}$ is finite.    
Given the analytic representation of the conformal blocks, 
we can series expand the sum (\ref{windowsum})
in the form 
\be
u^{\tau_0} \sum_{n=0}^{\tau_k-\tau_0} \sum_{m=0}^{\infty} d_{nm} x^{n} \bar{x}^{m}
\ee
and then resum it as 
\be
\label{onevbleresum}
x^{\tau_0} \sum_{n=0}^{\tau_k-\tau_0} x^n g_n(\bar{x})\,.
\ee
where the functions $g_n$ contain transcendetal functions of one-variable. Indeed 
the ansatz for these $g_n$ descends from the two-variable ansatz \eqref{G2uvresult}, upon performing the same series expansion as in \eqref{onevbleresum}.
We call an expression of the form (\ref{onevbleresum}) a \emph{one-variable resummation}.

The initial number of free coefficients grows with $p_1+p_2+p_3+p_4$, 
because of the denominator factors $(x-\bar{x})$ in \eqref{G2uvresult}, 
and obviously with the number of $su(4)$ channels. 
Cancelling $x=\bar{x}$ poles alone still leaves 
a large number of free coefficients.  
Imposing OPE predictions in and below window is indeed crucial to finally obtain the minimal loop functions. 

\begin{table}[h!]
\centering
\begin{tabular}{|c||c|c|c|}
	\hline
	correlator& $\substack{\rule{0pt}{0.25cm}\displaystyle \text{initial\ free\ coeffs.}\\ \rule{0pt}{.35cm}\displaystyle \text{in } h_{\Box},h_{u},h_{v},h_{1}}$  & 
	$\substack{\rule{0pt}{0.45cm}\displaystyle \text{after leading log matching}\  \\ \displaystyle	\text{\ and pole\ cancellation}  \\~ }$  & 
	$\substack{\rule{0pt}{0.45cm}\displaystyle  \text{after\ OPE\ predictions}\\  \rule{0pt}{0.35cm}
	\displaystyle \text{ in\ and\  below\ window} \\ ~}  $\\\hline
 	2222 & $1\times 378$ & 1 & 1 \\\hline
 	2233 & $1\times 496$ & 16 & 2 \\\hline
 	2244 & $1\times  579 $ & 20 & 2 \\\hline
 	3333 & $3\times 579 $ & 20	& 2	\\\hline
 	4444 & $6\times 946 $ & 68 & 4 \\\hline 
\end{tabular}
\caption{Number of tree-like free coefficients across the three steps of our algorithm.}
\label{tab:free_coeffs}
\end{table}

\subsubsection*{Ambiguities}

Imposing predictions in and below window fixes the majority of the free coefficients in the ansatz. 
A sample of this process is illustrated in Table~\ref{tab:free_coeffs}.  
The free parameters left are associated to a restricted class of tree-like functions, which we call \emph{ambiguities}. 
By construction,  these ambiguities do not contribute to the $\log^2u$ discontinuity in any channel, 
obey the correct crossing transformations by themselves, have no $x=\xb$ poles, 
and contribute only above window, i.e for twists $\tau\ge \tau^{\rm max}_{\vec{p}}$.  
Furthermore, we find the special feature that their SCPW coefficients have finite spin support, $l=0,1$.

The Mellin amplitude corresponding to the ambiguities is very simple, since it can be at most linear in the Mellin variables $(s,t)$, for two reasons. 
Firstly, it cannot be rational, as any additional pole would spoil our predictions in and below the window.
Therefore it has to be polynomial. 
Secondly, this polynomial cannot be higher order than linear, as it would generate tree-like terms with a higher degree denominator 
than allowed by our ansatz~\eqref{G2uvresult} for the minimal one-loop function $\mathcal{H}_{\vec{p}}^{(2)}$. 

For a generic correlator without any crossing symmetries, we can parametrise the full set of ambiguities by
\begin{align}\label{eq:ambiguity_general}
	\mathcal{H}_{\vec{p}}^{(2)}\Big|_{\text{ambiguity}} =\frac{ u^{ \frac{p_3-p_4}{2} } }{ v^{\frac{p_2+p_3}{2} } }  
	\oint u^{\frac{s}{2}}v^{\frac{t}{2}} \ \Gamma_{\vec{p}}\ \sum_{i=0}^{\kappa_{\vec{p}}\,-2}\sum_{j=0}^{\kappa_{\vec{p}}\,-2 -i}\left(\alpha_{\vec{p}}^{(1,ij)}+\alpha_{\vec{p}}^{(s,ij)}s+\alpha_{\vec{p}}^{(t,ij)}t\right)\ssigma^i\ttau^j,
\end{align}
where $\kappa_{\vec{p}}$ is the degree of extremality \eqref{deg_extremality}, 
and ${\Gamma}_{\vec{p}}$ is the combination of Mack's Gamma functions 
\be\label{MACK}
{\Gamma}_{\vec{p}}= \Gamma\left[\tfrac{p_1+p_2-s}{2}\right] \Gamma\left[\tfrac{p_3+p_4-s}{2}\right]
				\Gamma[\tfrac{p_1+p_4-t}{2}]\Gamma\left[\tfrac{p_2+p_3-t}{2}\right]
				\Gamma\left[\tfrac{p_1+p_3-U}{2} \right]\Gamma\left[\tfrac{p_2+p_4-U}{2}\right].
\ee
In \eqref{MACK} we have introduced an auxiliary Mellin variable $U$ which makes crossing symmetry manifest, and it is defined as
\be
s+t+U=p_1+p_2+p_3+p_4-4.
\ee


Thus, for a generic correlator, we find $\frac{3(\kappa_{\vec{p}}-1)\kappa_{\vec{p}}}{2}$ undetermined ambiguities. In cases in which the correlator 
has some crossing symmetry, we have to count only crossing symmetric combinations. 

Let us construct explicitly the ambiguities for the correlators we will discuss in the next sections:
\begin{itemize}
\item $\langle\mathcal{O}_2\mathcal{O}_2\mathcal{O}_2\mathcal{O}_2\rangle$. 
The only fully crossing symmetric combination one can build is the constant Mellin amplitude $1$, so there can only be a single ambiguity: $\alpha_{2222}^{1}$.\footnote{
The corresponding function in position space is $\dbar{4444}$~\cite{Aprile:2017bgs}. The value of $\alpha_{2222}^{(1)}=60$ (in the conventions of~\cite{Aprile:2017bgs}) was found
by using a supersymmetric localisation computation~\cite{Chester:2019pvm}. 
Such a non-zero value breaks analyticity in spin for the twist 4 one-loop anomalous dimension at spin zero, in agreement with the argument from the Lorentzian inversion formula \cite{Alday:2017vkk}.}

\item $\langle\mathcal{O}_2\mathcal{O}_2\mathcal{O}_p\mathcal{O}_p\rangle$. 
This family of correlators is not fully crossing symmetric if $p>2$.
The remaining invariance can be understood as invariance under $t\leftrightarrow U$.
As a result, we are left with two out of three ambiguities, 
\be\label{amb_22pp_discussion}
\alpha_{22pp}^{1},\qquad {\rm and}\qquad \alpha_{22pp}^{2}s.
\ee

\item $\langle\mathcal{O}_3\mathcal{O}_3\mathcal{O}_3\mathcal{O}_3\rangle$. 
This correlator admits up to linear terms in $\ssigma$, and $\ttau$, but crossing symmetry only allows two (fully symmetric) ambiguities, 
\be\label{amb_3333_discussion}
\alpha_{3333}^{1}(1+\ssigma+\ttau)\qquad {\rm and} \qquad \alpha_{3333}^{2}(s+U\ssigma+t\ttau).
\ee 
A correlator with $\kappa=3$ but no crossing symmetries would admit $9$ ambiguities.

\item $\langle\mathcal{O}_4\mathcal{O}_4\mathcal{O}_4\mathcal{O}_4\rangle$. 
The full crossing symmetry of this correlator greatly reduces the number of ambiguities. 
With at most quadratic terms in $\ssigma$, and $\ttau$,  one can construct four independent ambiguities: 
two ambiguities with constant Mellin amplitudes 
\be\label{amb_4444_discusssion1}
\alpha_{4444}^{1}(1+\ssigma^2+\ttau^2)\qquad {\rm and}\qquad  
\alpha_{4444}^{2}(\ssigma+\ttau+\ssigma\tau)
\ee 
and two other ambiguities with linear terms 
\be\label{amb_4444_discusssion2}
\alpha_{4444}^{3}(s+U\ssigma^2+t\ttau^2),\qquad {\rm and}\qquad 
\alpha_{4444}^{4}(t\ssigma+U\ttau+s\ssigma\ttau).
\ee
A correlator with $\kappa=4$ but no crossing symmetries would admit $18$ ambiguities.
%
\end{itemize}

Notice that our analysis here is already in agreement with the observed number of ambiguities shown in Table~\ref{tab:free_coeffs}.

Since our bootstrap algorithm has a built-in position space implementation, 
it will be useful to rewrite the Mellin amplitude for the ambiguities in such a way that the comparison with the position space result is simple.  
This rewriting follows our organisation of the OPE into above threshold, in and below window, and it is explained in Appendix \ref{AppTree2}.

%
\subsection{$\langle \mathcal{O}_3 \mathcal{O}_3 \mathcal{O}_3 \mathcal{O}_3 \rangle$}\label{3333_oneloop}

We begin illustrating our bootstrap algorithm with $\mathcal{H}^{(2)}_{3333}$. 
The solution for the polynomial coefficients 
$h_{1},\ldots h_{11}, h_{\Box},h_{u},h_{v},h_0$ is listed in the {ancillary file.}
{For simplicity, the ancillary file contains $\mathcal{H}^{(2)}_{3333}$ with a particular value of the ambiguities.}

The 3333 correlator {has degree of extremality $3$} and full crossing symmetry.  
The long sector decomposes into three representations, $[000]$, $[101]$ and $[020]$, with threshold twist $\tau^{\max}=6$. 

The resummation of the $\log^2 u$ discontinuity can be obtained  from $\widehat{\mathcal{D}}_{3333}$ 
and $\Delta^{(8)}$ as explained in Appendix \ref{10d}. With this data we can then initiate the first step 
of our algorithm, by matching and imposing crossing symmetry of the ansatz.

In the second step of the algorithm we impose absence of $x=\bar{x}$ poles on the ansatz. 

Finally, we have to impose OPE predictions in and below window.
Here the window is empty, since all external charges are equals. 
This implies that upon projecting the ansatz onto the $\log^1 u$ stratum we have to set to zero
the one-variable expansion up to order $O(x^3)$. 
OPE predictions below window are instead non trivial: 

In $[000]$ the unitary bound is $\tau=2$, and no long supergravity states contribute, since these are all string states. A non trivial prediction 
comes in at twist $4$. Here there is only one double trace operator
${\mathcal K}_{22;4,l,[000]}$.
Using~\eqref{L2} we thus get a prediction for $L^{(2)\mathcal{H} }_{3333;4,l,[000]}$.~\footnote{The explicit  expression for $C^{(1)}_{33; 4,l,[000]}$ is in Appendix, given by \eqref{3pt_pplus5}.}
We have 
\bea
L^{(2)\mathcal{H} }_{3333;2,l,[000]} &=& 0 \label{twist2empty} \\
L^{(2)\mathcal{H} }_{3333;4,l,[000]} &=& \frac{9\times 4800}{ (l+1)(l+6) } \frac{((l+3)!)^2}{ (2l+6)! }\frac{1+(-1)^l}{2} \ .
\label{3333_twist4_000_analitic}
\eea
The one-variable resummation of \eqref{3333_twist4_000_analitic} input into~\eqref{analitic_prediction_uno} reads
\be\label{3333_analitic_000}
\mathcal{H}^{(2)}_{3333}\Big|_{\Upsilon_{[000]} \log^0 u}= 9\times 6! \times \frac{x^2}{\bar{x}^4 } \left[  5( {\bar x}-2){\bar x} {\rm Li}_1(  {\bar x}) + \tfrac{5}{3}( 6-6 {\bar x}+ {\bar x}^2 ) {\rm Li}^2_1( {\bar x} ) \right] + O( x^3)\ .
\ee

In the $[101]$ and $[020]$ sectors, the unitary bound is $\tau=4$. There are no predictions descending from the long sectors at tree level. Instead, 
this is the first case in which we need to consider the consequences of protected semishort operators at twist $4$, 
through our formula \eqref{newL2uno} and the results for $S_{4;l+2,[1]}$ and $S_{4;l+2}$ given in \eqref{s4l}.
More precisely, there is an $S_{4;l+2,[1]}$ for $[101]$, which implies 
\bea
L^{(2)\mathcal{H} }_{3333;4,l,[101]} &=& {}\frac{9\times 576}{ (l+2)(l+5) } \frac{((l+3)!)^2}{ (2l+6)! } \frac{1-(-1)^l}{2}
\eea
with corresponding one-variable resummation,
\be\label{3333_analitic_101}
\mathcal{H}^{(2)}_{3333}\Big|_{\Upsilon_{[101]} \log^0u }=
9\times 6! \times \frac{x^2}{\bar{x}^4 } \left[  3( {\bar x}-2){\bar x} + ( 6-6 {\bar x}+\tfrac{7}{5} {\bar  x}^2 ) {\rm Li}_1( {\bar x} ) + \tfrac{1}{5} ( {\bar x}-2){\bar x} {\rm Li}^2_1( {\bar x} ) \right] + O( x^3)
\ee
and an $S_{4,l+2,[0]}$ for $[020]$ which gives,
\bea
L^{(2)\mathcal{H} }_{3333;4,l,[020]} &=& {} \frac{9\times 288}{ (l+3)(l+4) } \frac{((l+3)!)^2}{ (2l+6)! } \frac{1+(-1)^l}{2} 
\eea
with one-variable resummation
\be\label{3333_analitic_020}
\mathcal{H}^{(2)}_{3333}\Big|_{\Upsilon_{[020]} \log^0u } =
9\times 6! \times \frac{x^2}{\bar{x}^4 } \left[  \tfrac{6}{5} {\bar x}^2 + \tfrac{3}{5} ({\bar x}-2){\bar x} {\rm Li}_1( {\bar x} ) + \tfrac{1}{10}{\bar x}^2 {\rm Li}^2_1( {\bar x} ) \right] + O( x^3)
\ee

There is an important logical distinction between $L^{(2)\mathcal{H} }_{3333;4,l,[101]}$ and $L^{(2)\mathcal{H} }_{3333;4,l,[020]}$ we should highlight.
The twist $4$ of $[020]$ lies at the bottom of multiplet recombination, in the sense that 
$\tau=2a+b+2$ with $b=2$ and $a=0$. This means that the corresponding SCPW does not get shifted 
by multiplet recombination in another $su(4)$ representation. In fact, our formula \eqref{newL2uno} makes 
explicit that there is no extra summation over $a$ that needs to be taken into account.
This is not the case for the twist $4$ of $[101]$, where instead the SCPW coefficient 
receives a contribution due to multiplet recombination of $[000]$ at twist $2$. 
However, there is no $S_{2;l+2,[0]}$ contribution, 
therefore $L^{(2)\mathcal{H} }_{3333;4,l,[101]}=S_{4;l+2,[1]}$ holds exactly.

Coming back to our ansatz, we match \eqref{3333_analitic_000}, \eqref{3333_analitic_101} and \eqref{3333_analitic_020}.
{ Recall that we had  20} free coefficients, i.e. coefficients not fixed by demanding absence of $x=\bar{x}$ poles. 
However after matching the OPE predictions below window we are left with only $2$ free coefficients.  
The functions they span are the final ambiguities. They come out in the form 
\bea
\mathcal{H}^{(2)}_{3333}\Big|_\text{ambiguity}&=&{u}^{3}\oint { u}^{s} { v}^{t }\ \Gamma[-s]^2\Gamma[-t]^2\Gamma\big[s+t+5\big]^2 \times\Big[ \notag\\
&& \qquad \quad
 (1 + \ssigma+ \ttau) { \beta}_{3333}^1
+ 
\big(s + \ttau t - \ssigma (5+s+t) \big) { \beta}_{3333}^2 \Big]
\label{Mellin_amb_3333}
\eea
By redefining the Mellin variables $s,t$ we obtain a perfect match with our previous discussion in \eqref{amb_3333_discussion}. 
Upon inspection, the SCPW of \eqref{Mellin_amb_3333} only contributes at spin $l=0$ for twist above threshold.

\subsection{$\langle \mathcal{O}_4 \mathcal{O}_4 \mathcal{O}_4 \mathcal{O}_4 \rangle$}\label{4444_oneloop}

The next correlator we study is 4444. The solution of our bootstrap problem, 
written up in the basis \eqref{G2uvresult}, is appended in the ancillary file. 
{For simplicity, the ancillary file contains $\mathcal{H}^{(2)}_{4444}$ with a particular value of the ambiguities.}

The 4444 correlator is fully crossing symmetric and {has degree of extremality $4$}. 
The long sector decomposes into six $su(4)$ channels: $[000]$, $[101]$, $[020]$ and $[202]$, $[121]$, $[040]$.
The threshold twist is $\tau^{\rm max}=8$. 

The leading $\log^2u$ resummation is obtained by acting with $\widehat{\mathcal{D}}_{4444}$ and $\Delta^{(8)}$, 
as explained in Appendix \ref{10d}. We then initiate the algorithm by matching, 
imposing crossing symmetry, and absence of $x=\bar{x}$ poles. 

We finally come to the OPE predictions in and below window.
Being the window empty, we project the ansatz onto the $\log^1 u$ stratum and we set to zero the 
one-variable expansion up to $O(x^4)$. Below window we find instead non trivial physics. 
For the representations $[000]$, $[101]$, $[020]$, the discussion is similar to that in 3333 
for twist $4$, and continues at twist $6$ by including predictions coming from the long sector at tree level. 
For $[202]$, $[121]$, $[040]$ we will have to consider non trivial multiplet recombination taking 
into account the predictions arising from the semishort sector. We proceed in order.  

In the singlet channel $[000]$, there is an empty twist $2$ sector, 
then the $1/N^2$ subleading three-point couplings $C^{(1)}_{44;4,l,[000]}$ and $\mathbf{C}^{(1)}_{44;6,l,[000]}$ 
give non trivial predictions at twist $4$ and twist $6$.~\footnote{Explicit expressions for  $C^{(1)}_{44;4,l,[000]}$ and 
$\mathbf{C}^{(1)}_{44;6,l,[000]}$ are given in \eqref{3pt_pplus5} and \eqref{C144twist6000}, respectively.}
Recall that there are two double trace operators in $R_{6,l,[000]}$, therefore the twist $6$ computation yields a vector three-point function. 
We find, from~\eqref{L2},
\bea
L^{(2)\mathcal{H} }_{4444;2,l,[000]} &=& 0 \\
L^{(2)\mathcal{H} }_{4444;4,l,[000]} &=& \frac{16\times 4800}{ (l+1)(l+6) } \frac{((l+3)!)^2}{ (2l+6)! }\frac{1+(-1)^l}{2} \\
L^{(2)\mathcal{H} }_{4444;6,l,[000]} &=& \frac{16\times 360 (2l+9)^2 (119+ (2l+9)^2 )}{ (l+1)(l+2)(l+7)(l+8) } \frac{((l+4)!)^2}{ (2l+8)! }\frac{1+(-1)^l}{2} 
\label{4444_twist4_000_analitic}
\eea
The corresponding one variable resummation is
\begingroup
\bea
\arraycolsep=1pt\def\arraystretch{2.2}
\begin{array}{rl}
&\mathcal{H}_{4444}^{(2)}\Big|_{\Upsilon_{[000]}\log^0u}=16\times 6!\Big(\\
&
{\displaystyle \frac{x^2}{\bar{x}^4 }}
\left[  5( {\bar x}-2){\bar x} {\rm Li}_1(  {\bar x}) + \tfrac{5}{3}( 6-6 {\bar x}+ {\bar x}^2 ) {\rm Li}^2_1( {\bar x} ) \right] + \\
&
{\displaystyle \frac{x^3}{\bar{x}^5 } }\Big[ 
 {\bar x}^2 ({\bar x}-2)\left( \tfrac{ 4 {\bar x}^2}{ {\bar x}-1}+205 \right)  
  + ( 3440-5590 {\bar x} + \tfrac{ 7484}{3}  {\bar x}^2 -244 {\bar x}^3 ) {\rm Li}_1^2 ({\bar x} ) \\
 &
 \quad 
 - 300 {\bar x}( 6- 6 {\bar x} + {\bar x}^2 ) {\rm Li}_2({\bar x} )
  - {\bar x}(1230-1210 {\bar x} +211 {\bar x}^2 ){\rm Li}_1({\bar x} )
  \Big] \Big) + O(x^4)
 \end{array}
 \eea
\endgroup

In $[101]$ and $[020]$ there are twist $4$ predictions coming from semishorts operators at the unitarity bound, similarly to the case of 3333 discussed previously.
In particular, there is an $S_{4;l+2,[1]}$ for $[101]$ and an $S_{4;l+2,[0]}$ for $[020]$, {computed in \eqref{s4_4444} }
In addition we have $1/N^2$ three-point couplings $C^{(1)}_{44;6,l,[101]}$ and 
$\mathbf{C}^{(1)}_{44;6,l,[020]}$ which give predictions at twist $6$. 
Notice that twist $6$ is the first twist available for a long rep $[020]$, but the latter has doubling of the operators, i.e. $\mu=2$.

The list of results in $[101]$ is
\bea
L^{(2)\mathcal{H} }_{4444;4,l,[101]} &=& \frac{16\times 1600}{ (l+2)(l+5) } \frac{((l+3)!)^2}{ (2l+6)! } \frac{1-(-1)^l}{2}\\
L^{(2)\mathcal{H} }_{4444;6,l,[101]} &=& \frac{16\times 7200(l+1)(l+8)}{ (l+3)(l+6) } \frac{((l+4)!)^2}{ (2l+8)! } \frac{1-(-1)^l}{2}
\eea
with resummation
\begingroup
\be
\arraycolsep=1pt\def\arraystretch{2.2}
\begin{array}{rl}
&
\mathcal{H}_{4444}^{(2)}\Big|_{\Upsilon_{[101]}\log^0u}=16\times \frac{5^2 6!}{3^2} \Big(\\
& 
{\displaystyle \frac{x^2}{ \bar{x}^4 } }
\left[  3( {\bar x}-2){\bar x} + ( 6-6 {\bar x}+\tfrac{7}{5} {\bar  x}^2 ) {\rm Li}_1( {\bar x} ) + \tfrac{1}{5} ( {\bar x}-2){\bar x} {\rm Li}^2_1( {\bar x} ) \right] -\\
& 
{\displaystyle \frac{x^3 }{ \bar{x}^5 }  } \Big[ \tfrac{4}{75} ( 1527 {\bar x} -3014){\bar x}^2 + \frac{ 9 ({\bar x}-2) {\bar x}^4 }{ 5({\bar x}-1) }  
+ \tfrac{16}{75}( 746 - 766 \bar{x} + 201 \bar{x}^2)\bar{x} {\rm Li}_1(\bar{x}) \\
&
\quad + \tfrac{1}{10} (4-176 \bar{x}+87 \bar{x}^2)\bar{x} {\rm Li}_1^2(\bar{x}) 
+\tfrac{2}{5}(4+4 \bar{x}-3\bar{x}^2 )\bar{x} {\rm Li}_2(\bar{x} ) \Big] \Big) + O(x^4)
\end{array}
\ee
\endgroup

The list of results in $[020]$ is
\begingroup
\arraycolsep=1pt\def\arraystretch{2.2}
\bea
L^{(2)\mathcal{H} }_{4444;4,l,[020]} &=& \frac{16\times 1152}{ (l+3)(l+4) } \frac{((l+3)!)^2}{ (2l+6)! } \frac{1+(-1)^l}{2}\\
L^{(2)\mathcal{H} }_{4444;6,l,[020]} &=& \frac{16\times 864( { 40817} +16702 (2l+9)^2 +81(2l+9)^4) }{80 (l+1)(l+4)(l+5)(l+8) } \frac{((l+4)!)^2}{ (2l+8)! } \frac{1+(-1)^l}{2}
\eea
\endgroup
with resummation
\begingroup
\be
\arraycolsep=1pt\def\arraystretch{2.2}
\begin{array}{rl}
&
\mathcal{H}_{4444}^{(2)}\Big|_{\Upsilon_{[020]}\log^0u}=
16\times 2^2 6! \Big( \\
&
 {\displaystyle \frac{x^2}{\bar{x}^4 } }
 \left[  \tfrac{6}{5} {\bar x}^2 + \tfrac{3}{5} ({\bar x}-2){\bar x} {\rm Li}_1( {\bar x} ) + \tfrac{1}{10}{\bar x}^2 {\rm Li}^2_1( {\bar x} ) \right] + \\
 &
 {\displaystyle \frac{x^3}{\bar{x}^5 } } \Big[ \tfrac{3}{5}(382-175 \bar{x} ) + \frac{ 243 (\bar{x}-2)\bar{x}^2 }{25 (\bar{x}-1) }-\tfrac{3}{5} (11358-11342 \bar{x} +1981 \bar{x}^2)\bar{x} {\rm Li}_1(\bar{x} ) \\
 &\quad \tfrac{1}{5^2} (164640-246960 \bar{x}+98 794\bar{x}^2-8667\bar{x}^3 ) {\rm Li}_1^2(\bar{x} ) -72 \bar{x}^3 {\rm Li}_2(\bar{x}) \Big]\Big)+O(x^4)
\end{array}
\ee
\endgroup

Finally we arrive at the representations $[202]$, $[121]$ and $[040]$. The unitary bound for all of them is twist $6$, 
and the semishort predictions were computed in \eqref{s6_4444}. 
In the 4444 correlator, the twist $6$ of $[040]$ lies at the bottom of multiplet 
recombination because $\tau=2a+b+2$ with $b=4$ and $a=0$.
The prediction for $L^{(2)\mathcal{H}}_{4444;6,l,[040]}$ is thus straightforward. 
The predictions for $[121]$ and $[202]$ involve further shifts, 
which we now describe. From \eqref{newL2uno} we find
\bea
L^{(2)\mathcal{H} }_{4444;6,l,[040]}&=&-S_{4444;6,l+2; [0] }\\
L^{(2)\mathcal{H} }_{4444;6,l,[121]}&=&+S_{4444;6,l+2; [1]}-S_{4444;4,l+3; [0]}\label{121twist6}\\
L^{(2)\mathcal{H} }_{4444;6,l,[202]}&=&-S_{4444;6,l+2; [2]}+S_{4444;4,l+3; [1]}\label{202twist6}
\eea
where in the last line we already implemented the absence of $S_{4444;2,l+2; [0]}$. 
Formulas \eqref{121twist6} and \eqref{202twist6} correctly 
include shifts to due to multiplet recombination at twist $4$ in $[020]$ and $[101]$, 
respectively. Let us give the explicit expressions here below. 

For $[040]$ we find
\be
L^{(2)\mathcal{H} }_{4444;6,l,[040]}=\frac{16\times 384 ( {29} + 3(2l+9)^2 ) }{(l+3)(l+6)} \frac{(l+4)!^2}{(2l+8)!}\frac{1+(-)^l}{2}
\ee
with resummation
\begingroup
\bea
\arraycolsep=1pt\def\arraystretch{2.2}
\begin{array}{rl}
&\mathcal{H}_{4444}^{(2)}\Big|_{\Upsilon_{[040]}\log^0u}=16\times 6! \times {\displaystyle \frac{x^3}{\bar{x}^5 } }\Big(\\
&
\big(208-\tfrac{16 \bar{x}^2}{5(\bar{x}-1)}  +\tfrac{112}{15} {\rm Li}_1^2(\bar{x}) \big) (2 - \bar{x})\bar{x}^2
- \tfrac{16}{3} (78- 78\bar{x} + 17 \bar{x}^2)\bar{x} {\rm Li}_1(\bar{x}) \Big) + O( x^4)
\end{array}
\eea
\endgroup

For $[202]$ we find
\bea
L^{(2)\mathcal{H} }_{4444;6,l,[202]}&=& 16 \times \left[\, \frac{2400 (l+2)(l+7) }{(l+3)(l+6)}- \left(\frac{1600}{(l+2)(l+5)}\right)_{l\rightarrow l+1}\, \right] \frac{(l+4)!^2}{(2l+8)!} \frac{1+(-)^l}{2}\notag\\
&=& \frac{16\times 200\left(-{83} + 3(2l+9)^2\right)}{(l+3)(l+6)}\frac{(l+4)!^2}{(2l+8)!}\frac{1+(-)^l}{2}
\eea
with resummation
\begingroup
\bea
\arraycolsep=1pt\def\arraystretch{2.2}
\begin{array}{rl}
&\mathcal{H}_{4444}^{(2)}\Big|_{\Upsilon_{[202]}\log^0u}= 16\times 6! \times {\displaystyle \frac{x^3}{\bar{x}^5 }}   \Big(\\
&\big(  \tfrac{25}{3} + \tfrac{5\bar{x}^2}{3(\bar{x}-1)} + \tfrac{35}{9} {\rm Li}_1^2 (\bar{x}) \big) \bar{x}^2(\bar{x}-2) +\tfrac{5}{9} (30-30\bar{x}+13\bar{x}^2 ){\rm Li}_1(\bar{x}) \Big) + O(x^4)
\end{array}
\eea
\endgroup

For $[121]$ we find
\begingroup
\arraycolsep=1pt\def\arraystretch{2.2}
\bea
L^{(2)\mathcal{H} }_{4444;6,l,[121]}&=&16\times\left[ \frac{ 72(401+174(2l+9)^2 + (2l+9)^4)}{(l+2)(l+4)(l+5)(l+7)}-\frac{1152}{(l+4)(l+5)}\right] \frac{(l+4)!^2}{(2l+8)!}\frac{1-(-)^l}{2} \notag\\
&=& \frac{16\times 72 (3+ (2l+9)^2) (167 + (2l+9)^2 )}{(l+2)(l+4)(l+5)(l+7)}\frac{(l+4)!^2}{(2l+8)!}\frac{1-(-)^l}{2}
\eea
\endgroup
with resummation
\begingroup
\bea
\arraycolsep=1pt\def\arraystretch{2.2}
\begin{array}{rl}
&
\mathcal{H}_{4444}^{(2)}\Big|_{\Upsilon_{[121]}\log^0u}= 16\times 6! \times {\displaystyle \frac{x^3}{\bar{x}^5 } }\Big(
\left( \tfrac{4 \bar{x}^3 (2-2\bar{x}+\bar{x}^2) }{5(\bar{x}-1)} - 2184\bar{x}(\bar{x}-1)+\tfrac{1874}{5} \bar{x}^3 \right)\\
&
\qquad -\big( 2184 - 3276 \bar{x} + \tfrac{7104}{5}\bar{x}^2-\tfrac{822}{5} \bar{x}^3) {\rm Li}_1(\bar{x}) + \tfrac{14}{5} (48 -48\bar{x}+7 \bar{x}^2) {\rm Li}_1^2(\bar{x}) \Big)\notag
 + O( x^4)
\end{array}
\eea
\endgroup

Incredibly all these predictions are consistent with the minimal ansatz~\eqref{G2uvresult} 
and uniquely fix the remaining coefficients, leaving only $4$ ambiguities. 
These are
\begin{gather}
\mathcal{H}^{(2)}_{4444}\Big|_\text{ambiguity}=
{u}^{4}\oint { u}^{s} { v}^{t }\ \Gamma[-s]^2\Gamma[-t]^2\Gamma\big[s+t+6\big]^2\Big[ \notag\\[.2cm]
\qquad
(1 + \ssigma^2+ \ttau^2) { \beta}_{4444}^1+\big(s + \ttau^2 t - \ssigma^2 (6+s+t) \big) { \beta}_{4444}^2 \\[.2cm]
\quad
+(\ttau+\ssigma+\ttau\ssigma) { \beta}_{4444}^3
+ \big(  \ttau(1+s+t)-\ssigma(5+t)-\ssigma\ttau (5+s)\big){ \beta}_{4444}^4\notag
\Big] 
\end{gather}
Again, we find perfect match with our previous discussion in \eqref{amb_4444_discusssion1}-\eqref{amb_4444_discusssion2}, after the redefinition of Mellin variables $s$ and $t$. 
The ambiguities have only spin $l=0$ support in any $su(4)$ channel, for twist above threshold.

\subsection{Next-to-Next-to-Extremal Correlators}
\label{one_loop_N2}

In this section we will consider four point correlators $\langle \mathcal{O}_{p_1} \mathcal{O}_{p_2} \mathcal{O}_{p_3} \mathcal{O}_{p_4} \rangle$
with external charges $\langle 2244\rangle $, $\langle 3335\rangle$ and $\langle4424\rangle$. These correlators are all N$^{2}$E.

For N$^{2}$E correlators there are no OPE predictions below window.
In particular, semishort predictions $S_{p_1p_2p_3p_4}$ vanish
because they are determined through \eqref{form} in terms of 
SCPW coefficients $S_{p(r)q(r)p_3p_4}$, where $p(r)+q(r)=p_{43}+2$, and
these correlators are next to extremal, thus completely vanishing
as a consequence of our definition of external single particles states. 
Because of the split $\mathcal{T}^{(2)}+\mathcal{H}^{(2)}$, it follows $L^{(2)\mathcal{H}}_{2+p_{43},[0p_{43}0]}=0$.

An additional peculiarity of $\langle 3335\rangle $ and $\langle 4424\rangle$, which generalises to other N$^2$E correlators as discussed later on in Section \ref{sec:N2E_gen_treelevel}, 
is the fact that the tree level functions $\mathcal{D}^{(1)}_{3335}$ and $\mathcal{D}^{(1)}_{4424}$ are proportional. 
This implies that, after taking into account a normalization, both correlators have the same one-loop $\log^2 u$ discontinuity.
Therefore, an ansatz having the correct crossing symmetries, and constructed 
by matching the $\log^2 u$ discontinuity and imposing absence of $x=\bar{x}$ poles, 
cannot distinguish between $\mathcal{H}^{(2)}_{3335}$ and $\mathcal{H}^{(2)}_{4424}$.
Very interestingly, this type of degeneracy is actually lifted at one-loop, 
because of the different OPE predictions in the window. 
This illustrates another important aspect of the OPE predictions in and below window.
In general, we expect the situation to be as follows:  
pairs of correlators which are degenerate at tree level will
have instead different minimal one-loop functions, distinguished by the OPE predictions in and below window.

For $\langle 3335\rangle $ and $\langle 4424\rangle$ we have $\log^1 u$ twist $6$ prediction (in the $[020]$ representation). 
Making manifest reciprocity symmetry, we write them in the form \eqref{test_M2}, 
\be\label{3335_4424_autofit}
(\tau=6)\qquad \left\{ \begin{array}{l}
\displaystyle
\frac{ Y^{(0)}_{3335} + Y^{(2)}_{3335 }(l+{\footnotesize\frac{9}{2}})^2 }{ (l+1)(l+4)(l+5)(l+8) } \frac{ (l+4)!(l+5)!}{(2l+8)!}\\[.4cm]
\displaystyle
\frac{ Y^{(0)}_{4424} + Y^{(2)}_{4424 }(l+{\footnotesize\frac{9}{2}})^2 }{ (l+1)(l+4)(l+5)(l+8) } \frac{ (l+4)!(l+5)!}{(2l+8)!}
\end{array}\right.
\ee
The values of the free $Y$ coefficient above, obtained from the OPE predictions, are
\be
\left\{
\begin{array}{l}
Y^{(0)}_{3335} = -4762800\\[.2cm]
Y^{(2)}_{3335}=\frac{4}{35} Y^{(0)}_{3335}
\end{array}\right. \qquad;\qquad 
\left\{
\begin{array}{l}
Y^{(0)}_{4424}=-4628736\\[.2cm]
Y^{(2)}_{4424 }=\frac{55}{2009} Y^{(0)}_{4424}
\end{array}\right.
\ee

We proceed as in previous sections. We
construct an ansatz which matches the leading logs, has the correct symmetries and no $x=\bar{x}$ poles. 
We then impose the OPE predictions in the window. 

The result for $\langle 3335\rangle$ and $\langle4424\rangle$ can be obtained in the following instructive way.
We initially normalize both correlators in a way that the leading logs are the same. Thus we construct {\it one} ansatz. 
Before imposing OPE predictions, this ansatz has six free coefficients. 
We now insist that the SCPW coefficients of the ansatz at $\tau=6$ have the form \eqref{3335_4424_autofit}, where we do not specify 
the values of $Y^{(0)}_{\vec{p}}$ and $Y^{(1)}_{\vec{p}}$ yet. 
This constraint returns a one-parameter ansatz with one additional  ambiguity.
We go back to the correct normalization for the correlators, and
we keep $Y^{(0)}_{\vec{p}}$ as the free parameter, isolating the tree-like function it multiplies. 
Then, we can write the minimal loop functions in the following form
\be
\mathcal{H}^{(2)}_{\vec{p}}= \mathcal{N}_{\vec{p}} \, \overline{\mathcal{H}}^{(2)} + \tfrac{1}{882} Y^{(0)}_{\vec{p}} u^2 \overline{D}_{4444}\qquad \vec{p}=3335||4424
\ee
where $\mathcal{N}_{4424}=128 $ and $\mathcal{N}_{3335}=135$. 
Because $Y^{(0)}_{3335}\neq Y^{(0)}_{4424}$, we find $\mathcal{H}^{(2)}_{3335}\neq\mathcal{H}^{(2)}_{4424}$. 
Differently from tree level, the minimal one-loop functions are distinct.

The ambiguity has the same functional form for both $\langle 3335\rangle $ and $\langle 4424\rangle $. Written in Mellin space it reads
\be\label{amb_3335}
\mathcal{H}^{(2)}_{3335||4424}\Big|_\text{ambiguity}=u^3 v \oint u^s v^t  \Gamma[-s]\Gamma[-s-1]\Gamma[-t]\Gamma[-t-1]
\Gamma[s+t+6]\Gamma[s+t+7]\, \beta_{\vec{p}}
\ee
Notice the combinations of Mack's $\Gamma_{\vec{p}}$ is the same for both $\langle 3335\rangle $ and $\langle4424\rangle$. 

The result for $\langle2244\rangle$ is more straightforward. 
In the window, we have both twist $4$ and $6$ predictions (in the $[000]$ representation). Again we write them in the form \eqref{test_M2}, namely 
\bea
\label{autofit_2244_1}
&&
(\tau=4)\qquad \frac{ X^{(0)}_{2244} }{ (l+1)(l+6)} \frac{ ((l+3)!)^2}{(2l+6)!} \\[.2cm]
\label{autofit_2244_2}
&&
(\tau=6)\qquad \frac{ Y^{(0)}_{2244} + Y^{(2)}_{2244 }(l+{\footnotesize\frac{9}{2}})^2 }{ (l+1)(l+2)(l+7)(l+8)} \frac{ ((l+4)!)^2}{(2l+8)!}
\eea
with predicted values,
\be
X^{(0)}_{2244}=-8\times1920,\qquad Y^{(0)}_{2244}=+8\times 176400, \qquad Y^{(2)}_{2244 }= \tfrac{76}{245}Y^{(0)}_{2244}.
\ee
In the orientation $\langle 2424\rangle$ the window is empty.

The bootstrap algorithm returns $\mathcal{H}_{2244}$ leaving only two ambiguities as we discussed in \eqref{amb_22pp_discussion}.  
In Mellin space we find
\be
\mathcal{H}^{(2)}_{2244}\Big|_\text{ambiguity}=u^4\oint u^s v^t  \Gamma[-s]\Gamma[-s-2]
\Gamma[-t]^2\Gamma[s+t+6]^2\left[ \beta^{1}_{2244} + \beta^{2}_{2244}\, s \right]
\ee

The minimal one-loop functions corresponding to $\langle 3335\rangle$ and $\langle4424\rangle$, and $\langle2244\rangle$,
can be downloaded from the attached ancillary files.  For $\langle 3335\rangle$ and $\langle 4424\rangle$ we only attached $\overline{\mathcal{H}}$.
In both cases, we have fixed some value of the ambiguities.

\def\namep{P}
\def\namem{m}

\section{Upgraded Tree Level Mellin Amplitudes} \label{NewRastelli}

In previous sections, we showed that the one-loop function $\mathcal{D}^{(2)}$ 
admits the split $\mathcal{D}^{(2)}=\mathcal{T}^{(2)}+\mathcal{H}^{(2)}$, 
where $\mathcal{H}^{(2)}$ encodes all the non trivial OPE predictions at $O(1/N^4)$ 
whereas $\mathcal{T}^{(2)}$ is a generalised tree-level function having no $\log^2$ contribution.  
Our final task is to bootstrap $\mathcal{T}$. 

The generalised tree level function $\mathcal{T}_{\vec{p}}$ 
is defined as the unique function, within the ansatz:
\bea
&&\label{ansatzTNewR}
\mathcal{T}^{}_{\vec{p}}
=
 \frac{  {\namep}_{\Box}\, \phi^{(1)}(x,{\bar x})   }{ \ (x-{\bar x})^{\mathbf{d}+2 } }  
 + \frac{ {\namep}_{v}\log(v) }{ \ (x-{\bar x})^{\mathbf{d}+1}  } +  \frac{1}{v^{  \kappa_{\vec{p}}-1  }} 
	\Big[ 
		\frac{  {\namep}_{u}\log(u) }{ \  (x-{\bar x})^{\mathbf{d}+1} }+ 
		\frac{  {\namep}_{1} }{ (x-{\bar x})^{\mathbf{d}-1} }\Big]\qquad
\\[.2cm]
&&
\qquad \mathbf{d}_{\vec{p}}=p_1+p_2+p_3+p_4-1,\qquad\qquad \kappa_{\vec{p}}={\rm min}(p_3,\tfrac{p_1+p_2+p_{3}-p_4 }{2})\qquad
\eea
such that:
\begin{enumerate}[(a)]
\item \label{con1}
	the threshold twist for the $\log(u)$ discontinuity is $\tau=\tau^{\rm max}$. 
\item \label{con2}
	the SCPW expansion below window {\it completely cancels} free theory contributions as described in~\eqref{LT} and ~\eqref{LT2}
\item \label{con3}
	there are no unphysical $x=\bar{x}$ poles in \eqref{ansatzTNewR}.
\end{enumerate}
The coefficient functions denoted by $\namep$ are polynomials in $x,\bar{x}$ and $\sigma,\tau$.  
As functions of $x,\bar{x}$ variables, these polynomials have a Taylor expansion of the form $x^n\bar{x}^m$ with $m+n\leq p_1+p_2+p_3+p_4$. 
The function $\mathcal{T}_{\vec{p}}$ is symmetric under $x\leftrightarrow \bar{x}$, 
therefore a given polynomial $P$ has the same symmetry as the transcendental function it multiplies. 
The $su(4)$ decomposition of $\mathcal{T}^{}_{\vec{p}}$ is obviously the same as for the full dynamical function $\mathcal{D}_{\vec{p}}$. 

Implementing condition~(\ref{con1}) implies 
\bea
{\namep}_{\Box}&=&O(u^{-\frac{p_{43}}{2}+\frac{ {\rm max}(  {p_1+p_2},\, {p_3+p_4}) }{2}} ) \\
{\namep}_{u}&=&O(u^{-\frac{p_{43}}{2}+\frac{ {\rm max}(  {p_1+p_2},\, {p_3+p_4}) }{2}} ) 
\eea

The above conditions in fact define a generalised tree function ${\mathcal T}_{\vec{p}}$  for {\em any} 
free $\cN=4$ theory, i.e. for all $N$. Indeed we can define 
it in terms of the coefficients ${A}^k_{\gamma }$ in front of each propagator structure in~\eqref{gen_free_theory0}, 
which we can leave completely arbitrary (other than the relations between them arising from imposing crossing symmetry).  
The polynomials $\namep$ in~\eqref{ansatzTNewR} become function of the free propagator coefficients, $\namep[\{A_\gamma^k\}]$. 
The precise value of these ${A}^k_{\gamma }$ does not affect any step of this algorithm. 
Furthemore, condition \eqref{con2} is overconstraining, 
and therefore the solution we find is unique, i.e. $\mathcal{T}_{\vec{p}}$ is unique.

Because of this uniqueness, we expect our function $\mathcal{T}_{\vec{p}}$ 
to reduce to known results at tree level when the propagator coefficients $A_\gamma^k$ 
take on their free theory values. Indeed, when the external charges are equal, 
our conditions are precisely those imposed in \cite{Dolan:2006ec}, and for arbitrary charges 
we expect to recover the tree-level correlators of Rastelli and Zhou~\cite{Rastelli:2016nze}.   
Notice that in position space the function of~\cite{Rastelli:2016nze} is described by 
the same ansatz as in \eqref{ansatzTNewR}, except for the change $\mathbf{d}_{\vec{p}}\rightarrow \mathbf{d}_{\vec{p}}-2$. 
(Various examples can be found in \cite{unmixing}  by rewriting the corresponding $\overline{D}$ representation.) 
Quite non trivially, $\mathcal{T}_{\vec{p}}$ {does} reduce to the function of~\cite{Rastelli:2016nze} 
when the coefficients $A^{k}_{\gamma}$ are truncated at $O(1/N^2)$. 
In fact, we find that all polynomials $\namep_{\Box,u,v,1}[\{A_\gamma^k\}]$ 
acquire an extra double zero $(x-\bar{x})^2$ when we restrict the $A^{k}_{\gamma}$ to their tree level value.

At tree level, the free theory coefficients $A_{\gamma}\big|_{1/N^2}$ are all 
proportional to each other, and thus satisfy linear relations. Therefore, we can 
understand the tree-level degeneration as the result of imposing on 
$\namep_{\Box,u,v,1}[\{A^{k}_{\gamma}\}]$ these tree-level linear relations. However,
the non-planar values of the $A^{k}_{\gamma}$ are not as simple, 
and the corresponding relations become non-linear.  

Similarly to the function of Rastelli and Zhou~\cite{Rastelli:2016nze} the most 
transparent representation of $\mathcal{T}_{\vec{p}}$ is given in Mellin space. 
We thus define the corresponding Mellin amplitude $\mathcal{M}[\mathcal{T}_{\vec{p}}](s,t)$ 
of the generalised tree similarly to that of the tree-level function of~\cite{Rastelli:2016nze}.
Amazingly all the generalised tree-level functions  $\mathcal{T}_{\vec{p}}$ -- 
defined by the above conditions $(a),(b),(c)$ -- can be written in this form 
with a simple rational Mellin amplitude with only simple poles. 

The specific form of $\mathcal{M}[\mathcal{T}_{\vec{p}}](s,t)$, i.e. finiteness and rationality,
translates into the observation that the entire function $\mathcal{T}_{\vec{p}}$ is determined uniquely 
in terms of the coefficient ${\namep}_{\Box}$ in from of the box function. This can be understood from the fact that
the box function contains a $\log u \log v$ term, which on the other hand arises only from a double pole in both 
$s$ and $t$ in the Mellin transform. 
More details about this statement are given in Appendix~\ref{AppTree2}.

In the next sections we make our discussion concrete by considering $\mathcal{T}_{3333}$ and $\mathcal{T}_{4444}$.
As a bonus of our definition of single-particle operators, we will also show that $\mathcal{T}_{\vec{p}}$ 
for next-to-next-to extremal correlators coincides with the function of Rastelli and Zhou.

\subsection{$\langle \mathcal{O}_3 \mathcal{O}_3 \mathcal{O}_3 \mathcal{O}_3 \rangle$}

The result for connected free theory was given in~\eqref{3333diagrams}. 
We rewrite it here below for convenience, 
\bea
&&
\langle \mathcal{O}_3 \mathcal{O}_3 \mathcal{O}_3 \mathcal{O}_3 \rangle_{\rm conn.\, free} =  \frac{ 9(N^2-4)^2(N^2-1)}{N^2} \times \bigg[ \notag\\
&&
9\left(u\ssigma+ \frac{u \ttau}{v} + u^2 \ssigma^2 + \frac{u^2 \ttau^2}{v^2}  +  \frac{ u^3 \ssigma^2\ttau}{v^2} + \frac{u^3\ssigma\ttau^2}{v^2} \right) + \frac{ 18(N^2-12) }{ (N^2-4) } \frac{u^2 \ssigma\ttau}{v} \bigg]. 
\rule{1cm}{0pt}
\eea
%
Crossing invariance of $\langle \mathcal{O}_3 \mathcal{O}_3 \mathcal{O}_3 \mathcal{O}_3 \rangle$
restricts the total number of connected coefficients $\{ {A}_{ 2 }^k, {A}_{ 4 }^k,{A}_{ 6 }^k \}$, in the generic sum over propagator structures \eqref{gen_free_theory0}, to only two independent ones. 
We have indeed 
\bea
\label{uno_3333}
{A}_2^0={A}_2^1={A}_4^0= {A}_4^2= {A}_6^1={A}_6^2&=&\frac{9}{(N^2-1)}A_0^0\\  
\label{due_3333}
{A}_{4}^1&=& \frac{ 18(N^2-12) }{ (N^2-4)(N^2-1) } A_0^0
\eea
where $A_0^0= \frac{ 9(N^2-4)^2(N^2-1)^2}{N^2}$.

The generalised tree level function in Mellin space is
\be
\mathcal{T}_{3333}=
{u}^{3}\oint { u}^{s} { v}^{t }\ \Gamma[-s]^2\Gamma[-t]^2\Gamma\big[s+t+5\big]^2 \mathcal{M}[\mathcal{T}_{3333}]\,.
\ee
with
\be\label{FAnction_3333}
\begin{array}{cll}
\mathcal{M}[\mathcal{T}_{3333}]=&   {\displaystyle  \frac{1}{(s+2)(t+1)(s+t+4)}   }      \big[  - {A}_2^0 \ + & \frac{1}{4}\ (s+2)\ ({A}_{4}^1 - 2 {A}_2^0)\ \big]\ + \\[.4cm]
			     &    {\displaystyle \frac{\ttau}{(s+1)(t+2)(s+t+4)} }       \big[  -{A}_2^0 \ + & \frac{1}{4}\ (t+2)\ ({A}_{4}^1 - 2 {A}_2^0) \ \big] \ + \\[.4cm]
                              &    {\displaystyle \frac{\ssigma}{(s+1)(t+1)(s+t+3)} }   \big[   -{A}_2^0 \ +& \frac{1}{4}(s+t+3)({A}_{4}^1 - 2 {A}_2^0) \big]\,.
\end{array}
\ee
Notice that  $\mathcal{M}[\mathcal{T}_{3333}]=\namem_{3333}^1(s,t)+ \ssigma \namem_{3333}^{\ssigma} (s,t)+ \ttau \namem_{3333}^{\ttau} (s,t)$.
Exploiting full crossing of $\mathcal{T}_{3333}$ it also useful to write
\be\label{rev_3333}
\mathcal{T}_{3333} = \frac{1}{u^2} \left( \mathcal{F}(u,v) + \ssigma u^5 \mathcal{F}(1/v,u/v) + \frac{\ttau u^5}{v^5} \mathcal{F}(v,u) \right),
\ee
with $\mathcal{F}$ such that  $\mathcal{F}(u,v)=\mathcal{F}(u/v,1/v)$.
The amplitude $\namem_{3333}^1(s,t)$ is indeed the Mellin transform of $\mathcal{F}$. Then, 
crossing invariance of $\mathcal{F}$ under $u\rightarrow u/v$ and $v\rightarrow 1/v$ corresponds to the identity
\be
\qquad \namem_{3333}^1(s,t)= \namem_{3333}^1(s,-t-s-5).\qquad
\ee
The other two identities which follow from \eqref{rev_3333} are
\bea
\namem_{3333}^{\ssigma}(s,t)=\namem^{1}_{3333}(-s-t-5,s),\qquad \namem^{\ttau}_{3333}(s,t)=\namem^{1}_{3333}(t,s).
\eea

In our notation, the amplitude of Rastelli and Zhou would correspond only 
to the term multiplied by ${A}_2^0$ in \eqref{FAnction_3333}. 
Indeed, the new contribution, proportional to ${A}_{4}^1 - 2 {A}_2^0$, 
vanishes when we plug in \eqref{uno_3333} and \eqref{due_3333}, and we expand at order $1/N^2$.
In the next example, i.e. the generalised tree-level correlator $\mathcal{T}_{4444}$, we will see similar features showing up,
and we will comment more in general about what is the pattern of $\mathcal{M}[\mathcal{T}]$.

\subsection{$\langle \mathcal{O}_4 \mathcal{O}_4 \mathcal{O}_4 \mathcal{O}_4 \rangle$}

There are in total $3+12$ propagator structures in free theory. The first three are disconnected and not relevant here.
We quote the result for connected free theory, 
\bea
&&
\langle \mathcal{O}_4 \mathcal{O}_4 \mathcal{O}_4 \mathcal{O}_4 \rangle_{\rm conn.\, free} =  \frac{ 16(N^2-9)^2(N^2-4)^2(N^2-1)^2}{(N^2+1)^2} \times \bigg[ 
\rule{3cm}{0pt}\notag\\[.1cm]
&&
\rule{2cm}{0pt}
\frac{16}{N^2-1}\left(u\ssigma+ \frac{u \ttau}{v} + u^3 \ssigma^3 + \frac{u^3 \ttau^3}{v^3}  +  \frac{ u^4 \ssigma^3\ttau}{v} + \frac{u^4\ssigma\ttau^3}{v^3} \right) +  \notag \\[.2cm]
&&
\rule{.2cm}{0pt}
8\left(\frac{27}{N^2(N^2-9)}+\frac{9}{N^2+1}-\frac{7N^2+4}{(N^2-4)(N^2-1)}\right)\left( u^2\ssigma^2 + \frac{u^2\ttau^2}{v^2} + \frac{u^4\ssigma^2\ttau^2}{v^2}\right)  +\notag\\[.2cm] 
&&
16\left(\frac{54}{N^2(N^2-9)}+\frac{18}{N^2+1}-\frac{16N^2+25}{(N^2-4)(N^2-1)}\right) \left( \frac{u^2\ssigma\ttau}{v} + \frac{u^3\ssigma^2\ttau}{v} + \frac{u^3\ssigma\ttau^2}{v^2}\right) 
\bigg].
\rule{1cm}{0pt}
\eea
Written as a sum over propagator structures as in \eqref{gen_free_theory0}, connected free theory is constrained by crossing symmetry to three independent classes, 
\bea
&&
{A}_{2}^{0}={A}_{2}^{1}={A}_6^0={A}_6^1={A}_8^1={A}_8^3=\frac{16}{N^2-1},\\
&& 
{A}_4^0={A}_4^2={A}_8^2=8\left(\frac{27}{N^2(N^2-9)}+\frac{9}{N^2+1}-\frac{7N^2+4}{(N^2-4)(N^2-1)}\right),\\
&&
{A}_6^1={A}_6^2={A}_4^1=16\left(\frac{54}{N^2(N^2-9)}+\frac{18}{N^2+1}-\frac{16N^2+25}{(N^2-4)(N^2-1)}\right).
\eea
We will consider 
$\{{A}_2^0, {A}_4^0,{A}_6^1\}$ as independent.

The generalised tree level function $\mathcal{T}_{4444}$ can be written conveniently in terms of just two independent functions $\widetilde{\mathcal{F}}$ and ${\mathcal{F}}$, in the following way, 
\bea
\label{treegen44441}
\mathcal{T}_{4444}&=&\frac{1}{u^2}\left[ \mathcal{F}(u,v) + \ssigma^2 u^6 \mathcal{F}(1/v,u/v) + \frac{\ttau^2 u^6}{v^6} \mathcal{F}(v,u) \right] \\
\label{treegen44442}
& & + \frac{1}{u^2} \left[\ssigma\ttau \widetilde{\mathcal{F}}(u,v) + \frac{\ssigma u^6}{v^6} \widetilde{\mathcal{F}}(v,u)+ \ttau u^6 \widetilde{\mathcal{F}}(1/v,u/v)  \right],
\eea
where both $\widetilde{\mathcal{F}}$ and $\mathcal{F}$ are invariant under $u\rightarrow u/v$ and $v\rightarrow 1/v$. 
Given the Mellin trasform, 
\be
\mathcal{T}_{4444}=
{u}^{4}\oint { u}^{s} { v}^{t }\ \Gamma[-s]^2\Gamma[-t]^2\Gamma\big[s+t+6\big]^2 \mathcal{M}[\mathcal{T}_{4444}]\,. 
\ee
with
\bea
\mathcal{M}[\mathcal{T}_{4444}]=\namem_{4444}^1+ \ssigma^2 \namem_{4444}^{\ssigma^2}+\ttau^2 \namem_{4444}^{\ttau^2}
+ \ssigma\ttau \namem_{4444}^{\ssigma\ttau}+ \ssigma \namem_{4444}^{\ssigma} + \ttau \namem_{4444}^{\ttau} 
\eea
we will specify $\namem_{4444}^{1}(s,t)$ and $\namem_{4444}^{\ssigma\ttau}(s,t)$, which are the Mellin transforms of ${\mathcal{F}}$ and $\widetilde{\mathcal{F}}$, respectively, 
and reconstruct $\mathcal{M}[\mathcal{T}_{4444}]$ 
%
by using symmetries, similarly to \eqref{treegen44441} and \eqref{treegen44442}.

The Mellin transforms of ${\mathcal{F}}$ and $\widetilde{\mathcal{F}}$ are
\be\label{FAnction_4444}
\begin{array}{clll}
\namem_{4444}^1=&   {  -\frac{  {A}_2^0 }{(s+3)(t+1)(s+t+5)}   }  & {- \frac{ \mathscr{L}_{4444}^1 }{ 2(s+2)(t+1)(s+t+5) } }&  {- \frac{ \mathscr{L}_{4444}^1+ \mathscr{L}_{4444}^2(s+1) }{ 6(s+1)(t+1)(s+t+5) } },\\[.3cm]
\namem_{4444}^{\ssigma\ttau}=&   {  -\frac{ {A}_2^0 }{(s+1)(t+2)(s+t+4)}   } & {+ \frac{ \mathscr{L}_{4444}^1 }{ 2(s+1)(t+1)(s+t+4) } }&  {+ \frac{ \mathscr{L}_{4444}^1 }{ 2(s+1)(t+2)(s+t+5) } } \\[.3cm]
 &  &  {- \frac{ \mathscr{L}_{4444}^1(s+3) }{ 3(s+1)(t+1)(s+t+5) } }   & {- \frac{ \mathscr{L}_{4444}^1-\mathscr{L}_{4444}^2 +(2\mathscr{L}_{4444}^1-\mathscr{L}_{4444}^2)(s+1) }{ 3(s+1)(t+2)(s+t+4) } }  ,
\end{array}
\ee
where we defined 
\beq \label{tlc}
 \mathscr{L}_{4444}^1={A}_{2}^0-{A}_{4}^0\qquad \mathscr{L}_{4444}^2=\tfrac{3}{2}{A}_{2}^0-{A}_{4}^0-\tfrac{1}{4} {A}_6^1
 \eeq
Finally,
\bea
\begin{array}{lll}
\namem_{4444}^{\ssigma}(s,t)&=&\namem_{4444}^{\ssigma\ttau}(t,s)\\[.2cm]
\namem_{4444}^{\ttau^2}(s,t)&=&\namem_{4444}^{1}(t,s)
\end{array}
\qquad
\begin{array}{lll}
\namem_{4444}^{\ttau}(s,t)&=&\namem_{4444}^{\ssigma\ttau}(-s-t-6,s)\\[.2cm]
\namem_{4444}^{\ssigma^2}(s,t)&=&\namem_{4444}^{1}(-s-t-6,s)\\ 
\end{array}
\eea

The terms in $\namem_{4444}^1$ and $\namem_{4444}^{\ssigma\ttau}$, proportional to $A_2^0$, give the amplitude of Rastelli and Zhou. 
The combinations $\mathscr{L}_{4444}^{i=1,2}$ in~\eqref{tlc} vanish at order $1/N^2$.

Let us highlight some new features of $\mathcal{M}[\mathcal{T}_{4444}]$ beyond tree level. 
Recall that the result of Rastelli and Zhou \cite{Rastelli:2016nze} can be obtained 
by considering an ansatz in Mellin space such that each monomial $\ssigma^i\ttau^j$ 
is accompanied by a {\it single} pole in the plane $(s,t)$. 
In comparison, the upgraded tree level amplitude has more structure than this.   
In particular, poles like $(s+2)(t+1)$ and $(s+1)(t+1)$, corresponding to powers of $u^2$and $u^3$ in the small $u$ expasion, and therefore corresponding to allowed twists below window, are also turned on. 
We see now that their residue is proportional to the linear constraints $\mathscr{L}_{4444}^{i=1,2}$, which indeed vanish at order $1/N^2$.
We also notice that by writing each pole in the form $\frac{1}{(s+n_1)(t+n_2)(s+t+n_3)}$ with integers $n_{i=1,2,3}$, the numerator is at most linear in $s$ and $t$. 
Therefore, we infer that the limit $s\rightarrow \beta s$ and $t\rightarrow \beta t$ with large $\beta$
scales like $O(\beta^{-2})$, i.e. one more power than the $O(\beta^{-3})$ of the tree level function of Rastelli and Zhou.

The case of $\mathcal{M}[\mathcal{T}_{4444}]$ exemplifies well what is the general pattern of $\mathcal{M}[\mathcal{T}_{\vec{p}}]$ in Mellin space. 
In fact we expect $\mathcal{M}[\mathcal{T}_{\vec{p}}]$ to be a rational function in which all allowed poles 
in the plane $(s,t)$ are turned on, eventually decorated by a non trivial numerator, 
which is nevertheless constrained by the large $s$ and $t$ behavior.
Similarly to our position space algorithm, 
the free coefficients in this ansatz will be fixed by demanding that the SCPW expansion
below window {\it completely cancels} free theory contributions as described in~\eqref{LT} and~\eqref{LT2}.

\subsection{Next-to-Next-to-Extremal Correlators}
\label{sec:N2E_gen_treelevel}

A next-to-next-to-extremal correlator is defined by $\kappa_{\vec{p}}=2$, 
i.e. a vector of external charges such that $p_3=2$ or $p_1+p_2+p_3-p_{4}=4$. 
There are only six propagator structures available, and indeed these correlators only contribute to 
a single $su(4)$ channel, namely $[0,p_{43},0]$. The definition of single particle states has 
two non trivial consequences. Firstly, it was proven in \cite{Aprile:2018efk} that 
the number of connected propagator structures actually reduces to three. 
Secondly, connected free theory is given by the {\rm exact} formula,
\be\label{free_n2conn}
  \mathcal{P}\, F_{\vec{p}}(N^2)\, p_1p_2p_3p_4 
  \left[    \big(1+\tfrac{p_{43}+p_{21} }{2} \big)  \frac{u\tau}{v}+  
										      \big(1+\tfrac{p_{13}+p_{42}}{2} \big) u\sigma +
									               \big(1+|\tfrac{p_{23}+p_{14}}{2}| \big) \frac{u^2\sigma\tau}{v}\right]
\ee
where $F_{\vec{p}}$ asymptotes $N^{(p_1+p_2+p_3+p_4-4)/2}$ in the large $N$ limit. For example, 
\begin{gather*}
F_{2244}=F_{3324}=\frac{\prod_{k=1}^{3} (N^2-k^2) }{ (N^2+1)}, \qquad F_{3335}=F_{3524}= \frac{\prod_{k=1}^{4} (N^2-k^2) }{ N(N^2+5)},\\
F_{4424}= \frac{\prod_{k=1}^{3} (N^2-k^2) (N^4-20N^2+9) }{ N(N^2+1)^2}. \qquad
\end{gather*}

Thus, for next-to-next-to-extremal correlators,  the non-planar result \eqref{free_n2conn} 
is the factorized product of the $1/N^2$ connected free theory, uplifted to all $N$ by the factor $F_{\vec{p}}(N^2)$. 
It follows that the all $N$ relative coefficients among the three propapator structures, 
was already captured by the $1/N^2$ result. Notice also that $F(N^2)$ manifestly vanishes when 
the number of colors $N$ is less than the charge of any of the external operators. Both these statements would be false 
if we replace our single particle operator $\mathcal{O}_p$ with the corresponding single 
trace half-BPS operator, thus dropping the admixture of multi-trace operators. 

The particular structure of connected free theory in \eqref{free_n2conn} implies the following exact relation on the SCPW coefficients, 
\be
L^{f}_{\vec{p},2+p_{43},[0p_{43}0]}= F_{\vec{p}}(N^2) \left[ {A}_{[l+2]}\Big|_{\tfrac{1}{N^2}} \right],\qquad L^{f}_{\vec{p},\tau\ge4+p_{43},[0p_{43} 0]}= F_{\vec{p}}(N^2) L^{(1)f}_{\vec{p},\tau,[0p_{43} 0]}.
\ee 
Therefore, for the purpose of constructing generalised tree-level functions, the defining
condition (b)  becomes  
\be
		L^{\mathcal{T}}_{\vec{p},2+p_{43},[0p_{43}0]}+F_{\vec{p}}(N^2) \left[ {A}_{[l+2]}\Big|_{\tfrac{1}{N^2}} \right]=0 
\ee
and by uniqueness we conclude that for next-to-next-to-extremal correlators the generalised 
tree level function $\mathcal{T}_{\vec{p}}$ equals the function of Rastelli and Zhou, 
properly normalized as in \cite{Aprile:2018efk}, multiplied by the factor $F_{\vec{p}}(N^2)$.   

Building on the property that $\mathcal{T}_{\vec{p}}$ is uniquely defined by the conditions 
(\ref{con1}),(\ref{con2}),(\ref{con3}) it is simple to see that next-to-next-to-extremal correlators sometimes have
an additional feature. In fact, there are sets of external charges 
$q_{i=1,..4}$ and $q'_{i=1,..4}$  such that the corresponding $1/N^2$ free theories will be proportional. 
For example, the following two families
\be\label{ex_equality_corr}
\rule{1.5cm}{0pt}
\begin{array}{l}
q_1=p+2 \\
q_2=p+2\\
q_3=2\\
q_4=2(q+1)
\end{array}
\rule{1.5cm}{0pt}
\begin{array}{l}
q'_1=q+2 \\
q'_2=q+2\\
\displaystyle q'_3=q'_4-2q\\
\displaystyle q'_4=q+({p+2})
\end{array}
\ee
Notice that both have the same value of the exponent $\mathbf{d}=p_1+p_2+p_3+p_4-1$, and both  
have the same threshold twist. This happens because the next-to-next-to extremality condition, which we can rewrite as 
$p_3+{\rm min}(0,\frac{p_1+p_2-p_3-p_4}{2})=2$, is achieved by the two different conditions on the minimum. Therefore,
\be
q'_3+{\rm max}\left(0,\frac{q'_1+q'_2-q'_3-q'_4}{2}\right)=q_3+{\rm max}\left(0,\frac{q_1+q_2-q_3-q_4}{2}\right).
\ee 
In the case $p=2$ and $q=1$, we obtain the correlators $4424$ and $3335$. 
Indeed, it is simple to verify that the corresponding tree level function from \cite{Rastelli:2016nze} are proportional to each other. 
This `degeneracy' is lifted at one-loop. As we explicitly showed  in Section \ref{one_loop_N2}, the minimal one-loop functions 
$\mathcal{H}^{(2)}_{3335}$ and $\mathcal{H}^{(2)}_{4424}$ 
are genuinely distinguished by OPE predictions in the window.

The constructions of $\mathcal{T}$ in this section, and the one of the minimal one-loop function in the previous section, 
conclude our journey through the determination of the dynamical one-loop function $\mathcal{D}^{(2)}=\mathcal{T}^{(2)}+\mathcal{H}^{(2)}$. 
The subject of the next section is instead inspired by the existence of the hidden tree-level symmetry found by \cite{Caron-Huot:2018kta}. 

\section{Explorations of the 10d symmetry at One-loop} \label{exploration10d}

A tree level correlator $\mathcal{D}^{(1)}_{pqrs}$ is obtained by acting with 
a differential operator $\widehat{\mathcal{D}}_{pqrs}$ on the stress-tensor four point correlator $\mathcal{D}^{(1)}_{2222}$.  
The existence of these operators is a consequence of the hidden tree level 10d conformal symmetry, unveiled in \cite{Caron-Huot:2018kta}. 
The structure of the anomalous dimensions \eqref{eta_anom_dim}, both numerator and denominator can also be understood in terms of this  hidden  symmetry.
In particular, the numerator is the eigenvalue of an 8-th order operator
which annihilates protected multiplets,  $\Delta^{(8)}$, and the structure of the denominator is in correspondence 
with that of the partial-wave decomposition of the $2\rightarrow 2$ flat space S-matrix of the type IIB axio-dilaton. 
Then, the hidden symmetry explains the residual degeneracies of the anomalous dimensions found in \cite{Aprile:2018efk}, and the proportionality of some 
the next-to-next-to-extremal tree level correlators, as we observed in Section \ref{sec:N2E_gen_treelevel}.


An interesting question to ask is whether an organising 10d principle persists at one-loop.  
We begin investigating this problem by showing that we can recast
the expression of our minimal one-loop functions 
by using the operators  
 $\Delta^{(8)}$ and $\widehat{\mathcal{D}}_{pqrs}$. 
We shall see that even though it is possible to achieve such a result, 
the way it happens departs slightly from the way 
we understood the physical properties of $\mathcal{H}^{(2)}_{\vec{p}}$
in Section \ref{structure_oneLoop} and \ref{minimal_loop_sec}.

\subsection{Leading Logs from $\Delta^{(8)}$ and $\widehat{\mathcal{D}}_{pqrs}$ }
\label{10d}

We introduce the operators $\Delta^{(8)}$ and $\widehat{\mathcal{D}}_{pqrs}$ by recalling few important facts. 

It was noticed in \cite{Aprile:2018efk} that the computation of the one-loop leading $\log^2 u$, i.e \eqref{ddisc_fa2}, could be reorganised and simplified drastically 
by introducing an 8-th order differential Casimir operator.  
It is indeed possible to rewrite the $\log^2 u$ discontinuity as
\bea
\label{ddisc_version1}
\mathcal{D}^{(2)}_{\vec{p}}\Big|_{\log^2 u}&=&
\sum_{a,b}\frac{\Upsilon_{[aba]}}{u^2}\left[{u^{-\frac{p_{43}}{2} }}\ \Delta_{[aba]}^{(8)}\ u^{+\frac{p_{43}}{2} }\right] { \mathcal{F} }^{(2)}_{\vec{p};[aba]} \\
\label{ddisc_version1B}
{ \mathcal{F} }^{(2)}_{\vec{p};[aba]} &=& 
\sum_{\tau\ge\tau^\text{max},l } \left(  {\bf M}^{(1)}_{\vec{\tau}} \frac{ \big( {\bf L}^{(0)}_{\vec{\tau}} \big)^{-1} }{\delta_{t(\tau)}^{(4)} \delta^{(4)}_{t(\tau)+l+1}   } {\bf M}^{(1)}_{\vec{\tau}} \right)_{(p_1p_2),(p_3p_4)} \frac{  \mathcal{B}^{\left( 2+\tfrac{\tau}{2} ,l\right)} }{u^{  \frac{ p_{43} }{2} } }
\eea
where the Casimir $\Delta_{[aba]}^{(8)}$ 
is precisely such that its eigenvalue is the numerator of the anomalous dimensions in \eqref{eta_anom_dim}, i.e. 
\bea
\delta^{(4)}_t&=& { 2}(t-1) (t+a) (t+a+b+1)(t+2a+b+2) \\
\Delta_{[aba]}^{(8)}\, \mathcal{B}^{2+\tau,l} &=& + \delta^{(4)}_{t} \delta^{(4)}_{t+l+1}\,   \mathcal{B}^{2+\tau,l}
\eea
It turned out that the resummation of ${ \mathcal{F} }^{(2)}_{\vec{p};[aba]}$ was remarkably simpler. (Notice that ${ \mathcal{F} }^{(2)}_{\vec{p};[aba]}$ here has a series expansion in integer powers of $u$)

A further improvement of \eqref{ddisc_version1} was achieved in \cite{Caron-Huot:2018kta}: 

Firstly, it is possible to repackage the action of the $su(4)$ dependent operators $\Delta^{(8)}_{[aba]}$, into a single compact operator 
\be \label{delta8}
\Delta^{(8)} =  \frac{x {\bar x} y {\bar y}}{(x-{\bar x})(y-{\bar y})}\prod_{i,j=1}^{2} \left(\mathbf{C}_{x_i}^{[+\alpha,+\beta,0]} -\mathbf{C}_{y_j}^{[-\alpha,-\beta,0]}\right) \frac{(x-{\bar x})(y-{\bar y})} {x {\bar x} y {\bar y}}
\ee
where $\alpha=p_{21}/2$, $\beta=p_{34}/2$ and $\mathbf{C}_{x}^{[\alpha,\beta,\gamma]}$ is the elementary $2d$ casimir
\bea\label{2dcasimir}
\mathbf{C}_{x}^{[\alpha,\beta,\gamma]}&=& x^2(1-x)\partial_{x}^2 +x(\gamma-(1+\alpha+\beta)x)\partial_{x} - \alpha\beta x\ .
\eea
Secondly, this $\Delta^{(8)}$ has the property that
\be
(u\sigma)^{-\frac{p_{43}}{2} }\Delta^{(8)} (u\sigma)^{\frac{p_{43}}{2} } \Big[ { \mathcal{F} }^{(2)}_{\vec{p};[aba]} \, \Upsilon_{a,b} \Big]\ =
\Upsilon_{[aba]}\left[{u^{-\frac{p_{43}}{2} } }\ \Delta_{[aba]}^{(8)}\ {u}^{+\frac{p_{43}}{2} }\right] { \mathcal{F} }^{(2)}_{\vec{p};[aba]}\ .
\ee
where $\mathcal{F}^{(2)}_{\vec{p};[aba]}$ {does not} depend on the $su(4)$ cross ratios. 
We can then commute $\Upsilon_{[aba]}$ and obtain the $\log^2 u$ discontinuity from the action 
of $\Delta^{(8)}$ on a prepotential $\mathcal{F}^{(2)}_{\vec{p}}$, namely
\be \label{delta8eq}
u^2 \mathcal{D}^{(2)}_{\vec{p}}\Big|_{\log^2 u} =  (u\sigma)^{-\frac{p_{43}}{2} }\Delta^{(8)} (u\sigma)^{\frac{p_{43}}{2} } \Big[  { \mathcal{F} }^{(2)}_{\vec{p}} \, \Big]\,,
\qquad \mathcal{F}^{(2)}_{\vec{p}}\equiv  \sum_{a,b} { \mathcal{F} }^{(2)}_{\vec{p};[aba]} \Upsilon_{[aba]}
\ee
Notice that the conjugated operator
$(u\sigma)^{-\frac{p_{43}}{2} }\Delta^{(8)} (u\sigma)^{+\frac{p_{43}}{2} } $ 
is invariant under the symmetry, $u\rightarrow u/v$ and $v\rightarrow 1/v$, which in our conventions holds when $p_{21}=0$.
We remark here that the expression of  $\Delta^{(8)}$ depends on the choice of external charges!

The remaining dependence on the external charges $\vec{p}$, can be absorbed into the action of the operators $\widehat{\mathcal{D}}_{\vec{p}}$.
These operators are defined at tree level by the relations
\be
\mathcal{D}^{(1)}_{\vec{p}}=
\widehat{\mathcal{D}}_{\vec{p}}\left( u^2\,\mathcal{D}^{(1)}_{2222} \right)\qquad \forall\, \vec{p}=(p_1p_2p_3p_4)
\ee
and are used at one-loop to compute 
\be\label{Simon_oneloop}
\mathcal{F}^{(2)}_{\vec{p}}= \widehat{\mathcal{D}}_{\vec{p}}\ \mathcal{F}^{(2)}_{2222}\,.
\ee
once $\mathcal{F}^{(2)}_{2222}$ is known \cite{Caron-Huot:2018kta}. 
The latter can be written in the form 
\bea
\label{Simon2222}
&\displaystyle \mathcal{F}^{(2)}_{2222}
=16 \left(\frac{x {\bar x}}{{\bar x}-x}\right)^7\Big[ \mathscr{F}_2(x,{\bar x})-\mathscr{F}_2({\bar x},x) \Big],\\[.2cm]
%
\mathscr{F}_2&=
\displaystyle
\left[ 1+\frac{x^2}{2} \left[ \sum_{i=0}^2 \left[\frac{{\bar x}-x}{x{\bar x}}\right]^{i+1} \frac{ x^i \partial_{1} x^i}{i!(5)_i} \right]\right]\Bigg( 
\frac{(x-2)(1-x)}{x^3}  -\frac{(1-x)^5}{8x^5} {\rm Li}_1^2(x)+ \notag
\\[.2cm]
&\displaystyle
+\frac{(1-(1-x)^5)}{4x^5}{\rm Li}_2(x)+\frac{(1-x)(7-7x+2x^2)}{8x^4} {\rm Li}_1(x)\Bigg). \notag
\eea
In writing \eqref{Simon2222} we have highlighted the max power of $(x-{\bar x})$ in the denominator. 
This has to be the same power of the tree-level function $\mathcal{D}^{(1)}_{2222}$ by construction.

In this paper we considered
\bea
\widehat{\mathcal{D}}_{2233}&=&\tfrac{3}{4\times 2}(4-u\partial_u) \notag\\
\widehat{\mathcal{D}}_{2323}&=&\tfrac{3}{4\times 2}(u\partial_u+v\partial_v) \notag\\
\widehat{\mathcal{D}}_{2244}&=&\tfrac{1}{4} (5-u\partial_u)(4-u\partial_u)  \notag\\[.1cm]
\widehat{\mathcal{D}}_{2424}&=&\tfrac{1}{4}  (1+u\partial_u+v\partial_v)(u\partial_u+v\partial_v)  \notag\\
\widehat{\mathcal{D}}_{3335}&=&\tfrac{ 3 \sqrt{15} }{4\times 4}(4-u\partial_u)v\partial_v(-u\partial_u -v\partial_v) \notag\\
\widehat{\mathcal{D}}_{4424}&=& \tfrac{2\sqrt{2}}{4}(4-u\partial_u)v\partial_v(-u\partial_u -v\partial_v) \notag\\
\widehat{\mathcal{D}}_{3333}&=&\tfrac{9}{4\times 4} \left[ (4-u\partial_u)^2+\frac{u\tau}{v}  (v\partial_v)^2 +u\sigma (u\partial_u+v\partial_v)^2\right]\notag\\
\widehat{\mathcal{D}}_{4444}&=& \tfrac{1}{4} \left[ (5-u\partial_u)^2+4\left[ \frac{u\tau}{v}  (v\partial_v)^2 +u\sigma (u\partial_u+v\partial_v)^2\right] \right]  (4-u\partial_u)^2  \notag
\\
& & 
\, + \tfrac{1}{4} \left[ \frac{(u\tau)^2}{v^2} (1-v\partial_v)^2(v\partial_v)^2 +(u\sigma)^2 (u\partial_u +v\partial_v)^2(1+u\partial_u +v\partial_v)^2 \right] \notag\\
& & \, \quad +\, \frac{u\tau}{v} (u\sigma) (v\partial_v)^2 (v\partial_v+u\partial_u)^2 
\label{operatori_pqrs_paper}
\eea
For these correlators we have verified explicitly that the prescription \eqref{Simon_oneloop} 
agrees with the more standard two variable CPW resummation obtained through \eqref{ddisc_fa2}. 
This amazing computation shows a very non trivial outcome of the ten-dimensional conformal symmetry.

More generally,
the operators $\widehat{\mathcal{D}}_{\vec{p}}$ have the unique form
\be\label{SimonDpqrs}
\widehat{\mathcal{D}}_{\vec{p}} = 
\sum_{\gamma=p_{43} }^{ {\rm min}(p_1+p_2,p_3+p_4)-4}
\left[ (u\ssigma)^{\frac{\gamma-p_{43}}{2}}  \sum_{k=0}^{  \frac{\gamma-p_{43}}{2} } \left( \frac{ g_{14} g_{23} }{ g_{13} g_{24} } \right)^k\mathrm{d}_{\vec{p}}^{\gamma k} \right]
\ee
where $\mathrm{d}_{\vec{p}}^{\gamma k}$ are some differential operators of degree $\frac{1}{2} \sum_{i}(p_i-2)$ 
in the letters $u\partial_u$ and $v\partial_v$. In fact, each monomial $\ssigma^i\ttau^j$ in 
the amplitude of Rastelli and Zhou corresponds to a propagator structure in \eqref{SimonDpqrs}. 
Finally, the sum over $\gamma$ contains at most propagator structures with $(\ttau/v)$ to the power $\kappa_{\vec{p}}-2$.

\subsection{A Pre-Amplitude study}

In this section we will explore the question: 
``Can we pull out of our minimal 
one-loop functions the operators  $\widehat{\mathcal{D}}_{pqrs}$ and $\Delta^{(8)}$?" %

Consider first the one-loop correlators $\langle 22pp\rangle$ for $p=2,3$ and $4$. 
(The cases $p=2,3$ have been determined in \cite{Aprile:2017bgs} and \cite{Aprile:2017qoy}, respectively.)
We have managed to rewrite these in terms of certain pre-amplitudes $\mathcal{L}^{(2)}_{22pp}$ such that the following equations are satisfied
\begin{align}
\mathcal{H}^{(2)}_{2222}=&\, \tfrac{1}{u^{2}}\Delta^{(8)} \mathcal{L}^{(2)}_{2222} +4 u^2  \overline{D}_{2 4 2 2} 
\label{delta8_2222}\\[.2cm]
%
\mathcal{H}^{(2)}_{2233}=&\,  \tfrac{1}{u^{2}} \Delta^{(8)} \mathcal{L}^{(2)}_{2233} - \tfrac{2u^2}{v^2} + 4u^2( \overline{D}_{1423}+ \overline{D}_{1432} ) + 24 u^2 \overline{D}_{2422} - 30 u^3 \overline{D}_{3522}
\label{delta8_2233}\\[.2cm]
%
\mathcal{H}^{(2)}_{2244}=&\, \tfrac{1}{u^{2}}\Delta^{(8)} \mathcal{L}^{(2)}_{2244} - \tfrac{8u^2}{v^2} + 24 u^3(\overline{D}_{2523}+ \overline{D}_{2532} )+ 40 u^2 \overline{D}_{2422} +48 u^3(  \overline{D}_{3522} -u \overline{D}_{4622})\notag\\
\label{delta8_2244}
\end{align}\\[-1.25cm]

The r.h.s of \eqref{delta8_2222}-\eqref{delta8_2244} consists of functions 
$\mathcal{L}^{(2)}_{22pp}$ such that $\Delta^{(8)}\mathcal{L}^{(2)}_{22pp}$ reproduces the minimal one-loop 
function on the l.h.s up to some tree level remainders. 
There is a single remainder for $p=2$, which happens to have the same structure of 
$\mathcal{D}^{(1)}_{2222}$.  For $p=3,4$, there is more structure in the remainders than the corresponding tree level correlators .  
The origin of these remainders goes together with $\Delta^{(8)}$, as we now explain.

The ansatz for the functions $\mathcal{L}^{(2)}_{22pp}$ has the same form the one for $\mathcal{H}_{\vec{p}}$ in \eqref{G2uvresult}, 
with the substitution $\mathbf{d}\rightarrow \mathbf{d}-8$, i.e.
\bea
\mathcal{L}^{(2)}_{22pp} &=&
				\frac{ \Wloop_{4-} +\Wloop_{3-} }{(x-{\bar x})^{\mathbf{d}_{\vec{p}}}} 
				+ \frac{\Wloop_{3+} 
				+ { \Wloop_{2+}} }{(x-{\bar x})^{\mathbf{d}_{\vec{p}}-1}} +
%
			\left[
						\frac{  \Wloop_{2-} }{ (x-{\bar x})^{\mathbf{d}_{\vec{p}}} }
						+ \frac{ \Wloop_{1v} + \Wloop_{1u}  }{(x-{\bar x})^{\mathbf{d}_{\vec{p}}-1} }+
							\frac{ \Wloop_{0}}{ (x-{\bar x})^{ \mathbf{d}_{\vec{p}}-3}} \right]\qquad
\eea
The basis of transcendental functions is unchanged
\bea\label{basis11}
\Wloop_{4-}&=&\hpiccolo_1\, \phi^{(2)}(x'_1,x'_2) + \hpiccolo_2\,  \phi^{(2)}(x,{\bar x}) + \hpiccolo_3\, \phi^{(2)}(1-x,1-{\bar x})
\notag\\[.3cm]
\Wloop_{3-}&=&  \hpiccolo_4\, x \partial_{x} \phi^{(2)}(x,{\bar x}) +\hpiccolo_5\, (x-1)\partial_{x}  \phi^{(2)}(1-x,1-{\bar x})  - (x \leftrightarrow {\bar x} )
\notag\\[.2cm]
\Wloop_{3+}&=& (x-{\bar x})\Big(  \hpiccolo_6\, \partial_v  \phi^{(2)}(x,{\bar x})+ \hpiccolo_7\, \partial_u  \phi^{(2)}(1-x,1-{\bar x})\Big) + \hpiccolo_8\, \zeta_3
\notag\\[.2cm]
\Wloop_{2+}&=&\hpiccolo_9 \log(u)\log(v)  +\hpiccolo_{10} \log^2 v +\hpiccolo_{11}\log^2 u
\eea
and
\beq\label{basis22}
\begin{array}{ll}
\Wloop_{2-}=\hpiccolo_{\Box}\, \phi^{(1)}(x,{\bar x})\qquad& \,\Wloop_{0}\,= \hpiccolo_0\\[.2cm]
\Wloop_{1u}\,=\hpiccolo_{u} \log u \qquad&  \Wloop_{1v}=\hpiccolo_{v} \log v\\
\end{array}
\eeq
Each coefficient function $\hpiccolo_{i=1,..,11,\Box,u,v,0}$ will be polynomial 
in the variables $x,{\bar x}$, since we are studying next-to-next-to-extremal correlators.

By construction, we make manifest the $\log^2 u$ discontinuity
\be\label{preddisc22pp}
\mathcal{L}^{(2)}_{22pp}\Big|_{\log^2 u}=\widehat{\mathcal{D}}_{22pp}\mathcal{F}^{(2)}_{2222}  
\ee
and impose $x_i\rightarrow x_i/(x_i-1)$ crossing symmetry, because this is a symmetry of the $\langle 22pp\rangle$  correlators, and it is a symmetry of $\Delta^{(8)}$. 
Then, we impose absence of $x=\bar{x}$ poles.

In our algorithm for the minimal one-loop function $\mathcal{H}_{22pp}$ we would cross the ansatz to the orientation ${2p2p}$ and match the corresponding $\log^2 u$ discontinuity. But
$\Delta^{(8)}$ depends on the orientation of the external charges, and we cannot proceed this way. Instead, we stay within the orientation $\vec{p}=22pp$, 
and apply $\Delta^{(8)}$ on $\mathcal{L}^{(2)}_{22pp}$. 
The resulting ansatz can now be understood as the starting point for the construction of our one-loop function,
In particular, after crossing to $\vec{p}=2p2p$, and matching the corresponding leading log, the weight four, three and two-symmetric coefficient functions are fixed.

Moving to the tree-like part, 
we encounter a major difference compared to our algorithm of Section \ref{minimal_loop_sec}: 
The action of $\Delta^{(8)}$ brings $\mathcal{L}^{(2)}_{22pp}$ outside the minimality of $\mathcal{H}^{(2)}_{22pp}$! 
We thus expect the presence of extra tree-level contributions with additional singular terms in $v$.\footnote{This is because 
weight four and three contributions in the preamplitude $\mathcal{L}_{\vec{p}}$, upon the action of $\Delta^{(8)}$, 
produce a cascade of tree level contributions with non minimal denominator. Some of these contributions in $\mathcal{L}_{\vec{p}}$ are fixed by matching the pre-amplitude $\log^2 u$ discontinuity}
%
%
Indeed, looking at the relations \eqref{delta8_2222}-\eqref{delta8_2244}, both $\Delta^{(8)} \mathcal{L}^{(2)}_{22pp}$ 
and the remainders have additional singular terms compared to our minimal one-loop function \eqref{G2uvresult}, 
but these cancel in the sum, in such a way that the r.h.s is indeed our minimal one-loop function.  

It is important to realise that the free coefficients we can play with, in order to 
obtain the remainders on the r.h.s of \eqref{delta8_2222}-\eqref{delta8_2244}, are the 
free coefficients in $\Delta^{(8)} \mathcal{L}^{(2)}_{22pp}$,  and the $\alpha_{i=1,\ldots}$ labeling the ambiguities of $\mathcal{H}_{22pp}$. 
Indeed, for the purpose of this section it is useful to think of the ambiguities as a sort of `gauge' parameters which we eventually fix to some particular value. 
(Of course, the most general form of the ambiguities on the r.h.s of \eqref{delta8_2222}-\eqref{delta8_2244} can be added in afterwards.)
Operationally, the idea is to fix free parameters 
in such a way that the resulting tree-like part in the difference $\mathcal{H}^{(2)}_{22pp}- u^{-2}\Delta^{(8)} \mathcal{L}^{(2)}_{22pp}$
has a lower power of $(x-\bar{x})$ in the denominator, compared to $\mathbf{d}_{\vec{p}}+8$.\footnote{In order to achieve this result it is useful to impose as many $x={\bar x}$ zeros as possible 
in the difference between $\mathcal{H}^{(2)}_{22pp}- u^{-2}\Delta^{(8)} \mathcal{L}^{(2)}_{22pp}$. Similarly, we impose as many $x=0$ zeros as possible. }

Being a differential operator, $\Delta^{(8)}$ has a kernel,
therefore any construction of $\mathcal{L}_{22pp}$ is only unique up to such a kernel.
The kernel does not show up in the  $\log^2 u$ discontinuity, 
because in that case the use of $\Delta^{(8)}$ is specified by the OPE, 
and in particular by the form of the anomalous dimensions \cite{Aprile:2018efk}. 

So far, the determination of $\mathcal{L}_{22pp}$ did not follow strictly the rules of our algorithm, especially regarding the minimality of our ansatz for $\mathcal{H}_{\vec{p}}$. 
However, it will be very surprising how the complexity of $\mathcal{L}_{22pp}$ is reduced in comparison to $\mathcal{H}_{22pp}$. 

The coefficient functions for $\mathcal{L}_{2222}$, listed here below in the basis \eqref{basis11}-\eqref{basis22}, are  ($Y_{\pm}\equiv 1\pm v$):
\be
\begin{array}{l}
\hpiccolo_1=+4u^4v^2(u-1+v)\,,\\[.2cm]
\hpiccolo_2=-4u^4(u+1-v)\,,\\[.2cm]
\hpiccolo_3=-4v(u^5-5u^4 Y_{+}-Y_{-}^4Y_{+}+5u^3(2Y_{+}^2-v)+uY_{-}^2(5Y_{+}^2-2v)-u^2Y_{+}(10Y_{+}^2-19v))\,,\\[.2cm]
\hpiccolo_4=-\tfrac{1}{3}u^3( u^4 +Y_{-}^4-4u^3Y_{+}+8uY_{-}^2Y_{+}-2u^2(3 Y_{-}^2-2v) )\,,\\[.2cm]
\hpiccolo_5=+\tfrac{1}{2}\hpiccolo_4 +\tfrac{1}{6}Y_-(5u^6-39u^5 Y_{+}+3Y_{-}^4( Y_{+}^2+4v)+u^4(73 Y_{+}^2-4 v) \\[.1cm]
\rule{2.5cm}{0pt}
- u Y_{-}^2 Y_{+} (19Y_{+}^2 + 32 v)-2u^3 Y_{+} (37Y_{+}^2-8v)+u^2 (51 Y_{+}^4 -56 v Y_{+}^2-88v^2))\,,\\[.2cm]
\hpiccolo_6=-\tfrac{1}{3}u^3 Y_{-}(u^2+Y_{-}^2+10u Y_{+})\,,\\[.2cm]
\hpiccolo_7=-\tfrac{5}{12}\hpiccolo_8+\tfrac{1}{6}(-4u^5 Y_{-}^2+3 Y_{-}^4(Y_{+}^2+4v)+u^4(19+48v+99v^2)-4 u Y_{-}^2 Y_{+}(4 Y_{+}^2+11v)+\\[.1cm]
\rule{2.5cm}{0pt}
-4 u^3(9+39 v+45 v^2+7v^3)+2u^2(17 Y_{+}^4+11 vY_{+}^2-64v^2))\,,\\[.2cm]
\hpiccolo_8=-\tfrac{2}{5}u^2(u^4+Y_{-}^4-2u(2+7v)( u^2 +Y_{-}^2)+u^2(6+24 v-94 v^2))\,,\\[.2cm]
\hpiccolo_9=-\tfrac{1}{3}( 14 u^5 -2 u^6-3 Y_{-}^5 Y_{+}-14u^4(2+v-3v^2)+2u Y_{-}^3(8 Y_{+}^2-5 v)+  \\[.1cm]
\rule{2.5cm}{0pt}  +u^3 Y_{-}(38 Y_{-}^2+135 v) -7 u^2 (5-v+v^3-5v^4) )\,, \\[.2cm]
\hpiccolo_{10}=-\tfrac{1}{3}v(7u^5 -3 Y_{-}^4(3+v)-u^4(37+35 v) + u Y_{-}^2(43+49 v+16 v^2) +\\[.1cm]
\rule{2.5cm}{0pt} +u^3(78+99v+38v^2) -u^2(82+60v+75v^2+35v^3) )\,,\\[.2cm]
\hpiccolo_{11}=-\tfrac{1}{3}u^4(2u^2-7Y_{-}^2-7u Y_{+})\,,
\end{array}
\ee
for weight four, three, and two symmetric, and 
\be
\begin{array}{l}
\widetilde{\hpiccolo}_{\Box}=+\tfrac{1}{180} (572 u^7 - 2939  u^6 Y_{+} +234 Y_{-}^6 Y_{+} - 3 u Y_{-}^4 (379 Y_{+}^2-4v) +u^5 (5295 Y_{+}^2-2320 v)  \\[.1cm]
\rule{1.55cm}{0pt} + 9u^2 Y_{-}^2 Y_{+}(193Y_{+}^2 -64 v) - 4 u^4 Y_{+}(974 Y_{+}^2 -2087 v) +2u^3 (67 Y_{+}^4-1322 v Y_{+}^2 +400 v^2)) \\[.2cm]
\widetilde{\hpiccolo}_{u}=-\frac{1}{5}u(29u^5-114u^4 Y_{+}-10 Y_{-}^4 Y_{+}+2u Y_{-}^2(31Y_{+}^2-34 v)\\[.1cm]
\rule{2.55cm}{0pt} +u^3 (193 Y_{+}^2 - 174 v) - 10 u^2 Y_{+}(16 Y_{+}^2-43 v))\\[.2cm]
\widetilde{\hpiccolo}_{v}=-\tfrac{1}{2}\widetilde{\hpiccolo}_{u}-\tfrac{1}{90}Y_{-} (317u^5-1115 u^4 Y_{+} -90 Y_{-}^4 Y_{+}+u Y_{-}^2( 569 Y_{+}^2-656 v)\\[.1cm]
\rule{2.55cm}{0pt}+2 u^3 (879Y_{+}^2-917 v)-u^2(1439 +451 v + 451 v^2 +1439 v^3))\\[.2cm]
\widetilde{\hpiccolo}_{0}=\tfrac{1}{45} u(47u^3-79u^2Y_{+}-15Y_{-}^2Y_{+}+u(77-4v+77v^2) )
\end{array}
\ee
for the weight two anti-symmetric, weight one and weight zero parts. The full tree-like function is given by combining
$\widetilde{\hpiccolo}_{i=\Box,u,v,0}$, and the following function
\bea\label{extra2222}
{\cal A}= 
\beta_{2222}\, u(2-u\partial_u)^2 \overline{D}_{1111}+ \mathcal{K}_{\Delta^{(8)} } 
\eea
The first term proportional to $\beta_{2222}$ will span the ambiguity of $\mathcal{H}^{(2)}_{2222}$. 
The function $\mathcal{K}_{\Delta^{(8)} }$ is the kernel of $\Delta^{(8)}$. 
A restricted choice of $\mathcal{K}_{\Delta^{(8)} }$ which is compatible with the symmetry $u\rightarrow u/v, v\rightarrow 1/v$, and the way we construct $\mathcal{L}^{(2)}_{2222}$, is
\bea\label{kernel_comp_2222}
\mathcal{K}_{2222}= k_1\, u \overline{D}_{1111}+ k_2\, (1+u+v) \overline{D}_{1111} + k_3 \zeta_3 + k_4 \log^2 v + k_5 (2\log u -\log v) + k_6
\eea

We will now present the results for $\mathcal{L}_{2233}$ and $\mathcal{L}_{2244}$, and to do so we will make 
use of the operators $\widehat{\mathcal{D}}_{pqrs}$. 
%
%
%
The result for $\mathcal{L}^{(2)}_{2233}$ is
\bea\label{rew_1loop_2233}
\mathcal{L}^{(2)}_{2233}-\widehat{\mathcal{D}}_{2233}\mathcal{L}^{(2)}_{2222}=
\Big[ \tfrac{1}{2} \Rem_3\, + \Rem_5\,(x-1)\partial_{x}\Big] \phi^{(2)}(1-x,1-{\bar x}) -(x\leftrightarrow {\bar x})+\notag\\
+ (x-{\bar x}) \Rem_7\,\partial_u\phi^{(2)}(1-x,1-{\bar x}) -32 \zeta_3 + \Rem_9\, \log(u)\log(v) + \Rem_{10}\, \log^2 v\,,
\eea
where 
\be
\begin{array}{ll}
\Rem_{3}=-\tfrac{2}{(x-{\bar x})^3}(u^3+u(7 Y_{+}^{2} +2 v)-5u^2Y_{+} -3Y_{+}^{3} )\\[.2cm]
\Rem_{5}=+\tfrac{1}{(x-{\bar x})^3}Y_{-}(7u^2-15uY_+ +8(Y^2_{+} - v))\\[.2cm]
\Rem_{7}=+\tfrac{1}{ (x-{\bar x})^2 }(5u^2-13u Y_{+} +8(Y_{+}^2 - v))\\[.2cm]
\Rem_{9}=+\tfrac{1}{(x-{\bar x})^2} Y_{-}(6Y_{-}-5u)\\[.2cm]
\Rem_{10}=+\tfrac{1}{2(x-{\bar x})^2}( 5u(5+3v)+60v-13u^2-12)
\end{array}
\ee
The $\log^2 u$ discontinuity of $\mathcal{L}_{2233}^{(2)}$ is captured by $\widehat{\mathcal{D}}_{2233}\mathcal{L}^{(2)}_{2222}$, as it should. Indeed 
the r.h.s of \eqref{rew_1loop_2233} has no overlap with the $\log^2 u$ projection. 
Surprisingly, $\widehat{\mathcal{D}}_{2233}\mathcal{L}^{(2)}_{2222}$ also captures relevant parts of the full $\mathcal{L}_{2233}^{(2)}$, but 
for the r.h.s of \eqref{rew_1loop_2233}.
The latter has no spurious poles by construction, and 
it can be shifted by the kernel of $\Delta^{(8)}$ without changing $\mathcal{H}_{2233}^{(2)}$, 
or \eqref{delta8_2233}. Notice that in $\mathcal{L}^{(2)}_{2222}$ we have included $\mathcal{K}_{2222}$ as given in \eqref{kernel_comp_2222}, 
and $\widehat{\mathcal{D}}_{2233}\mathcal{K}_{2222}$ now produces non-kernel functions proportional to $k_1$ and $k_2$. 
In particular, the r.h.s of \eqref{rew_1loop_2233} holds for specific values of $k_{i=1,\ldots 6}$.

The result for $\mathcal{L}_{2244}^{(2)}$ has the same level of complexity, 
once we use the property that $\widehat{\mathcal{D}}_{2244}$ and $\widehat{\mathcal{D}}_{2233}$ 
concatenate, i.e $\widehat{\mathcal{D}}_{2244}=\tfrac{2}{3}(5-u\partial_u)\widehat{\mathcal{D}}_{2233}$, 
and therefore we use the whole $\mathcal{L}^{(2)}_{ 2233}$ as starting point, rather than $\mathcal{L}^{(2)}_{2222}$. Then,
\bea\label{rew_1loop_2244}
\mathcal{L}^{(2)}_{2244}-\tfrac{2}{3}(5-u\partial_u)\mathcal{L}^{(2)}_{2233}=
\Big[ \tfrac{1}{2} \Rem_3\, + \Rem_5\,(x-1)\partial_{x}\Big] \phi^{(2)}(1-x,1-{\bar x}) -(x\leftrightarrow {\bar x})+\notag\\
 +(x-{\bar x}) \Rem_7\,\partial_u\phi^{(2)}(1-x,1-{\bar x}) - 88 \zeta_3 + \Rem_9\, \log(u)\log(v) + \Rem_{10}\, \log^2 v +\notag\\
+\, \Rem_{\Box}\, \phi^{(1)} (x,{\bar x})+ \Rem_{u} \log u+\Rem_{v} \log v\,,\rule{.25cm}{0pt}
\eea
where 
\be
\begin{array}{ll}
\Rem_{3}=-\tfrac{4}{3(x-{\bar x})^3}(u^3+17u(Y_{+}^{2} +2 v)-9u^2Y_{+} -9Y_{+}(Y_{+}^2+6v) )\\[.2cm]
\Rem_{5}=+\tfrac{2}{(x-{\bar x})^3}Y_{-}(6u^2-17uY_+ +(11Y^2_{+} + 16v))\\[.2cm]
\Rem_{7}=+\tfrac{2}{3(x-{\bar x})^2 }(7u^2-36 u Y_{+} +(33Y_{+}^2 + 48 v)\\[.2cm]
\Rem_{9}=+\tfrac{10}{(x-{\bar x})^2} Y_{-}(3Y_{-}-2u)\\[.2cm]
\Rem_{10}=+\tfrac{5}{(x-{\bar x})^2}( u(13+9v)+30v-7u^2-6)\\[.2cm]
\Rem_{\Box}=+\tfrac{72}{(x-{\bar x})^3}uv\\[.2cm]
\Rem_{u}=+\tfrac{36}{(x-{\bar x})^2}(u(Y_+-u))\\[.2cm]
\Rem_{v}=+\tfrac{18}{(x-{\bar x})^2}(Y_+-u)(Y_--u))
\end{array}
\ee
The use of $\mathcal{L}^{(2)}_{2233}$ as starting point, instead of $\mathcal{L}^{(2)}_{2222}$, 
has the effect of keeping  $(x-{\bar x})^3$ the maximal power of the denominator 
in the functions $\Rem$ given above. 
Again, the r.h.s of \eqref{rew_1loop_2244} holds for specific values of $k_{i=1,\ldots 6}$.


The construction of $\mathcal{L}^{(2)}_{22pp}$ depends on the initial orientation of the external charges we choose, 
since the latter goes along with $\Delta^{(8)}$ through the values of $p_{21}$ and $p_{34}$. 
In order to obtain $\mathcal{L}^{(2)}_{2p2p}$ we simply repeat the previous construction with minor modifications.

Consider the case of $\mathcal{L}^{(2)}_{2323}$ for illustration. 
We define $\mathcal{L}_{2323}$ by the equation which extracts $\Delta^{(8)}$ out of $\mathcal{H}_{2323}$,
\begin{align}
u^2\mathcal{H}^{(2)}_{2323}=&\ (u\sigma)^{-1/2}\Delta^{(8)} (u\sigma)^{+1/2}\mathcal{L}^{(2)}_{2323} -\frac{u^2(1+v)(3+u-v)}{2v^2}\notag \\
& \qquad+4 u^4( \overline{D}_{2431}+\overline{D}_{3421} )  + 24 u^4  \overline{D}_{2422} -30 u^4 \overline{D}_{3421} 
%
\label{delta8_2323}
\end{align}
Then we write $\mathcal{L}^{(2)}_{2323}$ as follows
\bea\label{rew_1loop_2323}
u\mathcal{L}^{(2)}_{2323}-u\widehat{\mathcal{D}}_{2323}\mathcal{L}^{(2)}_{2222}=
\Big[ \tfrac{1}{2} \Rem_3\, + \Rem_5\,(x-1)\partial_{x}\Big] \phi^{(2)}(1-x,1-{\bar x}) -(x\leftrightarrow {\bar x})+\notag\\
 +(x-{\bar x}) \Rem_7\,\partial_u\phi^{(2)}(1-x,1-{\bar x})  + \Rem_9\, \log(u)\log(v) + \Rem_{10}\, \log^2 v +\notag\\
+\, \Rem_{\Box}\, \phi^{(1)} (x,{\bar x})\,,\rule{.25cm}{0pt}
\eea
Notice a feature of $\mathcal{L}^{(2)}_{2323}$, which did not show up in $\mathcal{L}_{22pp}$:
This is the presence of the relative $1/u$ factor between the r.h.s and the l.h.s of \eqref{rew_1loop_2323}. Indeed the l.h.s is
\be
\begin{array}{ll}
\Rem_{3}=+\tfrac{1}{2(x-{\bar x})^3} (u^3-u^2(5v+8)+u(13+28v-5v^2) +3(3v^3+8v^2-9v-2) )\\[.2cm]
\Rem_{5}=+\tfrac{1}{(x-{\bar x})}(1-v) \Rem_{7} +\tfrac{1}{(x-{\bar x})^3}uY_{-}(17-18v+v^2+u^2-2u(v+9))\\[.2cm]
\Rem_{7}=-\tfrac{1}{8(x-{\bar x})^2 }(41+15v-57v^2+v^3+u^2(v+17)-2u(29+6v+v^2) )\\[.2cm]
\Rem_{9}=-\tfrac{1}{8(x-{\bar x})^2} (2u^3-u^2(51+5v)+12u(5+11v)-3Y_{-}^3) \\[.2cm]
\Rem_{10}=-\tfrac{1}{16(x-{\bar x})^2}(Y_{-}(209+17v(v-2))+2u^3+u^2(197-13v)-u(408+388v-28v^2))\\[.2cm]
\Rem_{\Box}=+\tfrac{1}{8(x-{\bar x})}(3u^2-3Y_{-}^2- u(31+5v))
\end{array}
\ee
and is not possible to use the freedom in the construction of $\mathcal{L}_{\vec{p}}$ to reabsorb the extra $1/u$.

Determining $\mathcal{L}^{(2)}_{pqrs}$ for other N$^2$E correlators can be done in a similar way. 
For such correlators, we can also reverse the procedure and bootstrap directly $\mathcal{H}_{pqrs}$, 
by making a simpler ansatz for $\mathcal{L}^{(2)}_{pqrs}$, apply $\Delta^{(8)}$, and complete it with a tree like function, 
Indeed, for a large value of $p_1+p_2+p_3+p_4$, 
i.e. a large denominator power $(x-\bar{x})^\#$, therefore a large number of initial free coefficients in the polynomial ansatz for $h_{i=1,\ldots }$ in \eqref{G2uvresult}, 
the use of $\Delta^{(8)}$ in combination with $\widehat{\mathcal{D}}_{pqrs}$ reduces considerably the complexity of the computation.

As far as we investigated, pulling $\Delta^{(8)}$ out of $\mathcal{H}_{pqrs}$ is possible even for multi-channel correlators.
In a first instance, this problem reduces to a collection of single channel computations, because  
$\Delta^{(8)}$ acts diagonally on the $su(4)$ harmonics $\Upsilon_{[aba]}$.  
However,  some extra care is needed in defining the ansatz for $\mathcal{L}^{(2)}_{pqrs}$, since the latter does not manifestly obey the rules of $\mathcal{H}_{pqrs}$. 
For example, already in the case of $\mathcal{L}^{(2)}_{2323}$ we have 
found the need of $1/u$ terms in the coefficient functions. This behavior is generic in multi-channel correlators.

Despite discrepancies, it would be fascinating to take full advantage of $\Delta^{(8)}$ and $\widehat{\mathcal{D}}_{pqrs}$ at one-loop. 
Perhaps, understanding the fate of the hidden 10d symmetry at one-loop for generic correlators, 
would provide a major insight on our construction of the preamplitudes $\mathcal{L}^{(2)}_{pqrs}$, thus on our bootstrap program. 
We leave this for a future work.

Before concluding, we mention a special property of $\langle 2222 \rangle$ at one-loop, or ``How to bootstrap $\mathcal{H}^{(2)}_{2222}$ without really trying !"
In fact, only in this case it is possible to carry out the following procedure on the ansatz \eqref{basis11} for $\mathcal{L}^{(2)}_{2222}$: 
impose $x\rightarrow x/(x-1)$ crossing symmetry, impose absence of $x=\bar{x}$ poles, 
apply $\Delta^{(8)}$, and impose the remaining crossing invariance under $x\rightarrow 1/x$. 
Without any reference to the $\log^2 u$ discontinuity, the above procedure returns a function with only three independent coefficients. 
If we furthermore impose analiticity in spin of the $\log^2 u$ discontinuity, we end up with two independent coefficients. 
One is multiplying a weight 4 anti-symmetric function, and the other one is multipling a tree-like function. By looking at the minimality of the weight 4 anti-symmetric function, we immediately 
conclude that the result is a linear combination of $\mathcal{H}^{(2)}_{2222}-4 u^2  \overline{D}_{2 4 2 2} $ and $u^2\overline{D}_{4444}$.


\section{Conclusions}


In this paper we have given a general algorithm for computing all one-loop quantum gravity four-point amplitudes in IIB supergravity on AdS${}_5\times$S${}^5$. 
It works for arbitrary external states, i.e. arbitrary KK modes on the five-sphere, and has been tested explicitly 
for $\langle 2244\rangle $, $\langle3333\rangle$,$\langle4444\rangle$,$\langle3335\rangle$ and $\langle4424\rangle$. 
These results are available in the ancillary files.

The amplitudes we studied 
are dual to four-point correlators of single-particle half-BPS operators, which we have properly identified,  in $\cN=4$ SYM with gauge group $SU(N)$,
in the regime of strong 't Hooft coupling, at order $1/N^4$ in  the large $N$ limit.
Our bootstrap program has its foundations in 
the detailed understanding of the spectrum of two-particle operators,
and OPE arguments on the CFT side. 
We first determine well-defined pieces of the one-loop correlator 
by extracting all relevant data from many four-point tree-level correlators. Then we
rearrange this data to determine the combination which appear at one-loop, and sum up the result. 
These pieces of the result  are fed into an ansatz for the full function, 
which yields the final result upon demanding no unphysical (euclidean) singularities, i.e. no poles when $x \rightarrow \bar x$. 

Our algorithm generalises in a non-trivial way 
the one developed in our previous works~\cite{Aprile:2017bgs,Aprile:2017qoy}, 
where the explicit results for $\langle 2222 \rangle $ and $\langle 2233 \rangle$ were obtained.
There are indeed a substantial number of new features which emerge. 
In fact, the one-loop amplitude for a correlator of KK modes with arbitrary weights, 
cannot be fixed just by the knowledge of the $\log^2 u$ discontinuity,
as was the case for $\langle 2222 \rangle $ and $\langle 2233 \rangle$.
%
%
Fortunately there are  more pieces of the one-loop amplitude which are determined 
via tree-level data. 
These come from window and below window OPE analysis, as illustrated in Figure \ref{fig1}, and explained in Section \ref{sec:OPEbeyond}.
They contribute to the single-log and no-log pieces of the one-loop amplitude, respectively. 
With the addition of this information the one-loop correlators are then completely determined up to well-understood ambiguities with finite spin support.

Two novel and note-worthy features of generic one-loop amplitudes,
which originate from the analysis of CFT data in and below the window, are: 1) 
the natural splitting of the one-loop dynamical function into two independent pieces, $\mathcal{D}^{(2)}=\mathcal{T}^{(2)}+\mathcal{H}^{(2)}$,
and 2) the need of a proper understanding of semishort operators contributions at order $1/N^4$, and multiplet recombination.

The splitting of $\mathcal{D}^{(2)}=\mathcal{T}^{(2)}+\mathcal{H}^{(2)}$ introduces what we call the ``generalised tree-level amplitude" $\mathcal{T}$. 
It is uniquely defined in terms of free theory%
\footnote{Indeed it can be defined in terms of any generalised free theory with arbitrary coefficients in front of the propagator structure}. 
It is  tree-like, since it is built out of polylogarithms with maximal weight 2, and does not contribute to the $\log^2 u$ discontinuity of $\mathcal{H}^{(2)}$.
It has no $\log u$ contributions below threshold, and crucially it completely cancels the contribution of recombined free theory below window. 
Within the ansatz \eqref{ansatzTNewR} this generalised tree-level amplitude has a unique solution. 

All the novel and interesting OPE dynamics is neatly combined into $\mathcal{H}^{(2)}$, which we call the ``minimal one-loop function". 
This function has to match the double logarithmic discontinuity and below threshold predictions (excluding those arising from free theory), as explained in Section \ref{sec:one_loop_corr}. 
In a very non-trivial way this is always possible within the minimal ansatz \eqref{G2uvresult}.
A much wider ansatz would be needed if $\mathcal{T}$ was not included in the first place.

The semi-short sector on its own has a completely independent story, which we determine at order $1/N^4$ purely from free theory correlators. 
This is done in Section \ref{free}, where we also clarify some important issues about the way the protected sector actually contributes in the $1/N$ expansion. 
Then, through multiplet recombination, we obtain predictions for $\mathcal{H}^{(2)}$ below window at the unitarity bound. 
What is truly remarkable is the fact that this independent input is consistent with our construction of $\mathcal{H}^{(2)}$, which from the very beginning descends from the double logarithmic discontinuity. 
In this sense, it would be interesting to have a fuller understanding of the meaning of the generalised tree-level amplitude and the split $\mathcal{D}^{(2)}=\mathcal{T}^{(2)}+\mathcal{H}^{(2)}$.

The remaining ambiguities have a clear description in Mellin space, and simply span 
the space of arbitrary linear functions in the Mellin variables $s,t$, consistent with crossing symmetry, as described around ~\eqref{eq:ambiguity_general}.
These ambiguities are manifest in our algorithm, because we have chosen the minimal ansatz. Widening our ansatz would allow in principle more freedom.
However, we expect true stringy ambiguities to be in one-to-one correspondence 
with polynomial Mellin amplitudes of some degree~\cite{0907.0151,Alday:2015eya}. 
Indeed, a polynomial Mellin amplitude of order $r$ has at most a spin $r$ contribution in its SCPW expansion, 
and counts as a contribution in a bulk 10d effective action which involves $\partial^{2r} R^4$.
The case $r=0$, i.e. the $R^4$ term,  appears at one-loop order with  $(\alpha')^{-1}$. 
Dimensional analysis then  implies that the $\partial^{2r} R^4$ term comes with coefficient $(\alpha')^{-1+r}$. 
 In summary
\begin{align}\label{form_conc}
	{\cal M}(s,t)=O(s^{r-a}t^a) \quad \leftrightarrow \quad  \text{max spin: } r \quad \leftrightarrow \quad  \partial^{2r} R^4 \quad \leftrightarrow \quad (\alpha')^{-1+r} \ .
\end{align}
The minimal ansatz naturally gives ambiguities that contribute up to order $(\alpha')^0$ only.  
Widening it would include ambiguities corresponding to higher order $\alpha'$ corrections \eqref{form_conc}.%
\footnote{At one-loop there are also stringy corrections which can be obtained from tree-level string corrections. See~\cite{Alday:2018kkw} for initial work in this direction.}
Although within our bootstrap program we cannot fix the value of the ambiguities,
it may be possible that a combination of different techniques will. 
For example, by using localisation techniques \cite{Chester:2019pvm} obtained the value of the $\langle 2222 \rangle$ ambiguity.

Finally we come to the question of ten-dimensional conformal symmetry~\cite{Caron-Huot:2018kta}.  
This symmetry implies a beautiful structure for the leading logs at any order. At one-loop the double 
logarithmic discontinuity for any correlator can be written as $\Delta^{(8)}$ acting on a much simpler object. 
Furthermore this object can be simplified further by pulling out the differential operators $\widehat{\mathcal{D}}_{pqrs}$. %
In section~\ref{exploration10d} we have examined the extent to which this hidden structure  transfers to the full one-loop function itself. 
We have found that the one-loop amplitudes can be written as  $\Delta^{(8)}$ acting on a pre-amplitude $\mathcal{L}^{(2)}_{pqrs}$  up to a tree-like remainder.
Although the resulting pre-amplitude cannot be written in the way the double logarithmic discontinuity can, 
i.e. directly as $\widehat{\mathcal{D}}_{pqrs}\mathcal{L}^{(2)}_{2222}$, 
we have found in a number of examples that the difference 
$\mathcal{L}^{(2)}_{pqrs}-\widehat{\mathcal{D}}_{pqrs}\mathcal{L}^{(2)}_{2222}$ can be put in a very simple form, 
i.e. \eqref{rew_1loop_2233}, \eqref{rew_1loop_2244} and \eqref{rew_1loop_2323}.
This pattern may hint at more structure in the result than is currently apparent, and we hope to investigate 
this possibility in the future. In a similar vein there are hints that the ten-dimensional symmetry also controls the leading 
order corrections in $\lambda^{-\frac{1}{2}}$ at tree level \cite{Drummond:2019odu}. 
Perhaps it can be used to study these four-point functions much more generally.

Future directions include a more detailed investigation of the Mellin space representation of our one-loop functions, 
which would extend the analysis of $\langle 2222\rangle$ in \cite{Alday:2018kkw}, 
as well as the possibility of pushing our bootstrap program to two loops. It would be fascinating if the results we obtain in the large $N$ expansion could be compared (possibly taking into account also the $\alpha'$ corrections) to the results based on integrable methods \cite{Basso:2017khq,Coronado:2018cxj,Bargheer:2019kxb,Basso:2019diw,Bargheer:2019exp}.
We also emphasise that our fresh new look at free $\cN=4$ SYM, especially our 
understanding of single particle operators and generalised tree-level functions,
suggests a different way to approach the mysterious six-dimensional $(2,0)$ theory, which 
has been recently studied from a holographic perspective in several papers~\cite{1507.05637,1712.08570,1805.00892,1902.00463}. 
%

\section*{Acknowledgements}
We thank Simon Caron-Huot for discussions about related topics.
FA is partially supported by the ERC-STG grant 637844- HBQFTNCER. JMD and HP are supported by ERC grant 648630-IQFT. PH is supported by STFC under consolidated grant ST/P000371/1 and the European Union's Horizon 2020 research and innovation programme under the Marie Sk\l odowska-Curie grant agreement No. 764850 SAGEX.
PH and FA acknowledge hospitality at the `Centro de Ciencias de Benasque' 
during the workshop `Gauge Theories, supergravity and superstrings 2019'.
FA thanks the organisers of the Parallel Session: String \& Mathematical Physics 
of the `Quantum Field Theory Meets Gravity' workshop at DESY, for warm hospitality. 
FA would like to thank Francesco Sanfilippo 
for invaluable conversations about coding and algorithms.


\appendix


\section{Superblocks}\label{superblocks}


Here we give the explicit definition of  the superblocks $\mathbb{S}_{\gamma,\underline{\lambda}}^{p_1p_2p_3p_4}$ following~\cite{Doobary:2015gia}.
They  are defined by a determinantal formula. 
Let us introduce first the function
\be\
F^{\alpha\beta\gamma\ula}\ =\  (-1)^{ p + 1} \frac{(x-y)(x-{\bar y})({\bar x}-y)({\bar x}-{\bar y})  }{ (x-{\bar x})	(y-{\bar y})}
\det \left( \begin{array}{cc}
	F^X_{\underline\lambda}&R\\[.1cm]
	K_{\underline \lambda}   &   F^Y
\end{array} \right),
\ee
where the determinant is taken on the $(p+2)\times(p+2)$ matrix  { (where $p=\min(\alpha,\beta)$) }
\begin{align}
(F^X_{\underline \lambda})_{ij}&=\Big([x_i^{\lambda_j  - j}
{}_2F_1( \lambda_j + 1 - j+\alpha, 
\lambda_j + 1 - j+\beta;  2 \lambda_j + 2 - 2 j+\gamma; 
x_i)] \Big)_{  
	1\leq i
	\leq 2,\,
	1\leq j\leq p
}\, \notag \\
(F^Y)_{ij}&=\Big((y_j)^{i - 1}
{}_2F_1(i -\alpha, i -\beta; 
2 i -\gamma; y_j)  \Big)_{
	1\leq i\leq p,\,
	1\leq j\leq 2}\,
\notag\\
(K_\ula)_{ij}&=\Big( -\delta_{i;\,j{-}\lambda_j}\Big)_{
	1\leq i\leq p,\, 1\leq j  \leq p}
\notag\\
(R)_{ij}&=\left(\tfrac{1}{x_i-y_j}
\right)_{1 \leq i \leq 2\,, 1\leq j \leq 2}\,.
\end{align}
(The brackets in the definition of $F_{\underline{\lambda}}^{X}$ mean deletion of the singular terms in the Taylor expansion in $x_i$ around $x_i=0$ when $\lambda_j < j$ and we have defined here $x_i=(x,{\bar x})$ $y_j=(y,{\bar y})$ in the matrix.) Then 
\be
\mathbb{S}_{\gamma,\underline{\lambda}}^{p_1p_2p_3p_4} = \mathcal{P}\, \left(\frac{x {\bar x}}{y {\bar y}}\right)^{\tfrac{1}{2}(\gamma-p_4+p_3)} F^{\alpha \beta \gamma \underline{\lambda}}\,,
\qquad \left\{\begin{array}{l} \alpha=\tfrac12(\gamma-p_{1}+p_2), \\[.2cm]  \beta=\tfrac12( \gamma +p_{3}-p_{4}). \end{array}\right.
\ee
The prefactor $\mathcal{P}$ is that of  (\ref{Pfactor}).


\section{Trees and Amplitudes}\label{sec:3pt-func}


\subsection{$1/N^2$ connected free theory} \label{AppTree1}

We quote a formula for connected free theory at order $1/N^2$, 
of a generic four-point function $\langle \mathcal{O}_{p_1}\mathcal{O}_{p_2} \mathcal{O}_{p_3} \mathcal{O}_{p_4} \rangle$.
The same formula is described in a different notation in \cite{Caron-Huot:2018kta}.  

Each propagator structure in free theory is labelled by monomials of the form 
$\mathcal{P}^{} \ssigma^{i-j}\ttau^j$ where $i,j$ belong to $T=\{ (i,j)\ |\ 0\leq i \leq \kappa_{\vec{p}},\, 0\leq j\leq i\}$, and the bound $\kappa_{\vec{p}}$
is precisely the degree of extremality.
The lattice of points described by $T$ is schematically depicted here below. 
We shall distinguish the three edges from the interior.

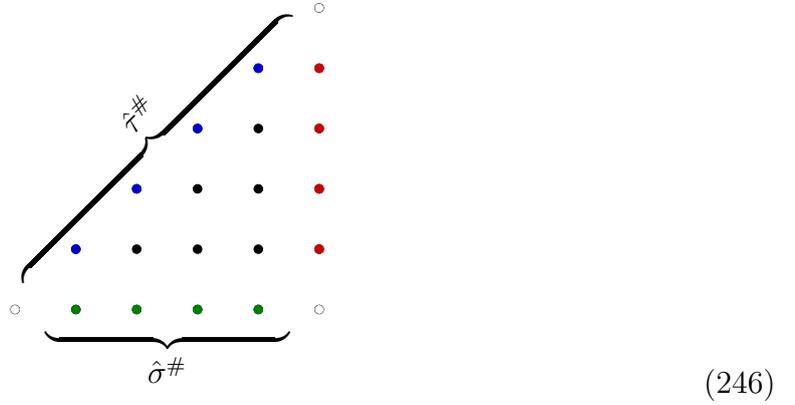
\beq
\begin{tikzpicture}[scale=.8]

\def\ini{-1}
\def\last{4}
\def\uno{0}
\def\tre{3}

\foreach \x in {\ini,...,\last}
	 \foreach \y in {\ini,...,\x}
 		   \filldraw[black] (\x,\y) circle (2pt);

	 \filldraw[white] (\ini,\ini) circle (1.8pt);
	  \filldraw[white] (\last,\last) circle (1.8pt);
	   \filldraw[white] (\last,\ini) circle (1.8pt);
	    
	   \node[rotate=45] at (\ini+2,\ini+3.25) {$\ttau^\#$};
	   \node[rotate=45] at (\ini+2.25,\ini+2.75) {$\overbrace{\rule{5cm}{0pt}}$};
	   
	    \node[] at (\ini+2.5,\ini-1) {$\ssigma^\#$};
	   \node[] at (\ini+2.5,\ini-.5) {$\underbrace{\rule{3.2cm}{0pt}}$};
	   
	 \foreach \x  in {\uno,...,\tre}
	 	\filldraw[green!50!black] (\x,\ini)   circle (2pt);  
		
	 \foreach \x  in {\uno,...,\tre}
	 	\filldraw[blue!80!black] (\x,\x)   circle (2pt);  		
				
	 \foreach \x  in {\uno,...,\tre}
	 	\filldraw[red!80!black] (\last,\x)   circle (2pt); 
		
\end{tikzpicture}
\label{triangolo}
\eeq

Vertices at the intersection of the edges correspond to disconnected diagrams, 
when they exists according to our definition of single particle states.
In \cite{Aprile:2018efk} we determined the value of the following 
connected propagator structure
\beq
 \frac{ \langle \mathcal{O}_{p_1} \mathcal{O}_{p_2} \mathcal{O}_{p_3} \mathcal{O}_{p_4}\rangle  }{ \mathcal{P}^{} }  \supset   \frac{ \sqrt{p_1p_2p_3p_4}}{N^2} \left[ 1+\tfrac{p_{43}+p_{21}}{2} \right] \frac{u\ttau}{v} 
\eeq
Looking at the diagram of $T$, we have determined the 
coefficient associated to the point $(1,1)$ on the diagonal edge of the triangle. 
From crossing on the other edges we find 
\bea
&&
\frac{ \langle \mathcal{O}_{p_1} \mathcal{O}_{p_2} \mathcal{O}_{p_3} \mathcal{O}_{p_4}\rangle  }{ \mathcal{P}^{} } \supset  
							 \frac{ \sqrt{p_1p_2p_3p_4} }{N^2}\Big[    \big(1+\tfrac{p_{43}+p_{21} }{2} \big) \sum_{k=1}^{\mathfrak{t}-1} \left( \frac{u\ttau}{v} \right)^k+ 
											  \big(1+\tfrac{p_{13}+p_{42}}{2} \big) \sum_{k=1}^{\mathfrak{t}-1} (u\ssigma)^k + \notag\\
&& \rule{5cm}{0pt}
\big(1+|\tfrac{p_{23}+p_{14}}{2}| \big) \sum_{k=1}^{\mathfrak{t}-1} (u\ssigma)^{k} \left( \frac{u\ttau}{v}\right)^{\mathfrak{t}+1-k} \Big]
\eea
By including propagator structure in the interior of $T$ we finally obtain the general formula 
\bea
&&
\frac{ \langle \mathcal{O}_{p_1} \mathcal{O}_{p_2} \mathcal{O}_{p_3} \mathcal{O}_{p_4}\rangle  }{ \mathcal{P}^{} }
					= \frac{ \sqrt{p_1p_2p_3p_4} }{N^2} \Big[    \big(1+\tfrac{p_{43}+p_{21} }{2} \big) \sum_{k=1}^{\mathfrak{t}-1} \left( \frac{u\ttau}{v} \right)^k+ 
											  \big(1+\tfrac{p_{43}-p_{21}}{2} \big) \sum_{k=1}^{\mathfrak{t}-1} (u\ssigma)^k + \notag\\
&&\rule{1.8cm}{0pt}  
											  \big(1+|\tfrac{p_1+p_2-p_3-p_4}{2}| \big)
											  \sum_{k=1}^{\mathfrak{t}-1} (u\ssigma)^{k} \left( \frac{u\ttau}{v}\right)^{\mathfrak{t}+1-k} 
											  +2 \sum_{T\backslash{\rm edges} } (u\ssigma)^{\#_1} \left(\frac{u\ttau}{v} \right)^{\#_2}\Big] \qquad
\eea

\subsection{Generalised Tree Level Amplitudes}\label{AppTree2}

The Mellin transform of generalised tree level functions 
has the same form of the Rastelli and Zhou integral. By using the conventions of \cite{Aprile:2018efk} we obtain,
\beq
\mathcal{T}_{}=
					\frac{ u^{ \frac{p_3-p_4}{2} } }{ v^{\frac{p_2+p_3}{2} } }\, 
					\oint { u}^{\frac{s}{2}} { v}^{\frac{t}{2} }\, {\Gamma}_{\vec{p}}\ 
						\mathcal{M}[\mathcal{T}](s,t),						\qquad p_{43}\ge p_{21}\ge 0,
\eeq

In order to make manifest how the pole structures of $\Gamma_{\vec{p}}$ 
relates to the OPE expansion in twists, and the log stratification of the tree level dynamical function,  
it is convenient to make manifest the location of the double poles in $s$ and $t$ and extract  
\bea\label{dd_poles}
\Gamma\big[ {\rm max}(\tfrac{p_1+p_2}{2},\tfrac{p_3+p_4}{2}) -\tfrac{s}{2} \big]^2 
								\Gamma\big[{\rm max}(\tfrac{p_1+p_4}{2},\tfrac{p_2+p_3}{2})-\tfrac{t}{2} \big]^2 
\eea
In our conventions $p_1+p_4\ge p_2+p_3$, therefore the max of the second $\Gamma$ in \eqref{dd_poles} is fixed. By
changing variables to
\bea
s\leftrightarrow \tfrac{s}{2}-{\rm max}(\tfrac{p_1+p_2}{2},\tfrac{p_3+p_4}{2}),\qquad 
t\leftrightarrow \tfrac{t}{2} - \tfrac{p_1+p_4}{2},
\eea
and introducing 
\be
P\equiv
\Big| \tfrac{p_1+p_2}{2}-\tfrac{p_3+p_4}{2}\Big|
\qquad Q\equiv \tfrac{p_4-p_3+p_1-p_2}{2},
\ee
we rearrange $\mathcal{T}^{}_{}$ into the form 
\bea
\mathcal{T}^{}_{}=
{
{u}^{{\rm max}(\tfrac{p_1+p_2-p_{43}}{2},p_3)} { v}^{\frac{p_{43}-p_{21} }{2} } }\oint { u}^{s} { v}^{t }\ \Gamma[-s]^2\Gamma[-t]^2\, 
\Big[ \tfrac{(-)^P}{(s+1)_P} \tfrac{ (-)^Q}{(t+1)_{Q} } \mathcal{M}[\mathcal{T} ](s,t)\Big] \label{T_mellin_form}
\\[.2cm]
\rule{1.5cm}{0pt}
\times 
\Gamma\big[2+s+t+{\rm max}(p_1,p_3+Q ) \big] \Gamma\big[2+s+t+{\rm max}(p_2+Q,p_4 ) \big]  \notag
\eea

The object highlighted in brackets $[\ldots]$ in the first line, which includes the amplitude and two Pochhammer symbols in the denominator, has only simple poles.
The sequence of double and simple poles splits in three sectors as follows 
\be\label{table_poles}
\begin{array}{c|c|c|c}
{\rm pole\ in}\ s & {\rm power\ of}\ u^{\#} & {\rm pole\ in}\ t & {\rm power\ of}\ v^{\#} \\
\hline
\rule{0pt}{.8cm}
\ \, \vdots &\ \vdots &\ \, \vdots &\ \vdots\\
\ 0 & p_3+{\rm max}(0,\tfrac{p_1+p_2-p_{3}-p_{4}}{2}) &\ 0 & \frac{p_{43}-p_{21}}{2} \\
\hline
-1 &\ \vdots &-1 &\ \vdots\\
\ \, \vdots &\ \vdots &\ \, \vdots &\ \vdots\\
-P & p_3+{\rm min}(0,\tfrac{p_1+p_2-p_{3}-p_{4}}{2}) & -Q &\ 0 \\
\hline
\rule{0pt}{.8cm}
\ \, \vdots &\ \vdots &\ \, \vdots &\ \vdots\\
-P^* & 1 & -Q^*& -p_3+1-{\rm min}(0,\tfrac{p_1+p_2-p_3-p_4}{2})
\end{array}
\ee
Notice the symmetric relation
\be
P^*-P=Q^*-Q=p_3-1+{\rm min}(0,\tfrac{p_1+p_2-p_{3}-p_{4}}{2})\ .
\ee

The first two sectors of the Table above contain information only from $\Gamma_{\vec{p}}$. We have
double poles of the form $\Gamma[-s]^2$ and $\Gamma[-t]^2$, which originates from $\Gamma_{\vec{p}}$ 
in the region where simple poles of the individual Gamma functions in \eqref{MACK} overlap. 
Then, we have remaining simple poles of the form ${(s+1)_P}{(t+1)_{Q} }$, in our notation. The
third sector instead arises only from $\mathcal{M}[\mathcal{T} ]$. 
A general subtlety in defining the (rational) mellin amplitude is due to the choice of contour of integration. 
This contour should separate poles in $s$ and $t$ from those in $s+t$, 
in order to have a well defined residue integration. This is achieved by 
rewriting $\mathcal{M}[\mathcal{T}]$ in the form, 
\be\label{example_simple_poles}
\mathcal{M}[\mathcal{T}]=
\sum_{\rm poles }\ \frac{ {m}^{p,q,r}(s,t) }{ (s+p)(t+q)(s+t+r) }\qquad r>-p-q\ .
\ee 
and paying attention that when $\mathcal{M}[\mathcal{T} ]$ is restricted to a single complex variable, 
let's say $(-n,t)$ for example, the only poles in $t$ that count, are those of the form $(t+m)$ in \eqref{example_simple_poles}, and those of the form $(s+t+r)\big|_{s=-n}$ are discarded.

In the special case of equal charges, $p_{i}=p$, the Mellin integral simplifies to
\bea\label{simple_pppp_ampl}
\mathcal{T}^{}_{pppp}=
{u}^{p}\oint { u}^{s} { v}^{t }\ \Gamma[-s]^2\Gamma[-t]^2\Gamma\big[s+t+p+2\big]^2 \times \mathcal{M}[\mathcal{T}_{pppp}]
\eea
Our new Mellin amplitudes \eqref{FAnction_3333} and \eqref{FAnction_4444}  
are presented in this way.

\subsubsection{Mellin-Barnes Integration}

We now perform the residue integration of \eqref{T_mellin_form} in detail. 

The computation will be organised as follows:
Firstly, we pick double poles in $s$ and double poles in $t$. 
Then, we pick two sequences of poles: we pick double poles in $s$ and simple poles in $t$, 
and we pick double poles in $t$ and simple poles in $s$. 
Finally we pick only simples poles in both $s$ and $t$. 
We will make the symmetry $s\leftrightarrow t$ visible.

The main input is a well known formula for Gamma function
\beq
\Gamma[-z]^2\Big|_{z\rightarrow k}\rightarrow\ \ \frac{1}{(k!)^2}  \partial_z\left(\frac{1}{k-z} \right) +\frac{1}{(k!)^2}\frac{\psi_{k+1}}{(k-z)}
\eeq

Proceeding in the order described above: ~\\[.5cm]
%
%
%
{\it Double Poles and Double Poles.}
The result of contour integration is 
\be
\begin{array}{c}
\displaystyle
(-)^{P+Q} \sum_{n,m\ge 0}\frac{{\rm u}^n}{(n!)^2}\frac{{\rm v}^m}{(m!)^2} \frac{ \Gamma\big[2+n+m+A \big] \Gamma\big[2+n+m+B\big] }{(n+1)_P(m+1)_Q}  \mathcal{M}[n,m]  \times \bigg(  \\[1cm]
\displaystyle
\left[ \log{\rm u} + \Psi_n [m]+ (\psi_{n+1}-\psi_{n+1+P} )+\frac{\partial_s \mathcal{M} }{ \mathcal{M}} \right] 
  \left[ \log{\rm v} + \Psi_m [n]+ (\psi_{m+1}-\psi_{m+1+Q} )+\frac{\partial_t \mathcal{M} }{\mathcal{M} }\right]   \\[1cm]
\displaystyle
 +\Big[ \Psi' + \frac{\partial_t\partial_s \mathcal{M} }{ \mathcal{M} }  - \frac{\partial_s  \mathcal{M} }{  \mathcal{M}  } \frac{\partial_t \mathcal{M} }{ \mathcal{M} } \Big] \bigg) 
\end{array}
\ee
For convenience we have defined $A={\rm max}(p_1,p_3+Q )$, $B={\rm max}(p_2+Q,p_4 )$ and the following two polygamma quantities,
\bea
\Psi_n[t]&=&\psi_{2+n+t+A}+\psi_{2+n+t+B}-2\psi_{n+1} \\
\Psi_m[s]&=&\psi_{2+s+m+A}+\psi_{2+s+m+B}-2\psi_{m+1} 
\eea
Notice that $\Psi'=\partial_t \Psi_n[t]=\partial_s\Psi_m[s]$ and the terms $\partial_s  \mathcal{M} \partial_t \mathcal{M}$ cancel out. 
~\\[.5cm]
%
%
%
{\it Simple Poles and Double Poles.}
The result of contour integration is 
\be
\begin{array}{c}
\displaystyle
(-)^{P+Q}\sum_{n\ge 0} \sum_{m=-Q^*}^{-1} \frac{{\rm u}^n}{(n!)^2}\frac{ {\rm v}^m  }{\Gamma[-m]^{-2} } \frac{  \Gamma\big[2+n+m+A \big] \Gamma\big[2+n+m+B\big] }{(n+1)_P} 
 \times\\[1cm]
%
\displaystyle
\rule{1cm}{0pt} 
\Big[ \log{\rm u} + \Psi_n [m]+ (\psi_{n+1}-\psi_{n+1+P} )+\frac{\partial_s \mathcal{M} }{  \mathcal{M}  } \Big]{\rm Res}\left[ \ \frac{ \mathcal{M} [n,t]}{\ (t+1)_Q}\ \right]_{t=m} 
\end{array}
\ee
and
\be
\begin{array}{c}
\displaystyle
(-)^{P+Q} \sum_{m\ge 0} \sum_{n=-P^*}^{-1} \frac{ {\rm u}^n}{ \Gamma[-n]^{-2} }\frac{ {\rm v}^m  }{(m!)^2} \frac{ \Gamma\big[2+n+m+A \big] \Gamma\big[2+n+m+B\big] }{(m+1)_Q} 
 \times\\[1cm]
\displaystyle
\rule{1cm}{0pt}
\Big[ \log{\rm v} + \Psi_m [n]+ (\psi_{m+1}-\psi_{m+1+Q} )+\frac{\partial_t \mathcal{M} }{ \mathcal{M} } \Big] {\rm Res}\left[\ \frac{\mathcal{M}[s,m]}{\ (s+1)_P}\ \right]_{s=n}
\end{array}
\ee
Notice that the leading $\log(u)$ discontinuity  is obtained
by picking, for each double poles in $s$, both double and simple poles in $t$. 
~\\[.5cm]
%
%
%
{\it Simple Poles and Simple Poles.}
The result of contour integration is 
\be
\begin{array}{c}
\displaystyle
(-)^{P+Q}\sum_{n=-P^*}^{-1} \sum_{m=-Q^*}^{-1} \frac{{\rm u}^n}{\Gamma[-n]^{-2}}\frac{ {\rm v}^m  }{\Gamma[-m]^{-2} } {  \Gamma\big[2+n+m+A \big] \Gamma\big[2+n+m+B\big] }
 \times\\[1cm]
%
\displaystyle
\rule{1.5cm}{0pt} 
{\rm Res}\left[ \ \frac{\mathcal{M}[n,t]}{\ (s+1)_P(t+1)_Q}\ \right]_{\substack{ s=n \\ t=m}} 
\end{array}
\ee

\subsubsection{Properties of the tree level Mellin Amplitude}

We record here an intriguing mathematical observation: 
the double poles location is encoded in the form of Mack's $\Gamma_{\vec{p}}$, 
and the residue on these poles depends on the Mellin amplitude $\mathcal{M}[\mathcal{T}_{\vec{p}}]$. 
Thus double poles are found to be in one-to-one correspondence with the polynomial ${\namep}_{\Box}$,
in the space representation of the function \eqref{ansatzTNewR}, 
because by residue integration, $\namep_{\Box}$ is the numerator of the term $\log(u)\log(v)\subset\phi^{(1)}$,  
\bea
&&
\frac{ {\namep}_{\Box}(x,\bar x,\ssigma,\ttau) }{ \ (x-{\bar x})^{ \mathbf{d}+1 } }=   
				{   u^{ {\rm max}(\tfrac{p_1+p_2-p_{43} }{2},\tfrac{p_3+p_4-p_{43}}{2}) }  }{ v^{\frac{p_{43}-p_{21}}{2} }  }\bigg[   
						\oint  u^s v^t \frac{\Gamma[-s]\Gamma[-t]}{\Gamma[s+1]\Gamma[t+1]} \times
						\rule{1.3cm}{0pt} \notag 
						\\[.2cm]
&&
\qquad\qquad\qquad\qquad
 \frac{ 
 \Gamma\big[2+s+t+A\big]
 \Gamma\big[2+s+t+B\big]}{
 (-)^{P+s+Q+t} (s+1)_P(t+1)_Q					}  
 \mathcal{M}[\mathcal{T}_{\vec{p}}][s,t] \bigg]\rule{.5cm}{0pt}
\label{masterPM}
\eea
with $P$, $Q$, $A$ and $B$ as in the previous section. 
%
%
Equation \eqref{masterPM} implies that the Mellin amplitude of $\mathcal{T}$ 
can be obtained from $\namep_{\Box}$ upon assuming $\Gamma_{\vec{p}}$, and viceversa. 
In particular, the conversion makes use of the formula (valid for any $\mathbf{d}$),
\bea
&&
\frac{{\rm u}^a\,{\rm v}^b }{(x-{\bar x})^{\mathbf{d}+1}}=\frac{\mathbf{d}!}{(2\mathbf{d})!} \oint  (-{\rm u})^s (-{\rm v})^t \ \frac{\Gamma[-s]\Gamma[-t]}{\Gamma[s+1]\Gamma[t+1]}\Gamma[s+t+X]^2\times  \rule{2cm}{0pt}  \notag \\[.2cm]
&& 
\quad  
\left[ (-)^{a+b} \frac{ (-s)_a (-t)_b}{ (s+1)_{\mathbf{d}-a} (t+1)_{\mathbf{d}-b}} (s+t+X)_{\mathbf{d}+1-a-b-X} (s+t+X)_{\mathbf{d}/2+1-a-b-X}\right]\qquad 
\label{Mellin-resum1}
\eea
applied to each monomial in $\namep_{\Box}$. The value of $X$ can be tuned 
afterwards by putting the final result in a canonical form. 
%

Summarizing, there is a bijection $\mathcal{M}[\mathcal{T}]\leftrightarrow \namep_{\Box}/(x-{\bar x})^{ \mathbf{d}_1-1 }$, which assumes
$\Gamma_{\vec{p}}$. This bijection also implies that the operators $\widehat{\mathcal{D}}_{pqrs}$ introduced in \cite{Caron-Huot:2018kta} and discussed around \eqref{SimonDpqrs}, can be simply obtained 
from $\namep_{\Box}$, thus reducing a computation about tree-level functions, 
to another one involving only rational functions.
It would be fascinating to know what is the uplifit of $\Gamma_{\vec{p}}$ at one-loop. 

\subsubsection{On tree level SCPW } \label{Apptree3prima} 

Having obtained the explicit tree-level four-point functions in position space 
as detailed in the previous section we then perform a SCPW decomposition 
to obtain the corresponding SCPW coefficients $M^{(1)}$.

As a function of twist and spin, ${M}^{(1)}_{\vec{p};\vec{\tau}}$ fits the ansatz 
\bea
{M}^{(1)}_{\vec{p};\vec{\tau}}&=& 
\displaystyle
\tfrac{ \left(\frac{2+2l+\tau+p_{43}}{2}\right)!\left(\frac{2+2l+\tau-p_{21}}{2}\right)! }{  (2+2l+\tau)!}
\tfrac{ \left(\frac{\tau+{\rm max}(p_1+p_2,p_3+p_4)}{2}\right)!\left(\frac{\tau+{\rm min}(p_1+p_2,p_3+p_4)}{2}\right)! }{  \tau!} \times\notag\\[.25cm]
& & \times \prod_{i=1}^{  {\rm max}(p_1+p_2,p_3+p_4)/2-\frac{p_{21}}{2} } \left( \tfrac{\tau-p_{21}}{2}-i+1\right) 
\prod_{j=1}^{  {\rm min}(p_1+p_2,p_3+p_4)/2-\frac{p_{43} }{2} } \left(\tfrac{\tau+p_{43}}{2} +j\right)^{-1} \notag\\[-.1cm]
& & \times\, \sum_{m=0}^{deg} \sum_{n=m}^{deg}\, \chi_{m,n}\, l^m \tau^{deg-n}
\label{fitcalM}
\eea
where $\chi_{m,n}$ are constants, depending only on $a,b$ and $\vec{p}$. The same ansatz works for $L^{(1)}_{\vec{p};\vec{\tau}}$

The degree of the polynomial is 
\be\label{degree_D1}
deg=-4+{\rm min}(p_1+p_2,p_3+p_4)-p_{43}+p_{21}=2(\kappa_{\vec{p}}-2)+p_{21},
\ee 
When $b$ is even, reciprocity symmetry implies that the polynomial in spin, in the last line \eqref{fitcalM}, is even in the variable $2l+\tau+ 3$. 
When $b$ is odd we find the following reduction.
For $l\in2\mathbb{Z}$, the polynomial in spin is of the form
\bea
&&
\sum_{m=0}^{deg} \sum_{n=m}^{deg}\, \chi_{m,n}\, l^m \tau^{deg-n} = ( l+\tau+2) \times \sum_{m=0}^{deg-1} \sum_{n=m}^{deg-1}\, \chi'_{m,n}\, l^m \tau^{deg-1-n}
\label{odd_b_M1L1}
\eea
Consequently, the polynomial for $l\in2\mathbb{Z}+1$ is obtained from the latter upon $\ell\rightarrow -\ell-\tau-3$.

Notice that the first factor in the second line of \eqref{fitcalM} vanishes for long superblock of twist
$\tau=p_{21},\ldots,p_{43},\ldots \tau_{\vec{p}}^{\rm max}-1$, where $p_{43}\ge p_{21}$. 

As function of spin, the analitic part of the tree level correlator fits into 
\bea
&&
\mathcal{B}^{(2+\tau,l)}\left[ u^{2+p_{43}} \mathcal{T}_{\vec{p},\, [aba]}\big|_{\frac{1}{N^2}} \right]
= \tfrac{\left(\frac{2+2l+\tau-p_{21}}{2}\right)!  \left(\frac{2+2l+\tau+p_{43}}{2}\right)! }{  (2+2l+\tau)!} \sum_{m=0}^{deg} \xi'_{m}(\tau,\vec{p},a,b)\, l^m\\
& &\qquad +\theta(\tau-\tau^{\rm max}) \tfrac{1}{2}\left[ \psi\left( \tfrac{4+2l+\tau-p_{21}}{2} \right) + \psi\left(\tfrac{4+2l+\tau+p_{43}}{2}\right) - 2\psi(3+2l+\tau) \right]{M}^{(1)}_{\vec{p},\tau,l,a,b} \notag
%
\eea
for some coefficients $\xi'_m$. The notation $\mathcal{B}^{(2+\tau,l)}\left[ f \right]$ means the CPW coefficient of $\mathcal{B}^{(2+\tau,l)}$ in the expansion of $f$.
If $\tau<\tau^{\rm max}$ we are in the window and we are computing $L^{(1)}_{\vec{p}}$. 
The derivative relation holds only for N${}^2$E correlators, and it is false otherwise.


\section{Some subleading three-point couplings} \label{Apptree3}


%
Let us begin from $\mathbf{C}^{(1)}_{(pp),4,l,[000]}$ and $\mathbf{C}^{(1)}_{(pp+1),5,l,[010]}$ with $p=3,4\ldots$, 
In both cases, we are studying \eqref{boot_V} in the simplified form, 
\bea
	C^{(1)}_{pp {\cK}_{4} } &=& 
							\left( C^{(0)}_{22 {\cK}_{4} }\right)^{-1} L^{(1)}_{22pp,\tau=4,l,[000]}\\
	C^{(1)}_{pp+1 {\cK}_{5} } &=&  
							\left( C^{(0)}_{23 {\cK}_{5} }\right)^{-1} L^{(1)}_{23pp+1,\tau=5,l,[010]} 
\eea
because only one long operator is exchanged. 
In $[010]$ we should also distinguish between even and odd spins. 
However, the knowledge of the even spin sector determines 
the odd sector through the reciprocity symmetry.  

Proceeding in the order, we can use results from \cite{unmixing} and \cite{Aprile:2017qoy}, to obtain
\bea
\label{2ppCPW}
\frac{ L^{(1)}_{22pp,\tau=4,l,[000]} }{2p} &=&
							\bigg[ \underbrace{(p-1)}_{\rm conn.\ free} +\frac{6p}{p-2}  \bigg] 
							\frac{((l+3)!)^2}{3(2l+6)!}= \frac{(p+1)(p+2)}{3(p-2)}\frac{((l+3)!)^2}{(2l+6)!}\\
\frac{ L^{(1)}_{23pp+1,\tau=5,l,[010]} }{\sqrt{6p(p+1)}}  &=&
							\bigg[ \underbrace{(p-1)}_{\rm conn.\ free} +\frac{4(2p+1)}{p-2}  \bigg] 
							(l+7)\frac{((l+4)!)^2}{5(2l+8)!}\notag\\[.2cm]
							&=&\quad 
								\frac{(p+2)(p+3)}{5(p-2)} (l+7)\frac{((l+4)!)^2}{(2l+8)!} \label{2ppplus1CPW}
\eea
These same expressions follow from the derivative relation, 
once analitically continued to the relevant twist. 
It is interesting to see how this happens: Consider first
\be\label{auxM122pp}
			{M}^{(1)}_{22pp,t=\tau/2,{\rm even}, [000]}=
								-\frac{2p}{(p-2)!}\left[\frac{2}{p!}(-)^p(t-p+1)_{2p} \frac{ (t!)^2 ((t+l+1)!)^2 }{ (2t)!(2t+2l+2)! }\right]
\ee
We should now take a $\partial_t$ of \eqref{auxM122pp}, 
and evaluate the result at $t=2$, which is above the regime of validity of the formula, 
i.e. $\tau^{\rm max}_{22pp}=2p$. Quite non trivially, the only term that will contribute comes from
$(t-p+1)_{2p}=(t-p+1)_{p-3}(t-2)(t-1)_{p+2}$. Indeed one effect of the derivative is 
to drop the factor $(t-2)$, which otherwise would make the result vanishing. Then, we find precisely \eqref{2ppCPW},
\be
			\partial_t  {M}^{(1)}_{22pp,t=\tau/2,{\rm even}, [000]} \Big|_{t=2}
									=\frac{2p(p+1)(p+2)}{3(p-2)}\frac{((l+3)!)^2}{(2l+6)!}
\ee 
Similarly, 
\bea
&&
			{M}^{(1)}_{23pp+1,t=\frac{\tau-1}{2},{\rm even},0,1}=\\[.2cm]
&&
\rule{.25cm}{0pt}
			-\frac{\sqrt{6p(p+1)}}{(p-2)!}\left[ \frac{2(-)^p}{(p+1)!} (t-p+1)_{2p+1} (l+2t+3) 
									  \frac{t!(t+1)!((l+t+2)!)^2}{(2t+1)!(2l+2t+4)!} \right] \rule{.5cm}{0pt} \notag
\eea
will have a contribution from $(t-p+1)_{2p+1}$ after taking the derivative 
at $t=2$, and the result will match non trivially \eqref{2ppplus1CPW}. 

The subleading three-point functions we looked for are
\bea\label{3pt_pplus5}
&&
				C^{(1)}_{pp {\cK}_4}= 
								\frac{p(p+1)_2 }{\sqrt{3}(p-2)} \frac{(l+3)!}{\sqrt{(2l+6)!} }\frac{1}{\sqrt{(l+1)(l+6)}}\left[ \frac{1+(-)^l}{2} \right]\\
&&
				C^{(1)}_{pp+1 {\cK}_5}=
								\frac{\sqrt{2 p(p+1)}(p+2)_2}{\sqrt{5}(p-2)}\frac{(l+3)!}{\sqrt{(2l+7)!}}
								\bigg[ \sqrt{\frac{(l+7)}{(l+1)} }\frac{1+(-)^l}{2} + \sqrt{\frac{(l+1)}{(l+7)} }  \frac{1-(-)^l}{2} \bigg] \notag\\
\eea

A first example with more than one operator is given by $\mathbf{C}^{(1)}_{(44),6,l,[000]}$. 
In the following, we refer more directly to the two operators labelled 
by $R_{6,l,[000]}$, with $\widetilde{\cK}_{6,1}$ and $\widetilde{\cK}_{6,2}$.
Then $\mathbf{C}^{(1)}_{(44),6,l,[000]}$ is the vector made by $C^{(1)}_{44 \widetilde{\cK}_{6,i=1,2} }$. 
The inputs we need from \cite{unmixing} are
\bea
		\mathbf{c}^{(0)}_{6,{\rm even},[000]}&=&
					\left[ \begin{array}{cc} +\sqrt{ \frac{l+2}{2l+9}} & +\sqrt{ \frac{l+7}{2l+9}} \ \\[.2cm]
						                           -\sqrt{ \frac{l+7}{2l+9}}  & +\sqrt{ \frac{l+2}{2l+9} }\  \end{array}\right]
						                         \\[.2cm]
		\mathbf{L}^{(0)}_{6,{\rm even},[000]}&=&
						{\frac{(l+1)(l+2)(l+7)(l+8) }{40}\left[ \begin{array}{cc} 
														\frac{4}{(l+2)(l+7)} & 0  \\ 0 & 1\end{array} \right]\frac{((l+4)!)^2}{(2l+8)!} }
\eea
Obtaining $L^{(1)}_{pp44,\tau=6,l,[000]}$ for $p=2,3,$ at twist $6$, 
we can finally solve \eqref{boot_V} in terms of the $3$-pt functions and get the result
\bea\label{C144twist6000}
C^{(1)}_{44 \widetilde{\cK}_{6,1} }= -6\sqrt{10} \,  \frac{ (2l+9) }{ \sqrt{ (l+2) } }\sqrt{ \frac{ (l+8)}{  (l+1)} }\frac{(l+4)!}{\sqrt{(2l+9)!}}\left[ \frac{(1+(-)^l)}{2} \right] \\[.2cm]
C^{(1)}_{44 \widetilde{\cK}_{6,2} }= +6\sqrt{10}\,  \frac{ (2l+9) }{\sqrt{   (l+7) } } \sqrt{ \frac{ (l+1)}{  (l+8)} }\frac{(l+4)!}{\sqrt{(2l+9)!}} \left[ \frac{(1+(-)^l)}{2} \right] 
\label{3pt_twist6_44}
\eea

We finish with our list of examples by computing $\mathbf{C}^{(1)}_{(44),6,l,[020]}$ 
and $\mathbf{C}^{(1)}_{(35),6,l,[020]}$ in $[020]$, in the even spin sector. We need the results
\bea
		\mathbf{c}^{(0)}_{6,{\rm even},[020]}&=&
				\left[ \begin{array}{cc} 
						+\sqrt{ \frac{l+5}{2l+9}} & +\sqrt{ \frac{l+4}{2l+9}} \ \\[.2cm] 
						+\sqrt{ \frac{l+4}{2l+9}}  & -\sqrt{ \frac{l+5}{2l+9} }\  \end{array}\right]
						\\[.2cm]
\mathbf{L}^{(0)}_{6,{\rm even},[020]}&=&
					{\frac{(l+1)(l+4)(l+5)(l+8) }{30}\left[ \begin{array}{cc} 
												\frac{2}{(l+4)} & 0  \\ 0 & \frac{9}{(l+5)} \end{array} \right]\frac{(l+3)!(l+4)!}{(2l+8)!} }
\eea
This case has $\mu=2$. (The odd spin sector in $[0,2,0]$ has $\mu=1$ and can be studied as in our first example.) 
The CPW data entering the r.h.s of \eqref{boot_V} for $\mathbf{V}_{44,8,l,[020]}$ 
and $\mathbf{V}_{35,8,l,[020]}$, is extracted from different correlators. In the first case 
we consider the CPW data 
\be
\frac{ L^{(1)}_{4424,6,l,[020]}}{\sqrt{ 128}}=
				\frac{84 }{5} \frac{(l+4)!(l+5)!}{(2l+8)!}, \qquad
\frac{ L^{(1)}_{4433,6,l,[020]}}{\sqrt{144}}=
				\frac{9 (80+ 3l(l+9)) }{5} \frac{((l+4)!)^2}{(2l+8)!}.
\ee
In the second case,  
\be
\frac{L^{(1)}_{2435,6,l,[020]}}{ \sqrt{120} }=
					\frac{7(32+l(l+9)) }{5 } \frac{(l+3)!(l+5)!}{(2l+8)!}, \quad
\frac{ L^{(1)}_{3335,6,l,[020]} }{ \sqrt{135} }=
					\frac{52}{15} \frac{(l+4)!(l+5)!}{(2l+8)!}.
\ee
For convenience we again refer to the operators labelled by $R_{6,{\rm even\, spin},[020]}$ 
with $\mathcal{\cK}_{6,1}$ and $\mathcal{\cK}_{6,2}$. Then, by using \eqref{boot_V} we obtain the results
\bea\label{C144twist6020}
C^{(1)}_{44 \mathcal{\cK}_{6,1} }= -6\sqrt{ \frac{6}{5} } \, \frac{  (9l+16) }{ \sqrt{ (l+5) } }\sqrt{ \frac{ (l+8)}{  (l+1)} }\frac{(l+4)!}{\sqrt{(2l+9)!}}\left[ \frac{(1+(-)^l)}{2} \right] \\[.2cm]
C^{(1)}_{44 \mathcal{\cK}_{6,2} }= +6\sqrt{ \frac{6}{5} }\, \frac{ (9l+65)  }{ \sqrt{  (l+4) } } \sqrt{ \frac{ (l+1)}{  (l+8)} }\frac{(l+4)!}{\sqrt{(2l+9)!}} \left[ \frac{(1+(-)^l)}{2} \right] 
\eea
and
\bea
C^{(1)}_{35 \mathcal{\cK}_{6,1} }= 21\sqrt{ 2 } \,  \frac{ (l+7) }{ \sqrt{ (l+4) } }\sqrt{ \frac{ (l+8)}{  (l+1)} }\sqrt{\frac{(l+3)!(l+5)!}{(2l+9)!}}\left[ \frac{(1+(-)^l)}{2} \right] \\[.2cm]
C^{(1)}_{35 \mathcal{\cK}_{6,2} }= 21\sqrt{ 2 }\,  \frac{ (l+2) }{ \sqrt{ (l+5) } } \sqrt{ \frac{ (l+1)}{  (l+8)} }\sqrt{\frac{(l+3)!(l+5)!}{(2l+9)!}}\left[ \frac{(1+(-)^l)}{2} \right] 
\eea


\section{Spin structure of $M^{(2)}$ and $L^{(2)}$} \label{App_structure_spin}


Given a pair of external charges $(p_1p_2)(p_3p_4)$ with $p_{43}\ge p_{21}$, consider a twist $\tau$
such that $\tau$ belongs to the window or below the window. Then,
the SCPW coefficients we extract from $M^{(2)}$ or $L^{(2)}$ have the form
\be\label{schema}
\sum_{(pq)\in R_{\tau,[aba]}} \frac{ \langle p_1 p_2pq \rangle_{conn.} \langle pq p_3p_4 \rangle_{conn}}{ \langle pq pq \rangle_{disc.} }
\ee
where $\langle \dots \rangle_{conn.}$ refers to $M^{(1)}$ or $L^{(1)}$ depending on the situation. 
The form of \eqref{schema} is actually too schematic, because it assumes $q-p\ge p_{21}$ and $p_{43}\ge q-p$, 
in the  summation over $pq\in R_{\tau,[aba]}$, and this is not always the case. 

In order to have better control over the summation, it is useful to visualize it geometrically. Consider first $M^{(2)}$. 
In the plane $(n,m)$, draw the two lines $p_1+p_2=n+m$ and $p_3+p_4=n+m$, and the rectangle $R_{\tau,[aba]}$.  
Since we are considering a $\tau$ in the window, a pair of external charges sits inside $R_{\tau,[aba]}$.

\be
\begin{tikzpicture}[scale=.54]
\def\prop{.5}

\def\shifthor{0}
\def\pttrE{(-\prop*1-\shifthor,\prop*7)}

\def\ptquattrO{(\prop*3-\shifthor,\prop*3)}
%

\draw[red] 					 (\prop*2-\shifthor,\prop*10) --	(\prop*6-\shifthor,\prop*6);
\draw[red]						 (-\prop*1-\shifthor,\prop*7)--	(\prop*2-\shifthor,\prop*10);
\draw[red]						 (\prop*3-\shifthor,\prop*3)--	(\prop*6-\shifthor,\prop*6);
\draw[red]						 (-\prop*1-\shifthor,\prop*7) -- 	(\prop*3-\shifthor,\prop*3);

\draw[blue]							 (\prop*2-\shifthor,\prop*4)  --(\prop*8-\shifthor,\prop*10);
\draw[blue]							 (\prop*1-\shifthor,\prop*5)  --(\prop*7-\shifthor,\prop*11);
\draw[blue]							 (\prop*8-\shifthor,\prop*10)  --(\prop*7-\shifthor,\prop*11);

\filldraw   								(\prop*7-\shifthor,\prop*11)  	circle (.07);
\draw   								(\prop*7-\shifthor,\prop*11)  	circle (.2);

%

\draw[-latex,gray, dashed]					(\prop*3-\shifthor,\prop*3) --(\prop*14-\shifthor,\prop*14);

\draw[-latex,gray, dashed]					(\prop*0-\shifthor,\prop*10) --(\prop*7-\shifthor,\prop*3);
\node[scale=.6] at (\prop*13-\shifthor,\prop*2.5)  {$n+m={\rm min}(p_1+p_2,p_3+p_4)$};

\filldraw   								(\prop*4-\shifthor,\prop*6)  	circle (.07);
\draw   								(\prop*4-\shifthor,\prop*6)  	circle (.2);

\draw[-latex,gray, dashed]					(\prop*5-\shifthor,\prop*13) --(\prop*11-\shifthor,\prop*7);
\node[scale=.6] at (\prop*17-\shifthor,\prop*6.5)  {$n+m={\rm max}(p_1+p_2,p_3+p_4)$};

\foreach \indeyc in {-1,0,1,2,3}
\foreach \indexc  in {0,1,2,3}
\filldraw   					 (\prop*\indexc+\prop*\indeyc-\shifthor, \prop*6+\prop*\indexc-\prop*\indeyc)   	circle (.07);

\end{tikzpicture}
\ee

In the figure, $R_{\tau,[aba]}$ is given in red, and each pair $(pq)\in R_{\tau,[aba]}$
is represented by a black dot. The rightmost edge of $R_{\tau,[aba]}$ lies 
on the line $n+m=\tau$. In fact we can foliate  $R_{\tau,[aba]}$ by the lines $n+m=\tau'$ 
for $\tau'=4+2a+b,\ldots \tau$. Running on any such line, the difference $m-n$ increases 
in the direction $+3\pi/4$ and decreases in the direction $-\pi/4$.  
The two pairs of external charges $p_1p_2$ 
and $p_3p_4$ are represented by a dot encircled.

There are at most three cases to be taken into account. For a pair 
$(pq)\in R_{\tau,[aba]}$ belonging to the line $p+q=\tau'$, we can have
\bea
{\rm (I)} \quad p_{21}\leq p_{43} \leq q-p \qquad
{\rm (II)} \quad  p_{21}\leq q-p \leq p_{43}\qquad
{\rm (III)} \quad  q-p \leq p_{21}\leq p_{43}
\eea
These are the three regions in which the blue rectangle divides $R_{\tau,[aba]}$. 

We shall now analyze the spin structure case by case, given that for 
any correlator $\langle q_1q_2q_3q_4\rangle$ we know the common factor, 
\be
 \tfrac{ \left(\frac{2+2l+\tau+q_{43} }{2}\right)!\left(\frac{2+2l+\tau-q_{21}}{2}\right)! }{  (2+2l+\tau)!}\
\ee
and the degrees in spin of all the SCPW involved. 

Let's assume without loss of generality that $p_1+p_2\leq p_3+p_4$.

Configuration (II) is the simplest, giving a contribution of the form 
$\frac{ \langle p_1 p_2pq \rangle_{conn.} \langle pq p_3p_4 \rangle_{conn}}{ \langle pq pq \rangle_{disc.} }$, in which
the factorials in disconnected free theory cancel with those in the numerator. 
The total degree of the numerator is
\bea
&
\overbrace{{\rm min}(p_1+p_2,p+q)-4-(q-p-p_{21}) }^{\langle p_1p_2pq\rangle}+\overbrace{ {\rm min}(p+q,p_3+p_4)-4-(p_{43}-q+p)}^{\langle pq p_3p_4\rangle} \notag\\
&
={\rm min}(p_1+p_2,p+q)-4-(p_{43}-p_{21})+(p+q-4)\equiv e^{\star}\qquad
\eea
We deduce that any configuration of type (II) contributes in the large spin limit as
\be
\frac{ \langle p_1 p_2pq \rangle_{conn.} \langle pq p_3p_4 \rangle_{conn}}{ \langle pq pq \rangle_{disc.} }\rightarrow
 \tfrac{ \left(\frac{2+2l+\tau+p_{43} }{2}\right)!\left(\frac{2+2l+\tau-p_{21}}{2}\right)! }{  (2+2l+\tau)!}  l^{e^{\star}- (p+q-2) }(1+O(1/l) ) 
\ee

Let's consider now a contribution from configurations of type (I), 
i.e. $\frac{ \langle p_1 p_2pq \rangle_{conn.} \langle p_3p_4 pq \rangle_{conn}}{ \langle pq pq \rangle_{disc.} }$.
In this case the factorials come out misaligned. In particular we find
\bea
\tfrac{ 
\left(\frac{2+2l+\tau-p_{21} }{2}\right)!   \left(\frac{2+2l+\tau+q-p }{2}\right)! \left(\frac{2+2l+\tau-p_{43} }{2}\right)! }{  \left(\frac{2+2l+\tau-p+q }{2}\right)! (2+2l+\tau)! }= 
\tfrac{ \left(\frac{2+2l+\tau+p_{43} }{2}\right)!\left(\frac{2+2l+\tau-p_{21}}{2}\right)! }{(2+2l+\tau)!}\left[ 
						\tfrac{\left(\frac{2+2l+\tau+q-p }{2}\right)! \left(\frac{2+2l+\tau-p_{43} }{2}\right)!  }{ \left(\frac{2+2l+\tau-p+q }{2}\right)! \left(\frac{2+2l+\tau+p_{43} }{2}\right)!}\right]
\eea
Notice that the term between square brackets is polynomial of degree $q-p-p_{43}\ge 0$, 
which is positive for type (I) configurations. There are small changes also in the 
computation of the degree of the numerator, i.e.
\bea
&
\overbrace{{\rm min}(p_1+p_2,p+q)-4-(q-p-p_{21}) }^{\langle p_1p_2pq\rangle}+
							\overbrace{ {\rm min}(p+q,p_3+p_4)-4-(q-p-p_{43} )}^{\langle p_3p_4 pq\rangle} \notag\\
&
=e^{\star} +2(p_{43}-q+p)\qquad
\eea
The result is that a configuration (I) in the large spin limit contributes as
\be
\frac{ \langle p_1 p_2pq \rangle_{conn.} \langle p_3p_4 pq \rangle_{conn}}{ \langle pq pq \rangle_{disc.} }\rightarrow
 \tfrac{ \left(\frac{2+2l+\tau+p_{43} }{2}\right)!\left(\frac{2+2l+\tau-p_{21}}{2}\right)! }{  (2+2l+\tau)!}  \frac{ l^{e^{\star}- (p+q-2) } }{ l^{q+p-p_{43} }}(1+O(1/l) ) 
\ee

The case of configurations of type (III) mirrors the case of configurations of type (I). 
Repeating the previous derivation with minor modifications we obtain 
that in the large spin limit a configurations of type (III) contributes as,
\be
\frac{ \langle pq p_1 p_2 \rangle_{conn.} \langle pq p_3p_4 \rangle_{conn}}{ \langle pq pq \rangle_{disc.} }\rightarrow
 \tfrac{ \left(\frac{2+2l+\tau+p_{43} }{2}\right)!\left(\frac{2+2l+\tau-p_{21}}{2}\right)! }{  (2+2l+\tau)!}  \frac{ l^{e^{\star}- (p+q-2) } }{ l^{p_{21}-q+p }}(1+O(1/l) ) 
\ee

The first conclusion we draw is that when summing over pairs in $(pq)\in R_{\tau,[aba]}$ 
such that  $p+q=\tau'$, the large spin limit is dominated by configurations of type (II), 
because compared to $l^{e^{\star}- (p+q-2)}$ both (I) and (III) are further suppressed.  

In order to decide which line $p+q=\tau'$, in the range $b+2a+4\leq \tau'\leq \tau$, gives the leading 
power law contribution, we further distinguish the two cases: $p_1+p_2\le \tau'$ 
and $4+2a+b\leq\tau'\le p_1+p_2$. 
We find
\be\label{estimates_collection_M2}
\begin{array}{cl}
{\rm if}\ p_1+p_2\le \tau'\ {\rm then}&\qquad e^{\star}-(\tau'-2)=(p_1+p_2)-(p_{43}-p_{21})-6\\[.2cm]
{\rm if}\ \tau'\le p_1+p_2\ {\rm then} &\qquad e^{\star}-(\tau'-2)=\tau'-(p_{43}-p_{21})-6
\end{array}
\ee
Thus, in the expression
\bea
M^{(2)}_{\vec{p};\tau,l,[aba]}&=& 
\tfrac{ \left(\frac{2+2l+\tau+p_{43}}{2}\right)!\left(\frac{2+2l+\tau-p_{21}}{2}\right)! }{  (2+2l+\tau)!}\times
\frac{ num_{M^{(2)}}(l) }{ den_{M^{(2)}}(l) }
\eea
the power law asymptotics of $num_{M^{(2)}}/den_{M^{(2)}}$, 
is $l^{-(p_{43}-p_{21})-6}\big[ l^{(p_1+p_2)}+ \ldots + l^{\tau'} +\ldots \big] $ 
and $l^{p_1+p_2}$ gives the leading contribution.  Since we know independently 
that the greatest denominator factor in the summation over $(pq)\in R_{\tau,[aba]}$ 
is given by disconnected free theory for $p+q=\tau$, we conclude that the degree in spin 
of $num_{M^{(2)}}$ has to be that of the greatest denominator, i.e $(\tau-2)$, plus $(p_1+p_2)-(p_{43}-p_{21})-6$.
This implies the general result 
\bea\label{proof_1}
{\rm degree}~num_{M^{(2)}}(l)&=& (\tau-4)+({\rm min}(p_1+p_2,p_3+p_4)-4-(p_{43}+p_{21}))\notag\\
&=& (\tau-4) + 2({\kappa}_{\vec{p}}-2) +p_{21}
\eea
after using the relation $\kappa_{\vec{p}}=\min\left(\tfrac12(p_1+p_2+p_3-p_4),p_3\right)$.

The details of the computation of $L^{(2)}_{\tau,(p_1p_2);(p_3p_4)}$ 
are very similar to those of $M^{(2)}$, but for an important detail. 
Since the twist $\tau$ belongs to the below window region, the rectangle $R_{\tau,[aba]}$ over which we are summing, is foliated by lines
$n+m=\tau'$ with $\tau'=4+2a+b,\ldots \tau<{\rm min}(p_1+p_2,p_3+p_4)$. Pictorially, the situation is as follows,
\be
\begin{tikzpicture}[scale=.54]
\def\prop{.5}

\def\shifthor{0}
\def\pttrE{(-\prop*1-\shifthor,\prop*7)}

\def\ptquattrO{(\prop*3-\shifthor,\prop*3)}
%

\draw[red] 					 (\prop*0-\shifthor,\prop*8) --	(\prop*4-\shifthor,\prop*4);
\draw[red]						 (-\prop*1-\shifthor,\prop*7)--	(\prop*0-\shifthor,\prop*8);
\draw[red]						 (\prop*3-\shifthor,\prop*3)--	(\prop*4-\shifthor,\prop*4);
\draw[red]						 (-\prop*1-\shifthor,\prop*7) -- 	(\prop*3-\shifthor,\prop*3);

\draw[blue]							 (\prop*2-\shifthor,\prop*4)  --(\prop*8-\shifthor,\prop*10);
\draw[blue]							 (\prop*1-\shifthor,\prop*5)  --(\prop*7-\shifthor,\prop*11);
\draw[blue]							 (\prop*8-\shifthor,\prop*10)  --(\prop*7-\shifthor,\prop*11);

\filldraw   								(\prop*7-\shifthor,\prop*11)  	circle (.07);
\draw   								(\prop*7-\shifthor,\prop*11)  	circle (.2);

%

\draw[-latex,gray, dashed]					(\prop*3-\shifthor,\prop*3) --(\prop*14-\shifthor,\prop*14);

\draw[-latex,gray, dashed]					(\prop*0-\shifthor,\prop*10) --(\prop*7-\shifthor,\prop*3);
\node[scale=.6] at (\prop*13-\shifthor,\prop*2.5)  {$n+m={\rm min}(p_1+p_2,p_3+p_4)$};

\filldraw   								(\prop*4-\shifthor,\prop*6)  	circle (.07);
\draw   								(\prop*4-\shifthor,\prop*6)  	circle (.2);

\draw[-latex,gray, dashed]					(\prop*5-\shifthor,\prop*13) --(\prop*11-\shifthor,\prop*7);
\node[scale=.6] at (\prop*17-\shifthor,\prop*6.5)  {$n+m={\rm max}(p_1+p_2,p_3+p_4)$};

\foreach \indeyc in {-1,0,1,2,3}
\foreach \indexc  in {0,1}
\filldraw   					 (\prop*\indexc+\prop*\indeyc-\shifthor, \prop*6+\prop*\indexc-\prop*\indeyc)   	circle (.07);

\end{tikzpicture}
\ee
Therefore, comparing with \eqref{estimates_collection_M2}, we now get for 
\bea
L^{(2)}_{\vec{p};\tau,l,[aba]}&=& 
\tfrac{ \left(\frac{2+2l+\tau+p_{43}}{2}\right)!\left(\frac{2+2l+\tau-p_{21}}{2}\right)! }{  (2+2l+\tau)!}\times
\frac{ num_{L^{(2)}}(l) }{ den_{L^{(2)}}(l) }
\eea
that  the leading asymptotic of $num_{L^{(2)}}/den_{L ^{(2)}}$ only counts twists in the region $\tau'<p_1+p_2$, as it should,
thus it is has the form $l^{-(p_{43}-p_{21})-6}\big[ l^{\tau} +\ldots \big]$.
We obtain the general result, 
\bea\label{proof_2}
{\rm degree}~num_{L^{(2)}}(l)= 2(\tau-4)-(p_{43}-p_{21}), \qquad
{\rm degree}~den_{L^{(2)}}(l)~=(\tau-2).
\eea

Formulas \eqref{proof_1} and \eqref{proof_2} summarize our proof of 
the spin structure of the SCPW coefficients $M^{(2)}$  and $L^{(2)}$. 

Finally, we distinguish between even and odd $b$. 

In the even $b$ case, even and odd spin cases go separately. Reciprocity implies that in both cases 
$num$ is an even polynomial of the variable $2l+\tau+3$,
\bea\label{test_M2_app}
M^{(2)}_{\vec{p};\tau,l,[aba]}&=& 
\tfrac{ \left(\frac{2+2l+\tau+p_{43}}{2}\right)!\left(\frac{2+2l+\tau-p_{21}}{2}\right)! }{  (2+2l+\tau)!}\times
\frac{ num_{M^{(2)}}(2l+\tau+3) }{ den_{M^{(2)}}(2l+\tau+3) }\\
\label{test_L2_app}
L^{(2)}_{\vec{p};\tau,l,[aba]}&=& 
\tfrac{ \left(\frac{2+2l+\tau+p_{43}}{2}\right)!\left(\frac{2+2l+\tau-p_{21}}{2}\right)! }{  (2+2l+\tau)!}\times 
\frac{ num_{L^{(2)} }(2l+\tau+3) }{ {den}_{L^{(2)}}(2l+\tau+3) }
\eea

In this odd $b$ case, the polynomials $num$ and $den$, for both $M^{(2)}$ and $L^{(2)}$ depend on the spin, whether it is even or odd. 
Picking
\bea\label{test_M2_appp}
M^{(2)}_{\vec{p};\tau,even,[aba]}&=&
\tfrac{ \left(\frac{2+2l+\tau+p_{43}}{2}\right)!\left(\frac{2+2l+\tau-p_{21}}{2}\right)! }{  (2+2l+\tau)!}\times
\frac{ num^{even}_{M^{(2)}}(l) }{ den^{even}_{M^{(2)}}(l) } \\
\label{test_L2_appp}
L^{(2)}_{\vec{p};\tau,even,[aba]}&=& 
\tfrac{ \left(\frac{2+2l+\tau+p_{43}}{2}\right)!\left(\frac{2+2l+\tau-p_{21}}{2}\right)! }{  (2+2l+\tau)!}\times 
\frac{ num_{L^{(2)} }(l) }{ {den}_{L^{(2)}}(l) }
\eea
the SCPW coefficients corresponding to odd spins is then obtained by considering
\be
\begin{array}{ll}
num^{\rm odd}_{M^{(2)}}(l)=num^{\rm even}_{M^{(2)}}(-l-\tau-3)\qquad &  den^{\rm odd}_{M^{(2)}}(l)=den^{\rm even}_{M^{(2)}}(-l-\tau-3),\notag\\[.2cm]
num^{\rm odd}_{L^{(2)}}(l)=num^{\rm even}_{L^{(2)}}(-l-\tau-3)\qquad& den^{\rm odd}_{L^{(2)}}(l)=den^{\rm even}_{L^{(2)}}(-l-\tau-3).
\end{array}
\ee
The factorials do not transform. 

Because of \eqref{odd_b_M1L1} and \eqref{facto_dfree} we can say that the ratio
$num^{even}(l)/den^{even}(l)$ for both $M^{(2)}$ and $L^{(2)}$ will always have a factor of $(l+\tau+2)$, 
thus reducing the degree of numerator and denominator. In particular, once $(l+\tau+2)$ is factored out,
the degree of the auxiliary numerator is down by $-2$ and that of the denominator is down by $-1$.

\subsection{Refining the One-Loop Ansatz with Reciprocity}

We conclude by illustrating the use of reciprocity symmetry in our bootstrap algorithm.

The starting point is the ansatz at the stage in which the leading logs have been matched, and there are no $x=\bar{x}$ poles. 
The idea is that whenever an OPE predictions in and below window is non trivial, rather than immediately input the prediction,
we first impose the correct spin structure of SCPW coefficients on the ansatz, by using 
\eqref{test_M2_app}-\eqref{test_L2_appp}. 
Recall that in \eqref{test_M2_app}-\eqref{test_L2_appp} 
we know the denominators. The corresponding numerators instead will 
be parametrized by a polynomial in $l$, according to the degree and the parity under $l\leftrightarrow -l-\tau-3$, 
as we understood in the previous section. We will leave the parameters in these polynomials free. 
We expect that imposing the spin structure solves a number of free parameters in the ansatz, and 
trade some of them for the new ones in the various numerators. We shall see in this way how much constraining is the spin structure of the SCPW alone.

\paragraph{3333.} As we saw in Section \ref{3333_oneloop}, there are non trivial OPE predictions below window at twist $4$ in all $su(4)$ channels. 
We impose twist $4$ SCPW of the form \eqref{test_L2} 
of the form
\be\label{con_recipro_3333}
\frac{ X_{[000]} }{ (l+1)(l+6) }, \qquad 
\frac{ X_{[101]} }{ (l+2)(l+5) }, \qquad 
\frac{ X_{[020]} }{ (l+3)(l+4) }, 
\ee

In the table below we report the results of imposing \eqref{con_recipro_3333} on the ansatz,
\vspace{-0.4cm}
	\begin{align}
	\begin{array}{|c||c||c|c|}
	\multicolumn{4}{c}{ 
		\begin{array}{l}
		\rule{0pt}{0cm}
		\end{array}
	}\\
	\hline
	\text{rep}                &  \text{twist }  & 
							\rule{0pt}{.7cm}  \substack{ \displaystyle \text{\# free coef. autofit} \\ \rule{0pt}{.35cm} \displaystyle \text{in the ansazt} \\~ }& 
							\substack{ \displaystyle \text{\# free coeff. left over}\\ \rule{0pt}{.35cm}  \displaystyle \text{in numerator} \\ ~ } \\\hline
%
	 [000]                                      	        &   4           			&3    				& 0                            	\\\hline
	 [020]	                                         &   4           			&3  				& 0                       	\\\hline
	 [101]	                                         &   4          				&1    				& 0                   		  \\\hline
%
	\end{array}
		\label{table}
	\end{align}

What happens here is that the constraint from reciprocity is so strong that
at the same time $X_{[000]}$, $X_{[101]}$, $X_{[020]}$ are fixed to their predicted values, 
and furthermore, the ansatz is left with no more free coefficients than the ambiguities, i.e. reciprocity fixes $\mathcal{H}^{(2)}_{3333}$ completely. 
However, we should highlight that the case of twist $4$, and so 3333, is actually very special because
all numerators entering \eqref{con_recipro_3333} have just zero degree in spin.

We shall see in the next example that reciprocity is still powerful but 
the ansatz will not be completely fixed.


\paragraph{4444.} This correlator exemplifies well the general story about reciprocity symmetry.
For given rep $[aba]$ there are non trivial OPE predictions below window, either $\tau=4$ or $\tau=6$.
We will impose that the ansatz has SCPW coefficients of the form 
\be\label{4444_recipro_1}
\begin{array}{c| c |c |c }
 \tau	    & [000] & [101] & [020]  \\ \hline
4  \rule{0pt}{0.6cm} 	&  \frac{ X_{[000]} }{ (l+1)(l+6) }  &  \frac{ X_{[101]} }{ (l+2)(l+5) } &  \frac{ X_{[020]} }{ (l+3)(l+4) } \\[.2cm]\hline
6  \rule{0pt}{0.8cm}	&  \frac{ Y^{(0)}_{[000]}+ Y^{(2)}_{[000]} (l+{\footnotesize\frac{9}{2}})^2 + Y^{(4)}_{[000]} (l+{\footnotesize\frac{9}{2}})^4  }{ (l+1)(l+2)(l+7)(l+8) }
   				 &  \frac{ Y^{(0)}_{[101]}+ Y^{(2)}_{[101]} (l+{\footnotesize\frac{9}{2}})^2 + Y^{(4)}_{[101]} (l+{\footnotesize\frac{9}{2}})^4  }{ (l+1)(l+3)(l+6)(l+8) }
  				  &  \frac{ Y^{(0)}_{[020]}+ Y^{(2)}_{[020]} (l+{\footnotesize\frac{9}{2}})^2 + Y^{(4)}_{[020]} (l+{\footnotesize\frac{9}{2}})^4  }{(l+1)(l+4)(l+5)(l+8) }
\end{array}
\ee
and 
\be\label{4444_recipro_2}
\begin{array}{c| c| c |c }
 \tau	    & [040] & [121] & [202]  \\ \hline
6 \rule{0pt}{0.8cm} &  \frac{ Y^{(0)}_{[040]}+ Y^{(2)}_{[040]} (l+{\footnotesize\frac{9}{2}})^2 + Y^{(4)}_{[040]} (l+{\footnotesize\frac{9}{2}})^4  }{ (l+3)(l+4)(l+5)(l+6) }
    &  \frac{ Y^{(0)}_{[121]}+ Y^{(2)}_{[121]} (l+{\footnotesize\frac{9}{2}})^2 + Y^{(4)}_{[121]} (l+{\footnotesize\frac{9}{2}})^4  }{ (l+2)(l+4)(l+5)(l+7) }
    &  \frac{ Y^{(0)}_{[202]}+ Y^{(2)}_{[202]} (l+{\footnotesize\frac{9}{2}})^2 + Y^{(4)}_{[202]} (l+{\footnotesize\frac{9}{2}})^4  }{(l+2)(l+3)(l+6)(l+7) }
\end{array}
\ee
with unknown coefficients in the numerators.

The actual OPE predictions give particular polynomials
in the various entries of the tables \eqref{4444_recipro_1} and \eqref{4444_recipro_2}. 
Comparing with the results in Section \ref{4444_oneloop}, we see that in some cases the rational functions simplify further. 
However, according to our discussion about 
the spin structure, \eqref{4444_recipro_1} and \eqref{4444_recipro_2} 
are the most general.

The way the ansatz is refined by imposing reciprocity is reported in the table below.
What happens is quite remarkable. 
\vspace{-0.4cm}
	\begin{align}
	\begin{array}{|c||c||c|c|}
	\multicolumn{4}{c}{ 
		\begin{array}{l}
		\rule{0pt}{0cm}
		\end{array}
	}\\
	\hline
	\text{rep}                &  \text{twist}  & 
							\rule{0pt}{.7cm}  \substack{ \displaystyle \text{\# free coef. autofitted} \\ \rule{0pt}{.35cm} \displaystyle \text{in the ansazt} \\~ }& 
							\substack{ \displaystyle \text{\# free coeff. left over}\\ \rule{0pt}{.35cm}  \displaystyle \text{in the numerator} \\ ~ } \\\hline
	 [000]                                       		  &   4            				&4   				& 0                            	\\\hline
	 	                                        		  &   6           				&4    				& 1                            	\\\hline
	[101]  	                                    	  &   4          				&4    				& 0                            	\\\hline
	 	                                    		  &   6          				&3    				& 1                            	\\\hline
	[020]  	                                    	  &   4          				&4    				& 0                            	\\\hline
	 	                                    		  &   6          				&3    				& 1                            	\\\hline	
	[040]  	                                    	  &   6          				&3    				& 1                            	\\\hline
	[121]  	                                    	  &   6          				&1    				& 0                            	\\\hline			
	[202]  	                                    	  &   6          				&2    				& 0                            	\\\hline
%
	\end{array}
		\label{table}
	\end{align}

Firstly, at twist $4$ reciprocity fixes completely the one variable resummation to its OPE prediction.
%
Secondly, at twist $6$, almost all free coefficients in the ansatz are fixed just by the symmetry. 
We can trade the 4 free coefficients left in the ansatz for those in \eqref{4444_recipro_1} and \eqref{4444_recipro_2}. For example, 
$Y^{(0)}_{[000]}$, $Y^{(0)}_{[101]}$, $Y^{(0)}_{[020]}$, $Y^{(0)}_{[040]}$. 
Summarizing, 
there is (only!) a 4 free-coefficients ansatz (with ambiguities) which can potentially describe $\mathcal{H}_{4444}^{(2)}$. 
By matching the value of these remaining free coefficients in the ansatz to the OPE predictions below window, we obtain $\mathcal{H}_{4444}^{(2)}$.


\end{document}